\documentclass[aps,prb,twocolumn,showpacs]{revtex4-1}
\usepackage{mathptmx}

\usepackage[T1]{fontenc}
\usepackage[latin9]{inputenc}
\usepackage{float}
\usepackage{amsmath}
\usepackage{amssymb}
\usepackage{graphicx}
\usepackage{esint}

\makeatletter
\@ifundefined{textcolor}{}
{%
 \definecolor{BLACK}{gray}{0}
 \definecolor{WHITE}{gray}{1}
 \definecolor{RED}{rgb}{1,0,0}
 \definecolor{GREEN}{rgb}{0,1,0}
 \definecolor{BLUE}{rgb}{0,0,1}
 \definecolor{CYAN}{cmyk}{1,0,0,0}
 \definecolor{MAGENTA}{cmyk}{0,1,0,0}
 \definecolor{YELLOW}{cmyk}{0,0,1,0}
}

\pdfoutput=1 

\usepackage{multirow}\usepackage{dcolumn}\usepackage{bm}\usepackage{ulem}\usepackage{amsfonts}\usepackage{color}\usepackage{colordvi}\usepackage{dcolumn}\usepackage{subfigure}

\usepackage[colorlinks,bookmarks=false,citecolor=blue,linkcolor=blue,urlcolor=blue,hyperfootnotes=true]{hyperref}

\makeatother

\begin{document}

\title{Pseudo-gap, charge order and pairing density wave
at the hot spots in cuprate superconductors }

\date{December 17, 2014}

\author{C. Pépin$^{1}$, V. S. de Carvalho$^{1,2}$, T. Kloss$^{3}$ and
X. Montiel$^{3}$}

\affiliation{$^{1}$IPhT, L'Orme des Merisiers, CEA-Saclay, 91191 Gif-sur-Yvette,
France }

\affiliation{$^{2}$Instituto de F\'{i}sica, Universidade Federal de Goiás, 74.001-970,
Goiânia-GO, Brazil}

\affiliation{$^{3}$International Institute of Physics, UFRN, Av. Odilon Gomes
de Lima 1722, 59078-400 Natal, Brazil}
\begin{abstract}
We address the timely issue of the presence of charge ordering at
the hot-spots in the pseudo-gap phase of cuprate superconductors in
the context of an emergent SU(2)-symmetry which relates the charge
and pairing sectors. Performing the Hubbard-Stratonovich decoupling
such that the free energy stays always real and physically meaningful
we exhibit three solutions of the spin-fermion model at the hot spots.
A careful examination of their stability and free energy shows that,
at low temperature, the system tends towards a co-existence of charge
density wave (CDW) and the composite order parameter made of diagonal
quadrupolar density wave and pairing fluctuations of Ref.\ {[}Nat.\ Phys.\ \textbf{9},
1745 (2013){]}.The CDW is sensitive to the shape of the Fermi surface
in contrast to the diagonal quadrupolar order, which is immune to
it. SU(2) symmetry within the pseudo-gap phase also
applies to the CDW state, which therefore admits a pairing density
pave counterpart breaking time reversal symmetry.
\end{abstract}
\maketitle

\section{Introduction}

The discovery of charge order (CDW) in the Pseudo-Gap (PG) phase in
non La-based Cuprate superconductors reached a steadily growing interest
in recent years. Initially observed by STM in Bi2212 \cite{Hoffman02}
and later in Bi2201 \cite{Hanaguri04,McElroy05,McElroy06,Kohsaka07,Wise08,He14,daSilvaNeto14},
the CDW phase was also observed in YBCO by quantum oscillation \cite{DoironLeyraud07,LeBoeuf07,Sebastian10,Laliberte11,Sebastian12}.
NMR \cite{Wu11,Wu13,Wu14} and sound experiments \cite{LeBoeuf13,Shekhter13}
confirmed the presence of CDW phase in YBCO while X-Ray bulk spectroscopy
\cite{LeTacon11,Chang12,Ghiringhelli12,Blackburn13b,Blanco-Canosa13,LeTacon14,Blanco-Canosa14}
has clearly characterized its checkerboard ordering with wave vector
along the x and y axes $Q_{x}=Q_{y}=0.33$ \cite{Harrison11,Harrison14}.
Deeper analysis concluded that CDW ordering wave vector is located
at the tips of the Fermi surface at the vicinity of the hot spots
\cite{Blanco-Canosa14,Vishik12,Comin14}.

An additional phase transition to a checkerboard CDW ordered phase
is observed at $T_{CDW}$ \cite{Xia08} below the PG line at $T^{\star}$
\cite{Alloul89,Warren89}, $T_{CDW}<T^{\star}$. Note that $T^{\star}$
coincides with the observation of loop current detected by neutron
\cite{Fauque06,Varma06,Baledent11,Bourges11,Sidis13}. In temperature-doping
phase diagram, the $T_{CDW}$-line has the typical form of a dome
\cite{Grissonnanche14} and its magnitude is compound dependent \cite{Vishik12,Comin14,Alloul14,Tabis14}
whereas the PG-line is rather universal \cite{Alloul14}.

One of the most difficult challenge of the field is to understand
how the recently observed CDW orders interferes with AF fluctuations
and whether it participates or not to the formation of the PG phase.
Although some alternative scenarios involving stronger Coulomb interactions
have been considered\cite{Chakravarty01,Kivelson03,KimLawler08,Allais14b,Allais14c,Atkinson14},
the proximity of the CDW ordering wave vector to the hot-spots is
a strong incentive to consider the Spin-Fermion (SF) theory, which
produced the most singular behavior at the hot-spots\cite{Millis93,abanov03,Abanov04}.
We follow here this route, keeping in mind that the SF model has been
the subject of intense recent scrutiny\cite{abanov03,Metlitski10,Metlitski10b,Efetov13,Sachdev13,Wang14,Einenkel14,Hayward14,Chowdhury14}. 

In the last year, it has become an increasing challenge
to the community to explain how to get an emergent CDW with the right
orientation of the wave vector. Indeed, when simple Random Phase Approximations
(RPA) are performed in this system, either actually in the charge
or spin channel, a maximum of intensity is obtained at $Q=\left(\mathbf{Q_{x}},\mathbf{Q_{y}}\right)$
wave vector located on the diagonal, while no peak is experimentally
observed in this direction. Some attempts to address this question
have considered that a pre-formed pseudo gap state consisting of short
range AF fluctuations or of a spin liquid do gap out the anti-nodal
part of the Fermi surface, letting behind some Fermi arcs\cite{Atkinson14,Chowdhury:2014ta}.
In the context of the three bands model, when the hopping to Cu $4s$-orbitals
is included, it is possible to rotate the wave vectors of the RPA
charge susceptibility align them with the crystalline axes. The ordering
wave vectors are situated at the ``tip of the Fermi arc''. Those
models though suffer from the consideration that whereas the ``tip
of the arcs'' is moving with temperature, the observed ordering wave
vector of the modulation is non dispersing\cite{Wise08}. Another
approach very similar to the one presented here considers directly
the bare electron Fermi surface for these models, and considers that
a CDW with the correct wave vector can be the low temperature order
of a pre-formed bound state breaking time reversal symmetry\cite{Wang14}.

In this paper we stay in the broad context of emergent
symmetries, where the d-wave SC state of high temperature superconductors
rotates to other symmetry sectors. The underlying idea is the old
idea of degeneracy of energy levels in quantum physics. When two energy
levels are degenerate, it can be accidental fact, but it can as well
signal that the two energy levels are related by a common symmetry.
This notion of emergent symmetry has been used in the past for cuprates
with the SO(5)-theory relating d-wave superconductivity to the
magnetic sector\cite{Eder:1998bs}\cite{Demler04}. The SU(2)-group
was used as well in relating the d-wave SC to the $\pi$-flux phases,
within gauge theoretic treatment of the t-J model\cite{Lee06}. Here
we use the same SU(2) symmetry group, rotating the d-wave SC order
to the charge sector. In Refs.\ [\onlinecite{Metlitski10a,Metlitski10b,Efetov13}],
this symmetry has been shown to be present in the Eight Hot Spots
(EHS) model where the Fermi velocity is linearized at the hot spots.
In Ref.\ [\onlinecite{Efetov13}], a PG state was identified as the primary
instability of the AFM QCP within this model.\textcolor{magenta}{{}
}In particular, it has been shown that \textcolor{blue}{the} underlying
SU(2) rotation produces a composite order parameter with a Quadrupolar
\textit{d}-wave component in the charge sector and preformed pairs
in the SC sector (QDW/SC). This short range order is a good candidate
for the PG gap phase since it breaks translational symmetry and is
thus able to produce a gap in the spectral functions. The ordering
wave vector though lies on the diagonal while experiments report charge
order at vectors $\mathbf{Q}_{x}$ and $\mathbf{Q}_{y}$ parallel
to the axes of the compounds. 

The goal of this paper is to examine whether a CDW
with wave vectors aligned with the crystalline axes can be stabilized
in the context of the EHS model. Although the EHS model is a very
idealized starting point for describing the physics of cuprate superconductors,
it has the merit to produce a microscopic model when the SU(2) symmetry
is verified at all energies. Curvature effects break the symmetry
in favor of SC pairing fluctuations while magnetic field breaks it
in favor of CDW charge order. Moreover, the SU(2) symmetry is realized
at only one point in the Fermi surface : at the hot spots. The understanding
of how SU(2) symmetry breaks when one goes from a description of hot
spots to a description of hot regions deserves a more detailed later
study. Here we focus still on the EHS model, with additional short-range
AF interaction. 

We introduce an original Hubbard-Stratonovich (HS) decoupling which
enables us to consider the CDW order (with its pairing
counterpart), and the QDW/SC order on the same footing. We find generically,
that pure QDW/SC order is stable while the pure CDW/PDW
solution is unstable.This
solution is in agreement with our previous findings, as well as with
several recent studies, which conclude that within the EHS model where
interactions are mediated by AFM paramagnons, the only instabilities
are pure d-wave diagonal orders and d-wave pairing states. In order
to get any new ( but weaker) instability we have to introduce an external
perturbation, which we do in the present work, in the form of short
range AF interaction which breaks orthorhombicity. We show then that
at lower temperature a third solution emerges in which QDW/SC and
CDW/PDW orders Co-Exist (CE solution). The transition
towards coexistence is found to be weakly first order. Our conclusion
is that the SF model supports the emergence of CDW with wave vectors
parallel to the axes, but in coexistence with a larger instability,
the QDW/SC order, which is a good candidate for the PG. The magnitude
of the CDW order depends on the details of the Fermi surface topology
while the QDW/SC order is insensitive to the shape of the Fermi surface.
Moreover, the underlying SU(2) symmetry of the PG
state enforces degenerate PDW counterpart to the CDW order. Since
the PDW lies at the hot-spot wave vector, it breaks TR symmetry, which
gives a natural explanation to observation that a Kerr signal has
been observed at the incipient CDW ordering transition.

\section{Method}

\subsection{Model}

Our starting point is the spin fermion model which can be described
through the Lagrangian $L=L_{\psi}+L_{\phi}$ where 
\begin{subequations}
\begin{align}
L_{\chi} & =\chi_{\mathbf{k}\sigma}^{*}\left(\partial_{\tau}+\epsilon_{k}+g\phi\sigma\right)\chi_{\mathbf{k}\sigma}\ ,\label{eq:1}\\
L_{\phi} & =\frac{1}{2}\phi D^{-1}\phi+\frac{u}{2}\left(\phi^{2}\right)^{2}\ .\label{eq:2}
\end{align}
\end{subequations} $L_{\chi}$ is the fermion Lagrangian
representing electrons with dispersion $\epsilon_{k}$ that are interacting
with a bosonic spin excitation $\phi$ described in
Lagrangian $L_{\phi}$ with the interacting magnitude
$g$. The spin-wave boson $\phi$ propagates through 
\begin{equation}
D^{-1}\left(\omega,\mathbf{q}\right)=\frac{\omega^{2}}{v_{s}^{2}}+\left(\mathbf{q-Q}\right)^{2}+m_{a} , \label{eq:3}
\end{equation}
where $v_{s}$ is the spin-wave velocity, $\mathbf{Q}$
is the AF ordering wave vector, and $m_{a}$ is the spin-wave boson
mass, which vanishes at the QCP.

We add to the original Lagrangian (\ref{eq:1}) and
(\ref{eq:2}) a small perturbation in the form of a short range Nearest
Neighbor (NN) super-exchange interaction 
\begin{align}
L_{C} & =\sum_{\left\langle i,j\right\rangle \sigma}\bar{J}_{i,j}\chi_{i\sigma}^{\dagger}\chi_{j-\sigma}\chi_{j-\sigma}^{\dagger}\chi_{i\sigma}
\end{align}
 where the notation $\left\langle i,j\right\rangle $ stands for NN
sites. In order to simplify the study we take a bi-partite modulation
of $\bar{J}_{i,j}$ which when we Fourier transforms gives 
\begin{align}
L_{C} & =2\sum_{\mathbf{k,}\mathbf{k^{\prime}},\sigma}\bar{J}_{\mathbf{k},\mathbf{k^{\prime}}}\chi_{\mathbf{k}\sigma}^{\dagger}\chi_{\mathbf{k+Q}-\sigma}\chi_{\mathbf{k^{\prime}+Q}-\sigma}^{\dagger}\chi_{\mathbf{k}^{\prime}\sigma}\label{eq:4}\\
\mbox{with } & \ \bar{J}_{\mathbf{k},\mathbf{k^{\prime}}}=\bar{J}_{x}\cos\left(k_{x}-k{}_{x}^{\prime}\right)+\bar{J}_{y}\cos\left(k_{y}-k_{y}^{\prime}\right)\ .\nonumber 
\end{align}
Typically we choose $\bar{J}_{x}\neq\bar{J}_{y}$
which breaks the $C_{4}$-symmetry of the lattice, which we relate
to a slight breaking of the orthorhombicity in real materials. $\mathbf{Q}=\left(\pi,\pi\right)$
is the AF modulation wave vector. 

\begin{figure}
\includegraphics[scale=0.4]{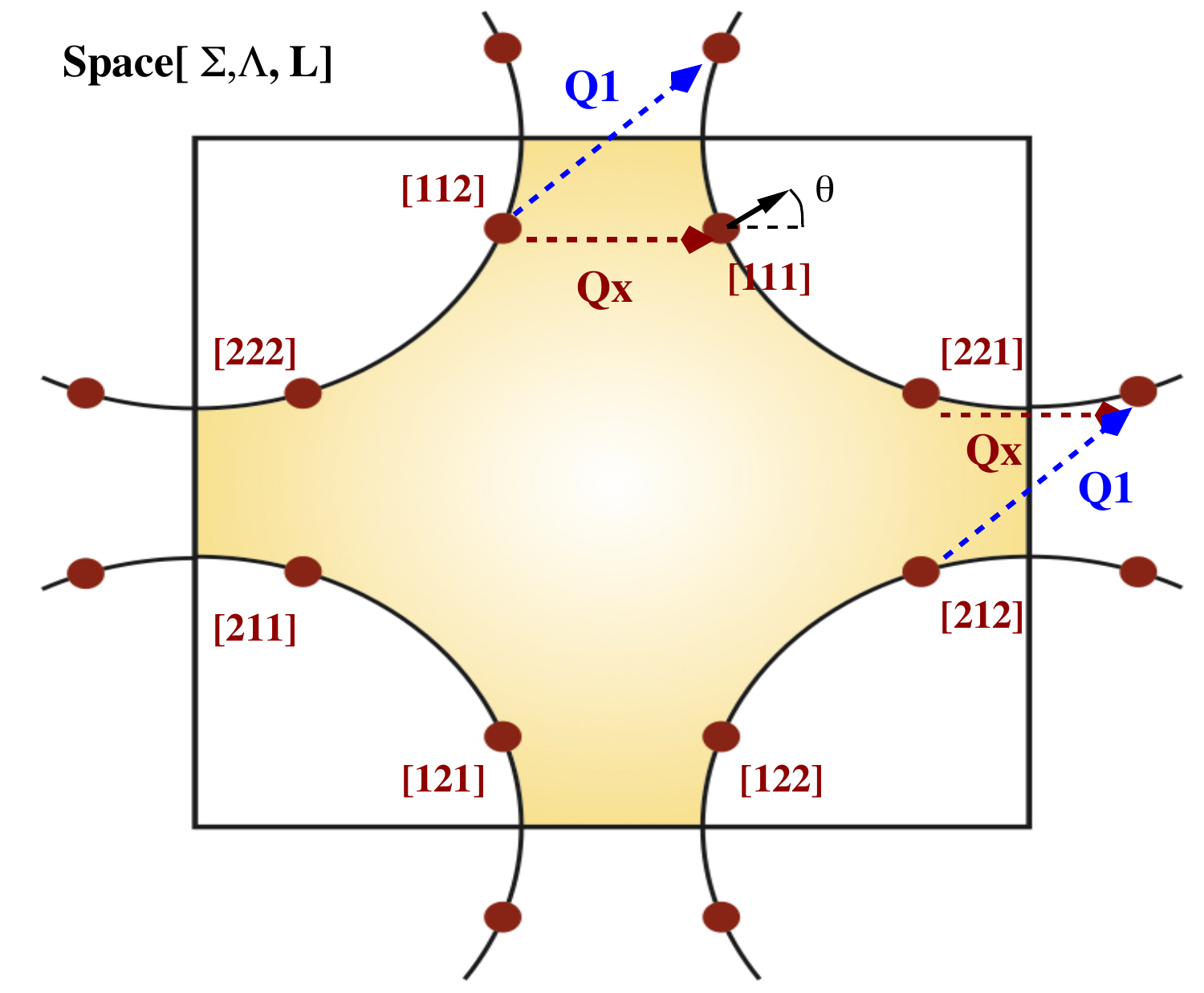} \vspace{-3ex}
 \caption{\label{Fig1} (Color online) Schematic Fermi surface of cuprate superconductors
in the first Brillouin zone of a square lattice. The eight hot spots
are depicted with red dots labeled with the $\left(\Sigma,\Lambda,L\right)$-space.
The angle $\theta$ is defined between the velocity vector at the
{[}111{]}-hotspot with the $x$-axis. The experimentally observed
CDW wave vector $\mathbf{Q}_{x}$ connects the $L_{2}$-sector, whereas
the diagonal QDW/SC vector $Q_{1}$ does not.}
\end{figure}

We further simplify the problem by restricting and
linearize the fermion dispersion represented in Fig.\ \ref{Fig1}
to the eight hot spots, which are the only points with critical scattering
through the paramagnons at $T=0$. Through such a
transformation the model is essentially projected onto the EHS model.
The hotspots are labeled in the $32\times32$ symmetry
space $\left(\Sigma,\Lambda,L,\tau,\sigma\right)$ in
which every sub-space is described by a Pauli Matrix.
To simplify the notation, we do not write in the paper the occurrence
of the identity Pauli matrices in the formulas. The first three tensor
products $\left(\Sigma,\Lambda,L\right)$ describe the symmetries
of the Brillouin Zone (BZ), with respectively $\Sigma$ for the permutation
of two hot spots inside a pair of hot spots, $\Lambda$ the permutation
of two pairs of hot spots inside a quartet and $L$ for the permutation
of the two quartets of hot spots, as depicted in
Fig.\ \ref{Fig1}. Finally, $\tau$ stands for the particle-hole
and $\sigma$ for the spin space.  The Fermi velocity is further linearized and written into
the matrix form $\hat{\epsilon}_{k}=v_{x}\hat{x}+v_{y}\hat{y}$, with
$\hat{x}=\left(k_{x}P_{\Sigma_{x}}-k_{y}P_{\Sigma_{y}}\right)\Lambda_{3}L_{3}$,
$\hat{y}=\left(k_{y}P_{\Sigma_{x}}-k_{x}P_{\Sigma_{y}}\right)\Lambda_{3}$,
$v_{x}=v\cos\theta$, $v_{y}=v\sin\theta$, with $\theta$ the angle
of the Fermi velocity with the $x$-axis (see Fig.\ \ref{Fig1}).
$P_{\Sigma_{x}}=\left(1+\Sigma_{3}\right)/2$ and $P_{\Sigma_{y}}=\left(1-\Sigma_{3}\right)/2$
are projection operators onto the first and second component of the
$\Sigma$-space. 

We naturally follow the notations of Ref.\ [\onlinecite{Efetov13}]
and introduce a $32\times32$ fermion vector with in the particle-hole
$\tau$-space 
\begin{align}
\psi=\frac{1}{\sqrt{2}}\left(\begin{array}{c}
\chi^{*}\\
i\sigma_{2}\chi
\end{array}\right)_{\tau}\ , & \psi^{\dagger}=\frac{1}{\sqrt{2}}\left(\begin{array}{cc}
-\chi^{t} & -i\sigma_{2}\chi^{\dagger}\end{array}\right)\ ,
\end{align}
 where $\chi^{t}$ is the simple transposition, and $\sigma_{2}=\left(\begin{array}{cc}
0 & -i\\
i & 0
\end{array}\right)_{\sigma}$ is the Pauli Matrix in the spin sector $\sigma$. A new conjugation
is further introduced with 
\begin{equation}
\bar{\psi}=\left(C\psi\right)^{t},\mbox{ with }C=\left(\begin{array}{cc}
0 & i\sigma_{2}\\
-i\sigma_{2} & 0
\end{array}\right)_{\tau}=-\tau_{2}\sigma_{2},
\end{equation}
with $\tau_{2}$ the Pauli Matrix in $\tau$-space.
Note that in absence of magnetic
field and of spin-flip phenomena, the degeneracy in the spin space
$\sigma$ afford us to focus only on the reduced $16\times16$ ($\Sigma,\Lambda,L,\tau$)
space.
We refer the reader to the SI of Ref.\ [\onlinecite{Efetov13}] for details.

Using Eq.\ (\ref{eq:2}) the spin-boson field $\phi$
is then formally integrated out of the partition function to get the
following effective partition function $Z=\int d\psi\exp\left(-S_{0}-S_{int}\right)$
with (here $x=\left(\mathbf{r},\tau\right)$ and the trace Tr is taken
over the $16\times16$ matrix space) \begin{subequations} 
\begin{align}
S_{0} & =\int dxdx^{\prime}\bar{\psi}_{x}g_{0}^{-1}\psi_{x'},\\
S_{int} & =\int dxdx^{\prime}\textrm{Tr}\left(J_{x-x'}\psi_{x}\bar{\psi}_{x^{\prime}}\Sigma_{1}\psi_{x^{\prime}}\bar{\psi_{x}}\Sigma_{1}\right),\label{eq:6}\\
& \mbox{ with }  \ J_{x-x'}=\frac{3g^{2}}{2}D_{x-x'}\label{eq:8-1}
\end{align}
\label{eq4} \end{subequations} The constant $J_{x-x^{\prime}}$
is chosen so that $C_{4}$ orthorhombic symmetry is broken ($J_{x}\neq J_{y}$,
Eq.\ (\ref{eq4})), which then allows us to make a distinction between
the two order parameters that we want to study. Note the factor $\Sigma_{1}$
in Eq.\ (\ref{eq:6}) which comes from the translation by the AFM wave
vector $\mathbf{Q=\left(\pi,\pi\right)}$ originating from the paramagnon
propagator in Eq.\ (\ref{eq:3}). There are many ways to decouple
the action of Eq.\ (\ref{eq4}) into the physically relevant hydro-dynamic
modes of the system. The issue we face here is that the various order
parameters that we want to study simultaneously have different symmetries.

\subsection{Order parameters}

The first order parameter that we consider is our
candidate for the PG phase, that we already considered
in few previous works\cite{Efetov13,Einenkel14,Meier13,Meier14}.
It is a composite order parameter constituted by a pairing d-wave
sector and a quadrupolar density wave sectors (QDW/SC). It writes
\cite{Efetov13} \begin{subequations} 
\begin{align}
{\hat{B}_{1}} & =B_{1}(\epsilon_{n},\mathbf{k}){\hat{U}}\quad\text{with}\\
{\hat{U}}_{\Lambda}= & i\left(\begin{array}{cc}
0 & {\hat{u}}_{\tau}\\
-{\hat{u}}_{\tau}^{\dagger} & 0
\end{array}\right)_{\Lambda},\qquad\hat{u}_{\tau}=\left(\begin{array}{cc}
\Delta_{-} & \Delta_{+}\\
-\Delta_{+}^{*} & \Delta_{-}^{*}
\end{array}\right)_{\tau} .
\label{eq:10}
\end{align}
\end{subequations} Herein, $\Delta_{-}$ is the QDW component and
$\Delta_{+}$ is the d-wave SC component of the order parameter.
They are defined as 
\begin{equation}
\Delta_{-}=\left\langle \psi_{\mathbf{k,\sigma}}^{\dagger}\psi_{\mathbf{k+Q_{1,2},\sigma}}\right\rangle ,\qquad\Delta_{+}=\left\langle \psi_{\mathbf{k},\sigma}\psi_{\mathbf{-k},-\sigma}\right\rangle ,\label{eq:11}
\end{equation}
with $\mathbf{Q}_{1,2}=\left(\mathbf{Q}_{x}\pm\mathbf{Q}_{y}\right)/2$.
Note that $\Delta_{-}$ can be re-written as $\Delta_{-}=\left\langle \psi_{\mathbf{k,\sigma}}^{\dagger}\psi_{\mathbf{-k,\sigma}}\right\rangle $
since at the hot spots, $\mathbf{Q}_{1,2}=2\mathbf{k_{hs}}$. In this
model where the Fermi surface is restricted to eight hot spots, the
QDW component describing the charge order is organized along the diagonal
vector, as a consequence of its pure d-wave symmetry.
The SU(2) symmetry of the model is enforced, in
this matrix representation, by the constraint $|\Delta_{+}|^{2}+|\Delta_{-}|^{2}=1$.
This constraint leads to reduce the two self-consistent mean-field
gap equations for $\Delta_{+}$ and $\Delta_{-}$
to only one self-consistent equation for the modulus
of the order parameter $B_{1}$. The study of the
temperature-doping phase diagram of the QDW/SC in the EHS model is
given in \cite{Efetov13}.

In addition to the QDW/SC order parameter, we
study the possibility for a CDW order parameter with
wave vector parallel to the $x,y$-axes defined as 
\begin{equation}
B_{2}=\left\langle \psi_{\mathbf{k,\sigma}}^{\dagger}\psi_{\mathbf{k+Q}_{x(y)},\sigma}\right\rangle .\label{eq:12}
\end{equation}
For reasons of simplicity we will consider exclusively
the wave vector $\mathbf{Q}_{x}$, which describes a ``stripe''
order parameter, for the study of a less pronounced nematicity, and comparison
with the checkerboard solution will be given in a forthcoming publication.
A CDW organized along the $\mathbf{Q}_{x}$ involves some off-diagonal
component in the L-sector. The CDW order parameter
writes in the ($\Sigma,\Lambda,L,\tau$) space as : 
\begin{equation}
\hat{B}_{2}=\left(\begin{array}{cc}
B_{2x} & 0\\
0 & B_{2y}
\end{array}\right)_{\Sigma}L_{2}\otimes\Lambda_{3}\otimes\hat{u}_{\tau}.\label{eq:13}
\end{equation}
 The notations $B_{2x}$ and $B_{2y}$ stand for
different amplitudes are allowed in the $P_{\Sigma_{x}}$ and $P_{\Sigma_{y}}$
sectors (corresponding respectively to the $\left(0,\pi\right)$
and $\left(\pi,0\right)$ sectors). The pure d-wave
case corresponds to $B_{2x}=-B_{2y}$ while the generic case corresponds
to a mixture of \textit{s}- and \textit{d}-wave symmetry around the
Fermi surface\cite{Fujita14}. Without any loss of
generality, the CDW matrix can handle an extra $\hat{u}_{\tau}$ -structure
in the particle-hole space, with the same definition as in Eq.\ (\ref{eq:10}).
This amounts rotate the CDW sector with the same SU(2) rotations as
for the QDW/SC cases, producing finite q-pairing in the particle-particle
channel, or PDW-pairing. This type of pairing has been suggested by
various groups in recent studies of the PG state of the cuprates \cite{Lee:2014ka}.
\begin{equation}
\Delta'_{-}=\left\langle \psi_{\mathbf{k,\sigma}}^{\dagger}\psi_{\mathbf{k+Q}_{x},\sigma}\right\rangle ,\qquad\Delta'_{+}=\left\langle \psi_{\mathbf{k},\sigma}\psi_{\mathbf{k+Q}_{x},-\sigma}\right\rangle .\label{eq:14}
\end{equation}
When solving self-consistently for $\hat{B}_{1}$ and $\hat{B}_{2}$,
the same small $\mathbf{k}$-space anisotropy must
be tolerated for $\hat{B}_{1}$ leading to 
\begin{equation}
\hat{B}_{1}=\left(\begin{array}{cc}
B_{1x} & 0\\
0 & B_{1y}
\end{array}\right)_{\Sigma}{\hat{U}}_{\Lambda}.
\end{equation}

\subsection{Hubbard Stratonovich (HS) decoupling}

The problem we face when decoupling Eq.\ (\ref{eq:7})
is that we would like to decouple the quadratic form naturally in
a symmetric way between one factor and its conjugate, but in order
to describe on the same footing the two order parameters $\hat{B}_{1}$
and $\hat{B}_{2}$, we must find a HS transformations which allows
an asymmetric decoupling in the $\bar{\psi}_{x}\psi_{x'}$ and $\Sigma_{1}\bar{\psi}_{x}\Sigma_{1}\psi_{x'}$
factors. In all generality, the HS decoupling can be written \begin{subequations}
\begin{align}
 & Z=\frac{\int{\cal D\left[\psi\right]}{\cal D}\left[Q_{a},Q_{b}\right]\ I[Q_{a,}Q_{b},\psi]\ \exp\left[-S_{0}-S_{int}\right]}{\int{\cal D}\left[Q_{a},Q_{b}\right]\ I[Q_{a,}Q_{b},0]}\ ,\label{eq5}\\
 & I\left[Q_{a},Q_{b,}\psi\right]=\nonumber \\
 & \qquad\exp\left[-J_{\mathbf{x-x'}}^{-1}\left(Q_{a}-i J_{\mathbf{x-x'}}\psi_{x}\bar{\psi}_{x'}\right)\left(Q_{b}-iJ_{\mathbf{x-x'}} \Sigma_{1}\psi_{x'}\bar{\psi}_{x}\Sigma_{1}\right)\right].
\end{align}
\end{subequations} In Eq.\ (\ref{eq5}) we must ensure that the
quadratic form in the exponential is always positive definite and
that the resulting free energy is real for any field $Q_{a,b}$. We
choose $Q_{a}=\Sigma_{1}Q^{\dagger}\Sigma_{1}$ and $Q_{b}=Q$. This
relation defines a new charge conjugation $Q_{a}=\bar{Q}$,
with 
\begin{equation}
\bar{Q}=\Sigma_{1}Q^{\dagger}\Sigma_{1}.\label{eg6}
\end{equation}
 In the remainder of the paper, this conjugation will be used instead
of the original charge conjugation. The notion of positive definite
form inside the path integral are thus defined with respect to the
conjugation Eq.\ (\ref{eg6}).

\subsection{Free energy}

After the HS decomposition and integrating out the
fermions, we get the following free energy :
\begin{align}
\frac{\Delta F}{T} & =\int \text{Tr} J^{-1}\left(Q_{a}Q_{b}\right)-\frac{1}{2}\int \text{Tr} \ln\left(g_{0}^{-1}+i\left(Q_{a}+\Sigma_{1}Q_{b}\Sigma_{1}\right)\right)\  \nonumber \\
 & =\int dxdx' \text{Tr} J_{x-x'}^{-1}\left(\bar{Q}_{x,x'}Q_{x',x}\right) \nonumber \\
 & -\frac{1}{2}\int dxdx' \text{Tr} \ln\left(g_{0}^{-1}+i\left(\bar{Q}_{x,x'}+\Sigma_{1}Q_{x,x'}\Sigma_{1}\right)\right)\ ;  \label{free1}
\end{align}
 $g_{0}^{-1}=i\epsilon+i\hat{\mathbf{v}}\cdot\mathbf{\nabla_{x}},$
and $Q=i\hat{B}_{1}+\hat{B}_{2}$. 
Projecting Eq.\  (\ref{free1}) onto the $\Sigma_{x}=\left(1+\Sigma_{3}\right)/2$
and $\Sigma_{y}=\left(1-\Sigma_{3}\right)/2$ axes, we get
\begin{align}
\frac{\Delta F}{T} & =\frac{\Delta F_{x}}{T}+\frac{\Delta F_{y}}{T} 
=\int \text{Tr} \, J^{-1}_{x-x'} \left(\hat{\bar{B}}_{x}\hat{B}_{x}+\hat{\overline{B}}_{y}\hat{B}_{y}\right)\label{free3} \nonumber \\
 & -\frac{1}{2}\int \text{Tr} \ \text{\ensuremath{\ln}}\left(g_{0x}^{-1}-\left(\hat{\overline{B}}_{x}+\hat{B}_{y}\right)\right) \nonumber  \\
 &-\frac{1}{2}\int \text{Tr} \ \ln\left(g_{0x}^{-1}-\left(\hat{\overline{B}}_{y}+\hat{B}_{x}\right)\right)\ ,
\end{align}
which can be re-cast into 
\begin{align}
\frac{\Delta F}{T} & =\text{Tr} \int dxdx'J_{x-x'}^{-1}\left[\hat{\overline{B}}_{x}\hat{B}_{x}+\hat{\overline{B}}_{y}\hat{B}_{y}\right]\label{free3-1} \nonumber  \\
 & -\frac{1}{2}\int dxdx' \text{Tr} \ln\left(g_{0x}^{-1}-\hat{b}_{x}\right)+\left(x\leftrightarrow y\right)\ ,
\end{align}
with $\hat{b}_{x}=\hat{\overline{B}}_{x}+\hat{B}_{y}$
and $\hat{b}_{y}=\hat{\overline{B}}_{y}+\hat{B}_{x}$ and $x=\left(\mathbf{r},\tau\right)$
and $x'=\left(\mathbf{r'},\tau'\right)$ denote the set of coordinates
whereas $\hat{B}_{x}$ and $\hat{g}_{x}$ denote the projection onto
the $\Sigma_{x}$ and $\Sigma_{y}$ axes.

In order to convince oneself that the new conjugation
is giving physically meaningful results, we write the MFE by differentiating
sequentially with respect to $\overline{B}_{x,x'}$ and $B_{x,x'}$.
We get
\begin{subequations}
\begin{align}
J_{x-x'}^{-1}\hat{B}_{x} & =-\frac{1}{2} \text{Tr} \left[\hat{g}_{x}\right], \label{eq24}\\
J_{x-x'}^{-1}\hat{\overline{B}}_{x} & =-\frac{1}{2} \text{Tr} \left[\hat{g}_{y}\right] , \\
J_{x-x'}^{-1}\hat{B}_{y} & =-\frac{1}{2} \text{Tr}  \left[\hat{g}_{y}\right], \\
J_{x-x'}^{-1}\hat{\overline{B}}_{y} & =-\frac{1}{2} \text{Tr} \left[\hat{g}_{x}\right] ,
\end{align}
\end{subequations}
with $\hat{g}_{x}=\left(g_{0x}^{-1}+\hat{b}_{x}\right)^{-1}$
and $\hat{g}_{y}=\left(g_{0y}^{-1}+\hat{b}_{y}\right)^{-1}$.
We see that in order for the MFE to have a solution,
we need to impose 
\begin{subequations}
\begin{align}
\hat{\overline{B}}_{x} & =\hat{B}_{y}, \label{rep1-1}\\
\hat{\overline{B}}_{y} & =\hat{B}_{x},\label{rep2}
\end{align}
\end{subequations}
which works as a condition of reality for the HS
fields. Later on we look for solutions of the MFE for fields real
with respect to the old conjugation $\bar{B}=\Sigma_{1}B\Sigma_{1}$, 
($B^{\dagger}=B$).

It turns out that the matrices structure inside
the free energy can be reduced using the trick that
for all $M=\left(\begin{array}{cc}
A & D\\
C & B
\end{array}\right)$
\begin{equation}
\det\left(M\right)=\det B\det\left(A-DB^{-1}C\right),\label{magic-1}
\end{equation}
where A,B,C,D are matrices. The intermediate steps
are summarized in Appendix \ref{sec:APPENDIX-A:-Reduction}. The final
result for the free energy is 
\begin{subequations} 
\begin{align}
F_{0} & =T\sum_{\epsilon}\!\int dxdx'J_{x-x'}^{-1}\!\left[\overline{B}_{1x}B_{1x}+\overline{B}_{1y}B_{1y}+\overline{B}_{2x}B_{2x}+\overline{B}_{2y}B_{2y}\right],\\
F_{x} & =-\frac{1}{2}T\sum_{\epsilon}\int dxdx'\big[-\ln\left(d_{x}-b_{2x}^{2}\right)+\ln\big(\left(\epsilon^{2}+b_{1x}^{2}\right)d_{x}^{2}\nonumber \\
 & +\left(vp_{x}\cos\theta d_{x}+vp_{y}\sin\theta\left(d_{x}-2b_{2x}^{2}\right)\right)^{2}\big)\big],\\
F_{y} & =-\frac{1}{2}T\sum_{\epsilon}\int dxdx'\big[-\ln\left(d_{y}-b_{2y}^{2}\right)+\ln\big(\left(\epsilon^{2}+b_{1y}^{2}\right)d_{y}^{2}\nonumber \\
 & +\left(vp_{x}\sin\theta d_{y}+vp_{y}\cos\theta\left(d_{y}-2b_{2y}^{2}\right)\right)^{2}\big)\big],\\
 & d_{x}=\epsilon^{2}+\left(vp_{x}\cos\theta-vp_{y}\sin\theta\right)^{2}+b_{1x}^{2}+b_{2x}^{2},\nonumber \\
 & d_{y}=\epsilon^{2}+\left(vp_{x}\sin\theta-vp_{y}\cos\theta\right)^{2}+b_{1y}^{2}+b_{2y}^{2},\nonumber \\
 & b_{ix}=\overline{B}_{ix}+B_{iy},\qquad b_{iy}=\overline{B}_{iy}+B_{ix},\qquad i=1,2.\nonumber 
\end{align}
\label{eq:8} \end{subequations} Eqs.\ (\ref{eq:8}) and the introduction
of the new conjugation Eq.\ (\ref{eg6}) constitute the main technical
tools of this paper which enable a controlled discussion of the co-existence
and interplay of the two order parameters $\hat{B}_{1}$ (QDW/SC)
and $\hat{B}_{2}$ (CDW).

To derive self consistency equations for the four order parameters
$\overline{B}_{x}$,$\overline{B}_{y}$ and $B_{x}$,$B_{y}$ we differentiate
the free energy with respect to $\overline{B}_{x}$,$\overline{B}_{y}$
and $B_{x}$,$B_{y}$ successively (cf Appendix\ref{sec:APPENDIX-B:-Derivation}).
In order to perform this task one must Fourier transform
equations of the type Eq.\ (\ref{eq24}):
\begin{equation}
J_{x-x'}^{-1}B_{x,x'}=-\frac{1}{2} \text{Tr} \left[g_{0}^{-1}+b_{x,x'}\right].\label{eq:34-1}
\end{equation}
 From Eq.\ (\ref{eq:8-1}) we see that $J_{x-x'}^{-1}$ depends on
the paramagnon propagator $D_{x-x'}$. We multiply
both sides by $J_{x-x'}$, and after Fourier transforming and integrating
over the\textbf{ q}, we get (cf
Appendix \ref{sec:APPENDIX-B:-Derivation})

\begin{equation}
B_{\mathbf{k,}\mathbf{k}\mathbf{+P}}=-\frac{1}{2} \text{Tr} \sum_{\mathbf{q}}J_{\mathbf{q}}\left[g_{0}^{-1}+b_{\mathbf{k},\mathbf{k}+\mathbf{P}+\mathbf{q}}\right].\label{eq:34-1-1}
\end{equation}
The orders introduced in Eq.\ (\ref{eq:11}) and
(\ref{eq:12}) correspond to $\mathbf{P=}\mathbf{Q}_{1,2}$ and $\mathbf{P}=\mathbf{Q}_{x,y}$
respectively while the short hand notation $B_{\mathbf{k}}=B_{\mathbf{k},\mathbf{k+P}}$
and $b_{\mathbf{k}}=b_{\mathbf{k},\mathbf{k+P}}$ have been used. 

The final result for the Mean-Field Equations (MFEs) is \begin{subequations}
\begin{align}
B_{1x}\left(\epsilon_{n}\right) & =4\gamma_{1}T\sum_{\epsilon'}A_{1x}\left(\epsilon_{n},\epsilon'_{n}\right)B_{1y}\left(\epsilon'_{n}\right),\label{mf1}\\
B_{1y}\left(\epsilon_{n}\right) & =4\gamma_{1}T\sum_{\epsilon'}A_{1y}\left(\epsilon_{n},\epsilon'_{n}\right)B_{1x}\left(\epsilon'_{n}\right),\label{mf2}\\
B_{2x}\left(\epsilon_{n}\right) & =4\gamma_{2}T\sum_{\epsilon'}A_{2x}\left(\epsilon_{n},\epsilon'_{n}\right)B_{2y}\left(\epsilon'_{n}\right),\label{mf3}\\
B_{2y}\left(\epsilon_{n}\right) & =4\gamma_{2}T\sum_{\epsilon'}A_{2y}\left(\epsilon_{n},\epsilon'_{n}\right)B_{2x}\left(\epsilon'_{n}\right),\label{mf4}
\end{align}
\label{mfeqs} \end{subequations} where $\gamma_{1}=3g_{1}^{2}/2$,
$\gamma_{2}=3g_{2}^{2}/2$ are coupling
constants which can be slightly different from each other due to the
breaking of anisotropy (cf Eq.\ (\ref{eq5}) and the parameters
$A_{1x}$, $A_{1y}$,$A_{2x}$, $A_{2y}$ are given by \begin{subequations}
\begin{align}
A_{1x} & =\sum_{\mathbf{q}}\frac{D_{\omega,\mathbf{q}}}{V}\left(\frac{d_{x}+2vq_{x}vq_{y}\cos\theta\sin\theta}{d{}_{x}^{2}+4vq_{x}vq_{y}\cos\theta\sin\theta d_{x}-4b_{2x}^{2}v^{2}q_{y}^{2}\sin^{2}\theta}\right),\\
A_{2x} & =\sum_{\mathbf{q}}\frac{D_{\omega,\mathbf{q}}}{V}\left(\frac{d_{x}+2vq_{x}vq_{y}\cos\theta\sin\theta-2v^{2}q_{y}^{2}\sin^{2}\theta}{d{}_{x}^{2}+4vq_{x}vq_{y}\cos\theta\sin\theta d_{x}-4b_{2x}^{2}v^{2}q_{y}^{2}\sin^{2}\theta}\right),\\
A_{1y} & =\sum_{\mathbf{q}}\frac{D_{\omega,\mathbf{q}}}{V}\left(\frac{d_{y}+2vq_{x}vq_{y}\cos\theta\sin\theta}{d{}_{y}^{2}+4vq_{x}vq_{y}\cos\theta\sin\theta d_{y}-4b_{2y}^{2}v^{2}q_{y}^{2}\cos^{2}\theta}\right),\\
A_{2y} & =\sum_{\mathbf{q}}\frac{D_{\omega,\mathbf{q}}}{V}\left(\frac{d_{y}+2vq_{x}vq_{y}\cos\theta\sin\theta-2v^{2}q_{y}^{2}\cos^{2}\theta}{d{}_{y}^{2}+4vq_{x}vq_{y}\cos\theta\sin\theta d_{y}-4b_{2y}^{2}v^{2}q_{y}^{2}\cos^{2}\theta}\right),\\
 & \text{with}\qquad D_{\omega,\mathbf{q}}=\left(\gamma\left|\omega\right|+q_{x}^{2}+q_{y}^{2}+m_{a}\right),\nonumber 
\end{align}
\label{coefs} \end{subequations} where $\frac{1}{V}\sum_{\mathbf{q}}\equiv\int\frac{d\mathbf{q}}{\left(2\pi\right)^{2}}$.
Note, that the form of the propagator $D_{\omega,\mathbf{q}}$ slightly changed compared to Eq.\ (\ref{eq:3}) since we consider Landau damping.
A closer look at Eq.\ (\ref{coefs}) shows that the right hand-side
of (\ref{mf1}) and (\ref{mf4}) is always lower than the r.h.s. of
resp. (\ref{mf2}) and (\ref{mf3}). In order for the two solutions
to exist simultaneously, it is enough to introduce a slightly different
coefficient $J_{2}$ in front of (\ref{mf2}) and
(\ref{mf4}), with $g_{2}\geq g_{1}$, which henceforth
will favor the $B_{2}$-type of decoupling. This
difference can be introduced through a small perturbation like the
breaking of the $C_{4}$ symmetry.

\begin{figure}
a)\includegraphics[scale=0.25]{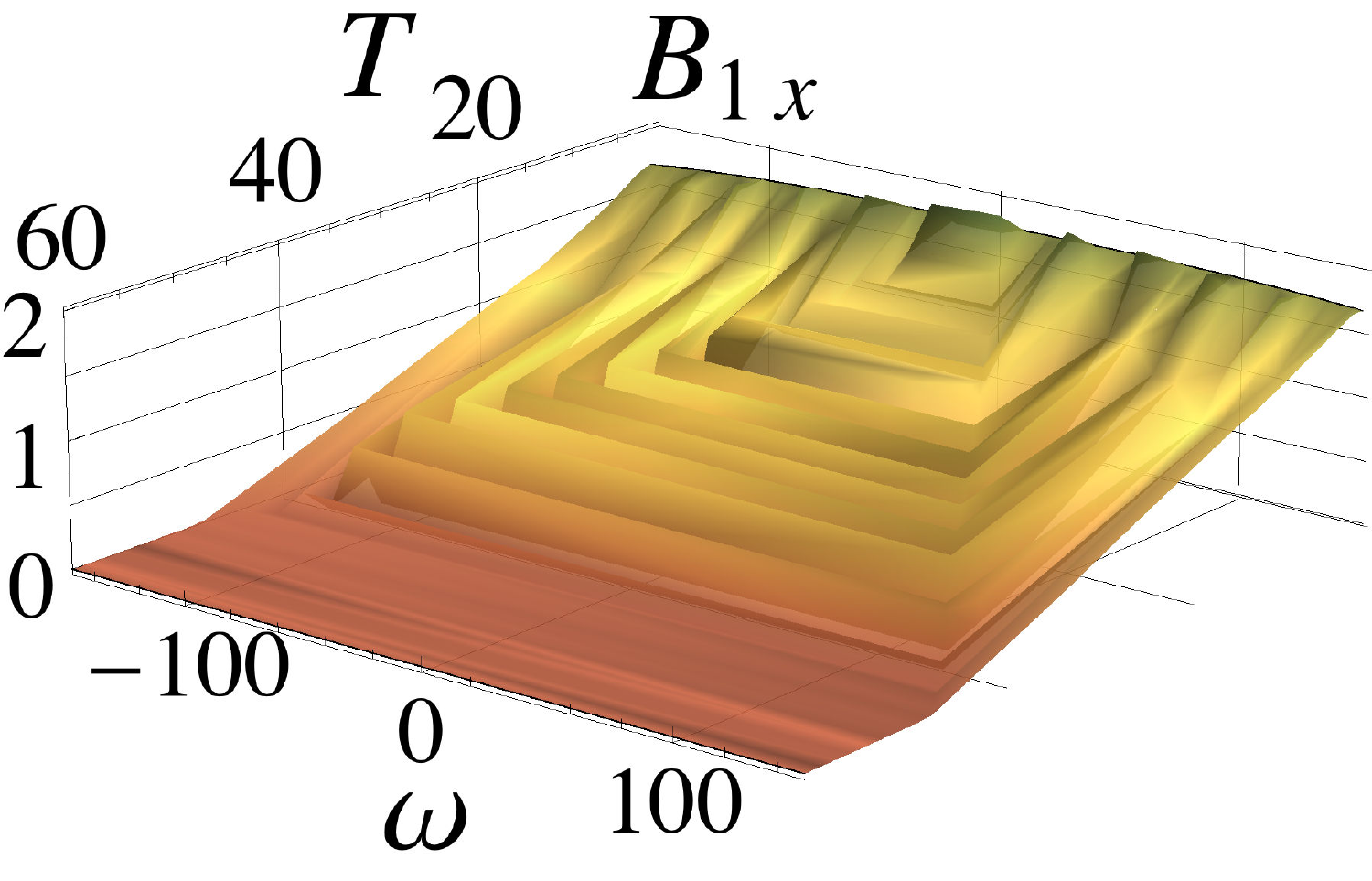} b) \includegraphics[scale=0.25]{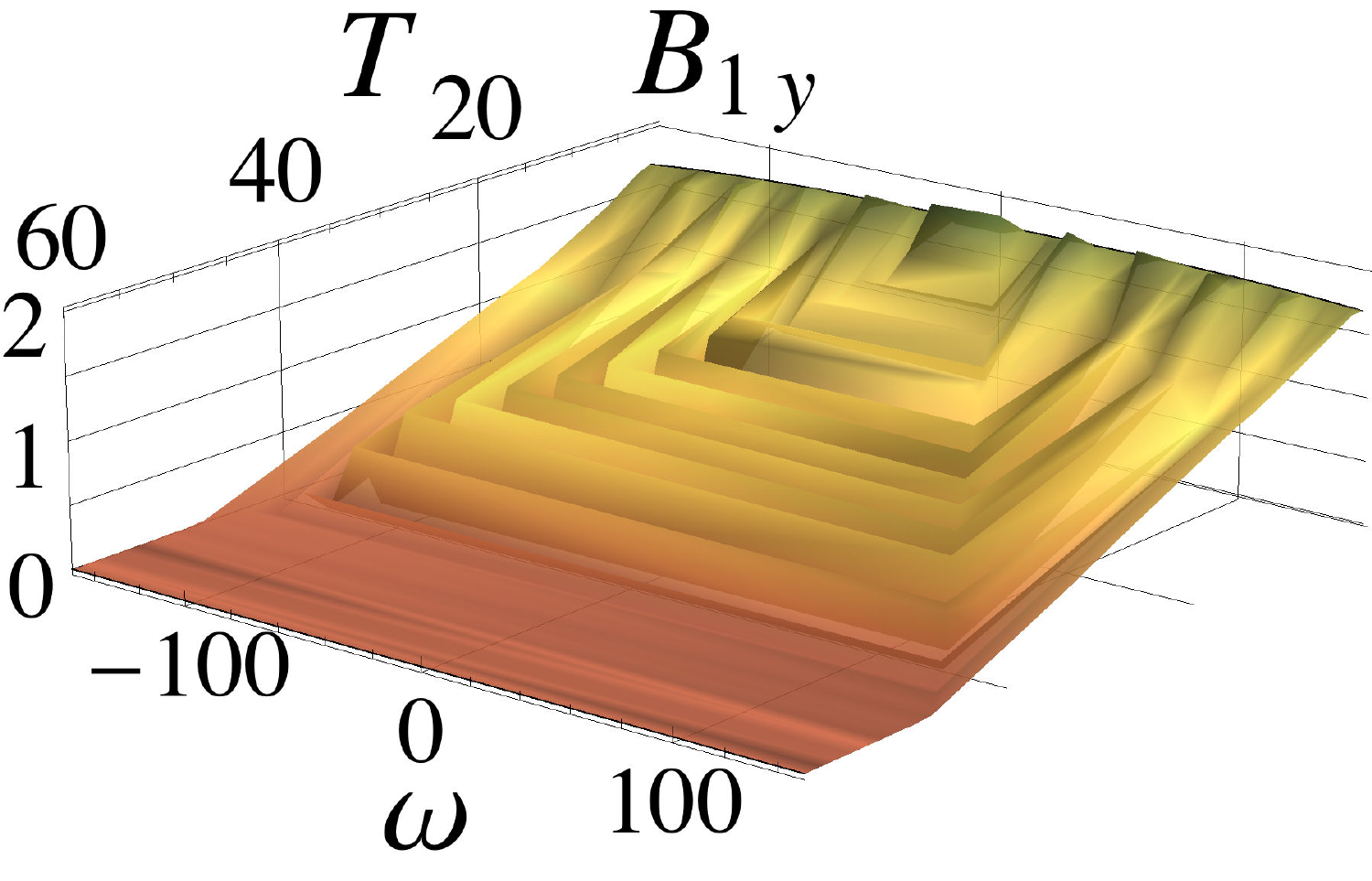}
c)\includegraphics[scale=0.25]{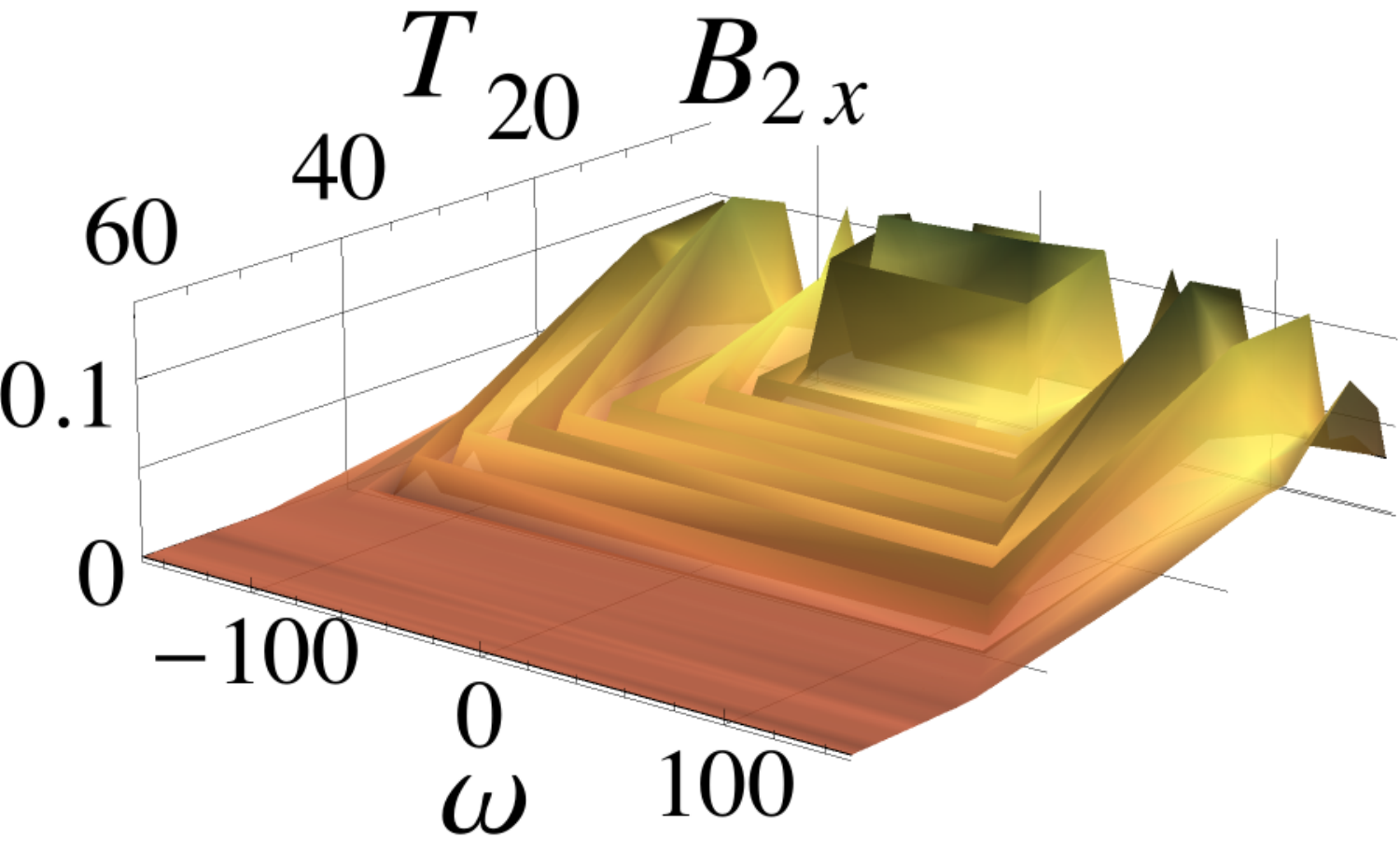} d) \includegraphics[scale=0.25]{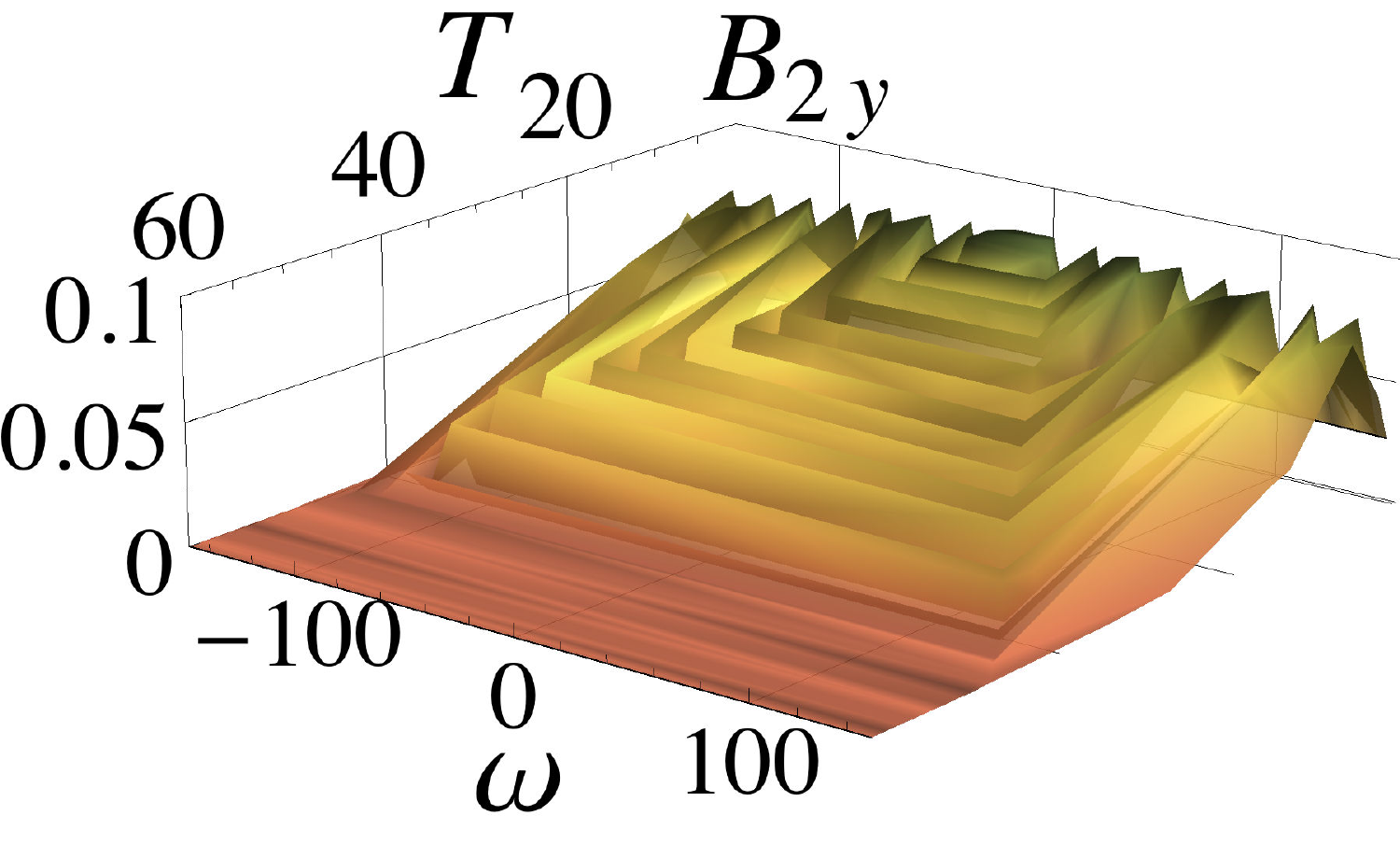}
\vspace{-2ex}
 \caption{\label{Fig2} (Color online) Typical form of the order parameters
in the CE phase as a function of frequency $\omega$ and temperature
$T$: $B_{1x/1y}$ (depicted in panel a) and b)) representing QDW/SC
and $B_{2x/2y}$ (depicted in panel c) and d)) representing CDW/PDW
order. Note that the CDW/PDW component is one order
of magnitude smaller than the QDW/SC solution. We take $g_{1}$ slightly
bigger than $g$ in order to stabilize the CDW sector. The actual
figure corresponds to $g_{1}=20$, $g_{2}=30$,
$v=6$, $m_{a}=0.1$, $\gamma=3$, $W=2\pi$, where
$W$ is the bandwidth of integration in \textbf{k}-space
and $\theta=0.1$.}
\end{figure}

\begin{figure}
a) \includegraphics[scale=0.24]{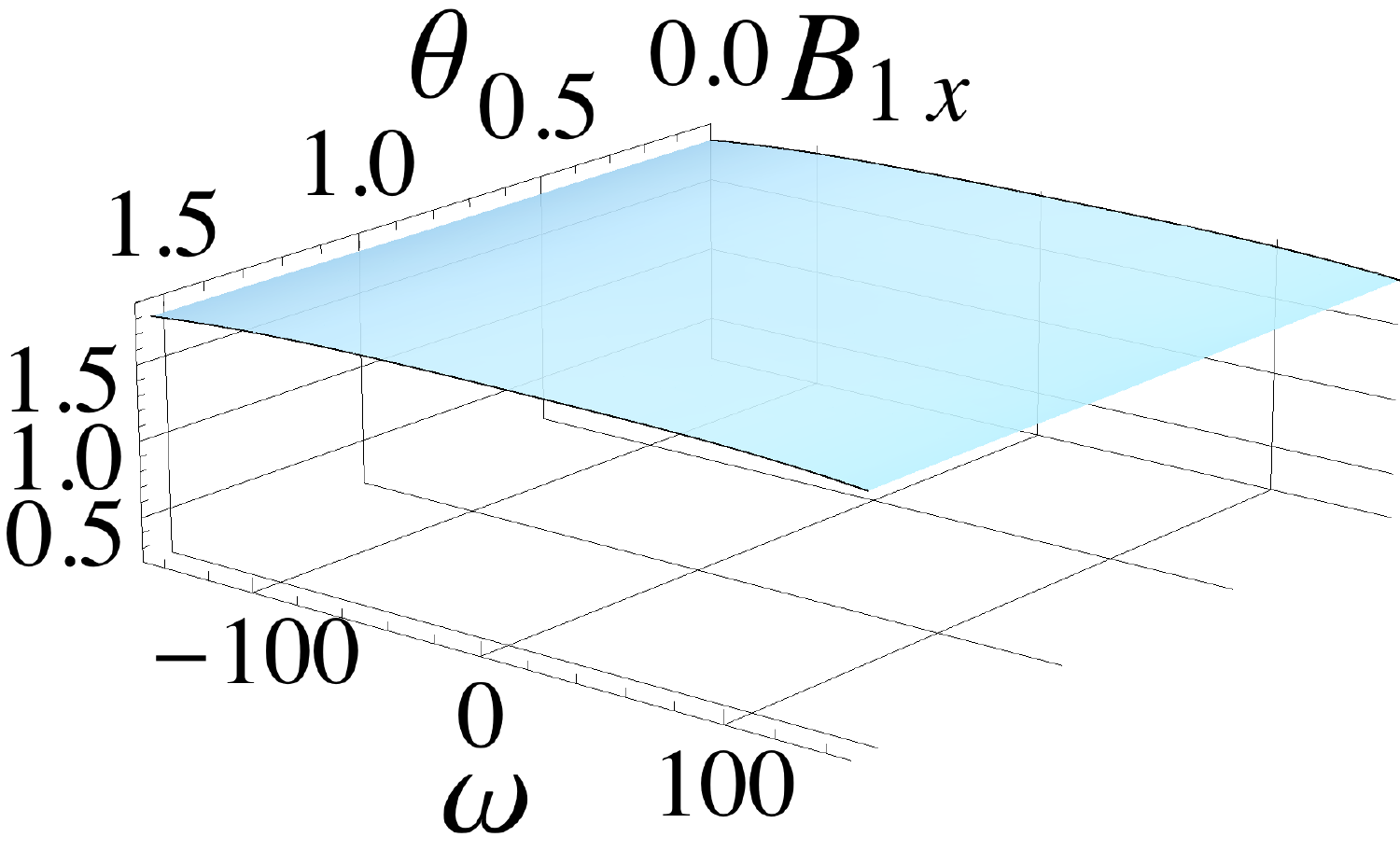} b) \includegraphics[scale=0.24]{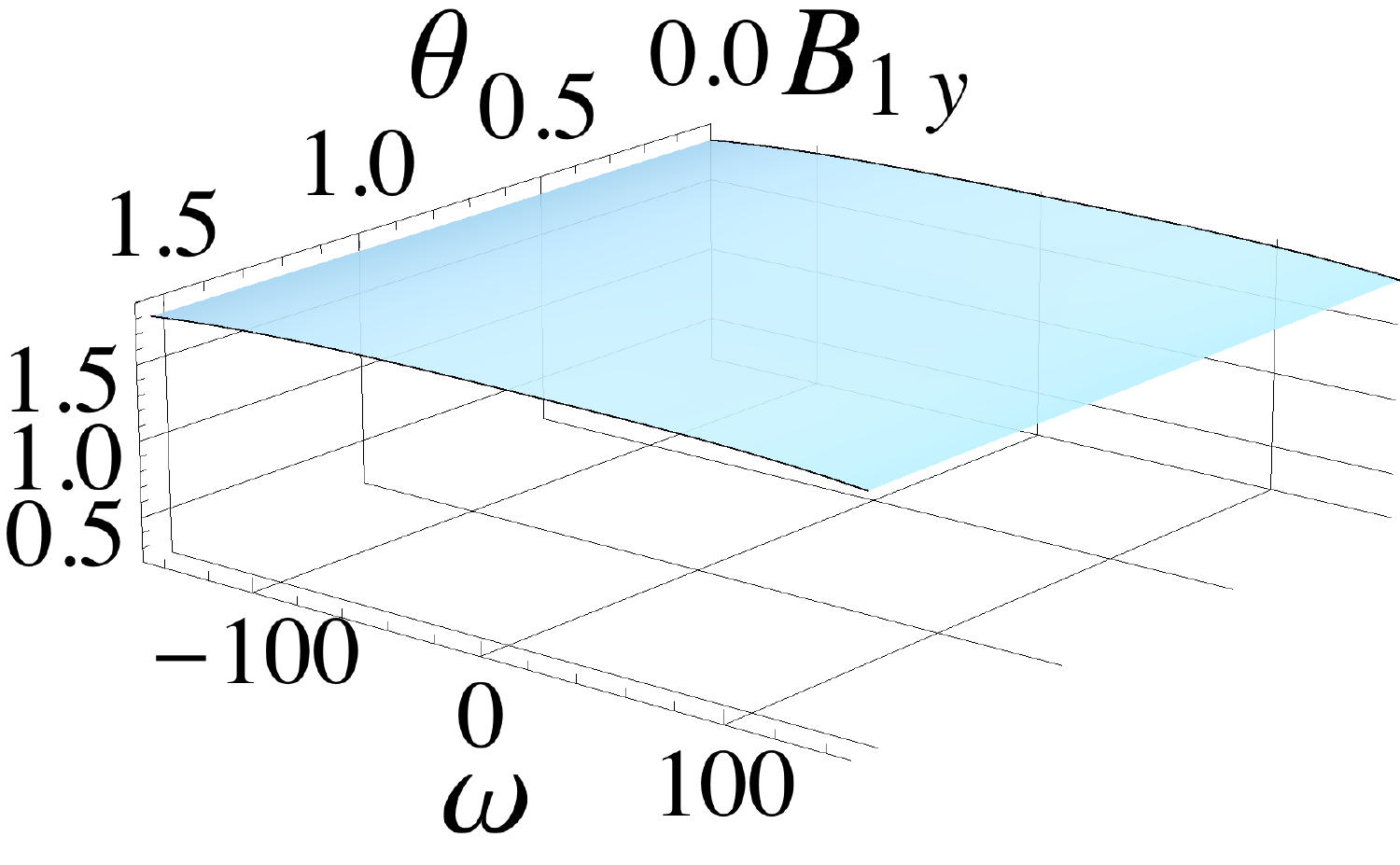}

c) \includegraphics[scale=0.24]{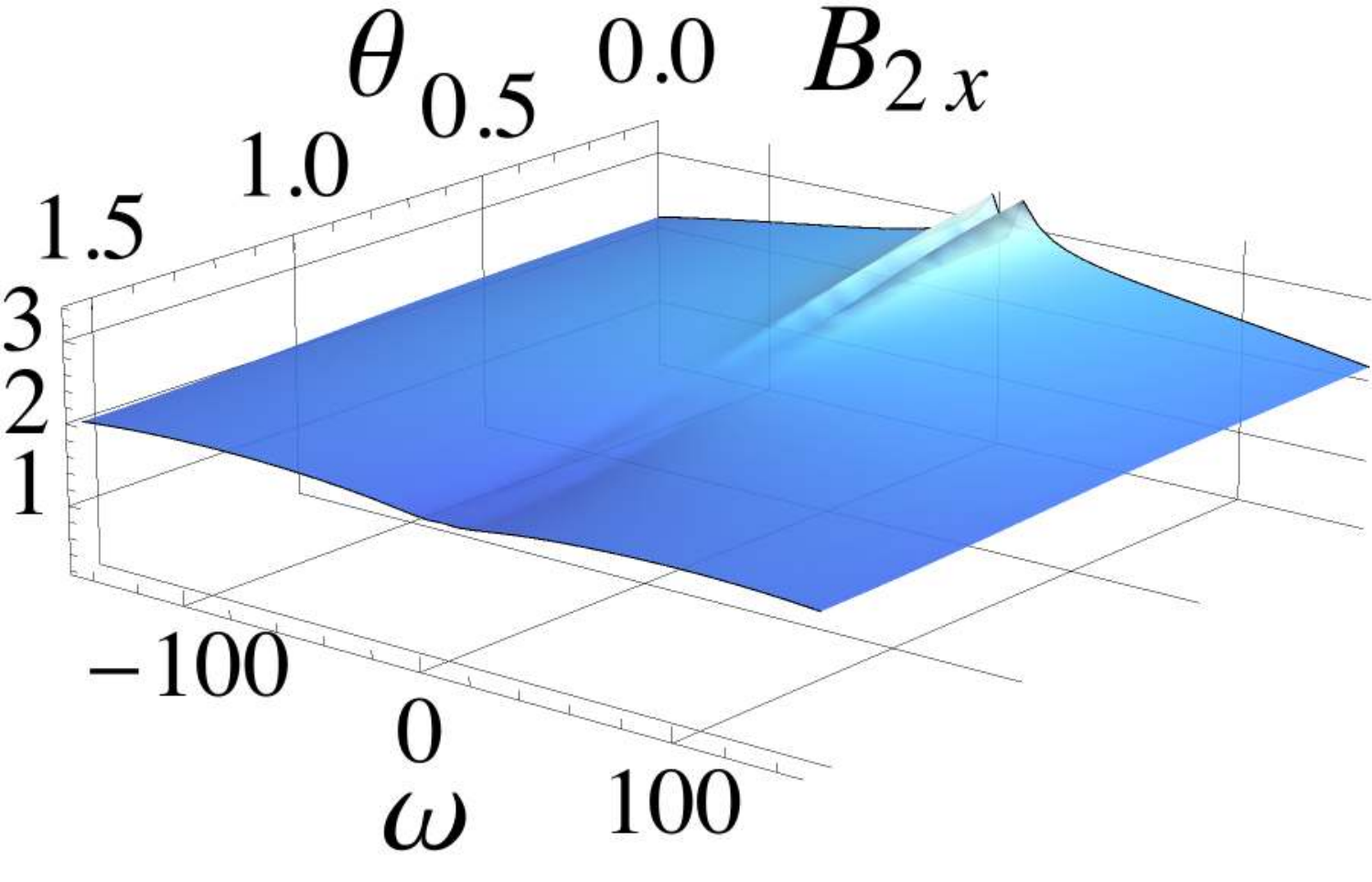} d) \includegraphics[scale=0.24]{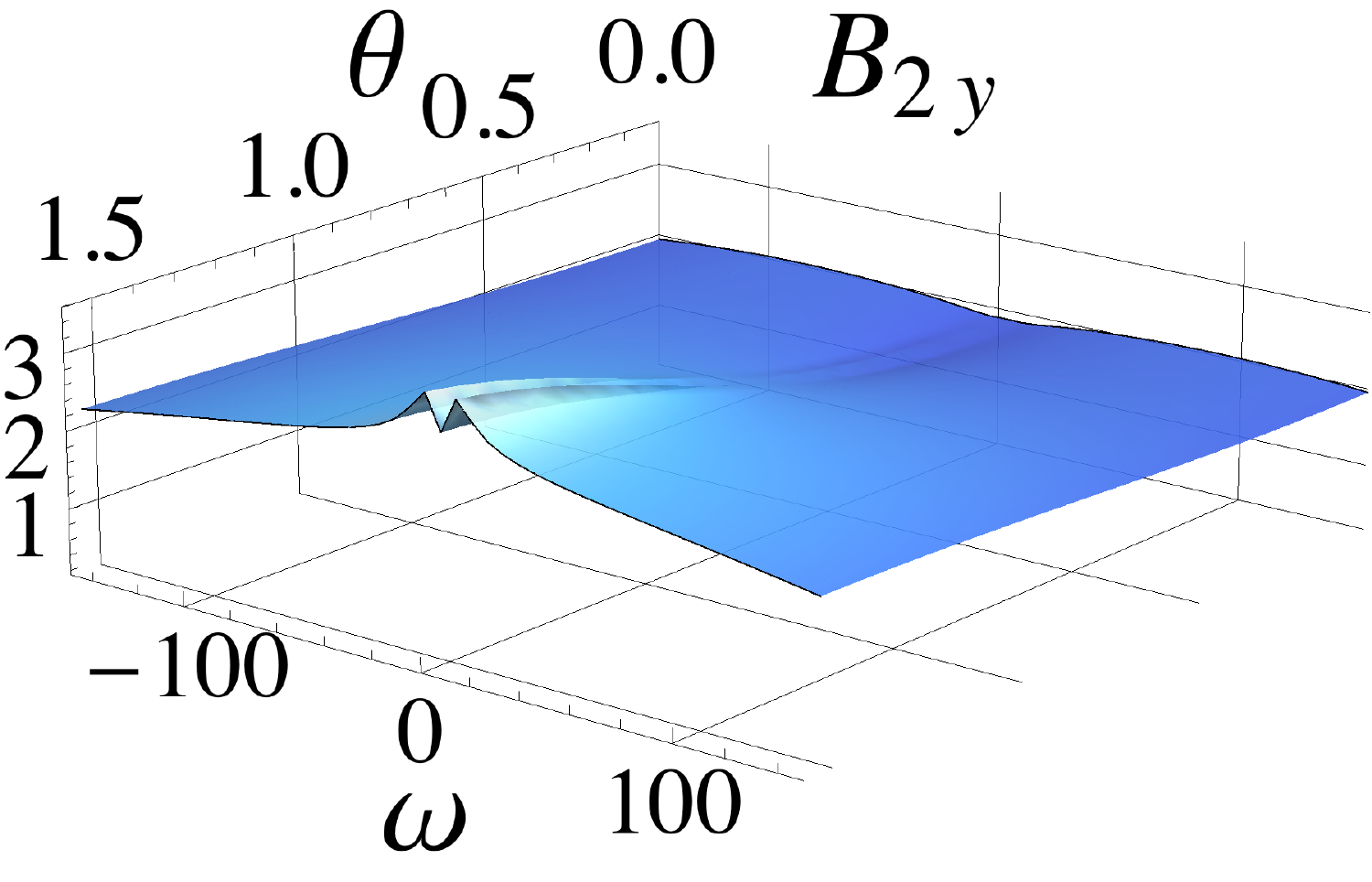}
\vspace{-2ex}
 \caption{\label{Fig3} (Color online) Typical form of the order parameters
in the CE phase as a function of frequency $\omega$ and velocity
angle $\theta$: $B_{1x/1y}$ (depicted in panel a) and b)) representing
QDW/SC and $B_{2x/2y}$ (depicted in panel c) and d)) representing
CDW/PDW order. Note that the CDW/PDW
component is one order of magnitude smaller than the QDW/SC solution.
The parameters are the same as in Fig.\ \ref{Fig2}. }
\end{figure}

The typical result of the MFEs for parameters $g_{2}\simeq g_{1}$
is given in Fig.\ \ref{Fig2}. 
We observe that three solutions are obtained, i) the pure QDW/SC solution
for which $B_{1}\neq0$ and $B_{2}=0$; ii) the pure CDW/PDW
solution for which $B_{2}\neq0$ and $B_{1}=0$; iii) the Co-Existence
(CE) solution where $B_{1}\neq0$ and $B_{2}\neq0$. Moreover, for
typical values of the coupling constants, solution i) and ii) have
similar magnitude, while for the CE solution $B_{2}\ll B_{1}$. The
dependence on the Fermi velocity angle $\theta$ which captures the
dependence of the solutions on the fermiology of the compounds is
depicted in Fig.\ \ref{Fig3}. We find that the pure QDW/SC solution
i) is insensitive to fermiology, whereas the pure CDW/PDW
solution ii) is more favorable when $\theta=0$, which correspond
to flat portions of the Fermi surface in the anti-nodal region. The
insensitivity of the QDW/SC solution to the value of $\theta$ stems
from the fact that, in our matrix framework $\left\{ \hat{B}_{1},\hat{\epsilon}_{k}\right\} =0$,
which is not the case for the CDW/PDW order. It
is here summarized in saying that the QDW/SC phase is much less sensitive
to the fermiology than then CDW/PDW phase\cite{Alloul14}.

\subsection{Stability analysis}

\begin{figure}
$a_{1}$) \includegraphics[scale=0.24]{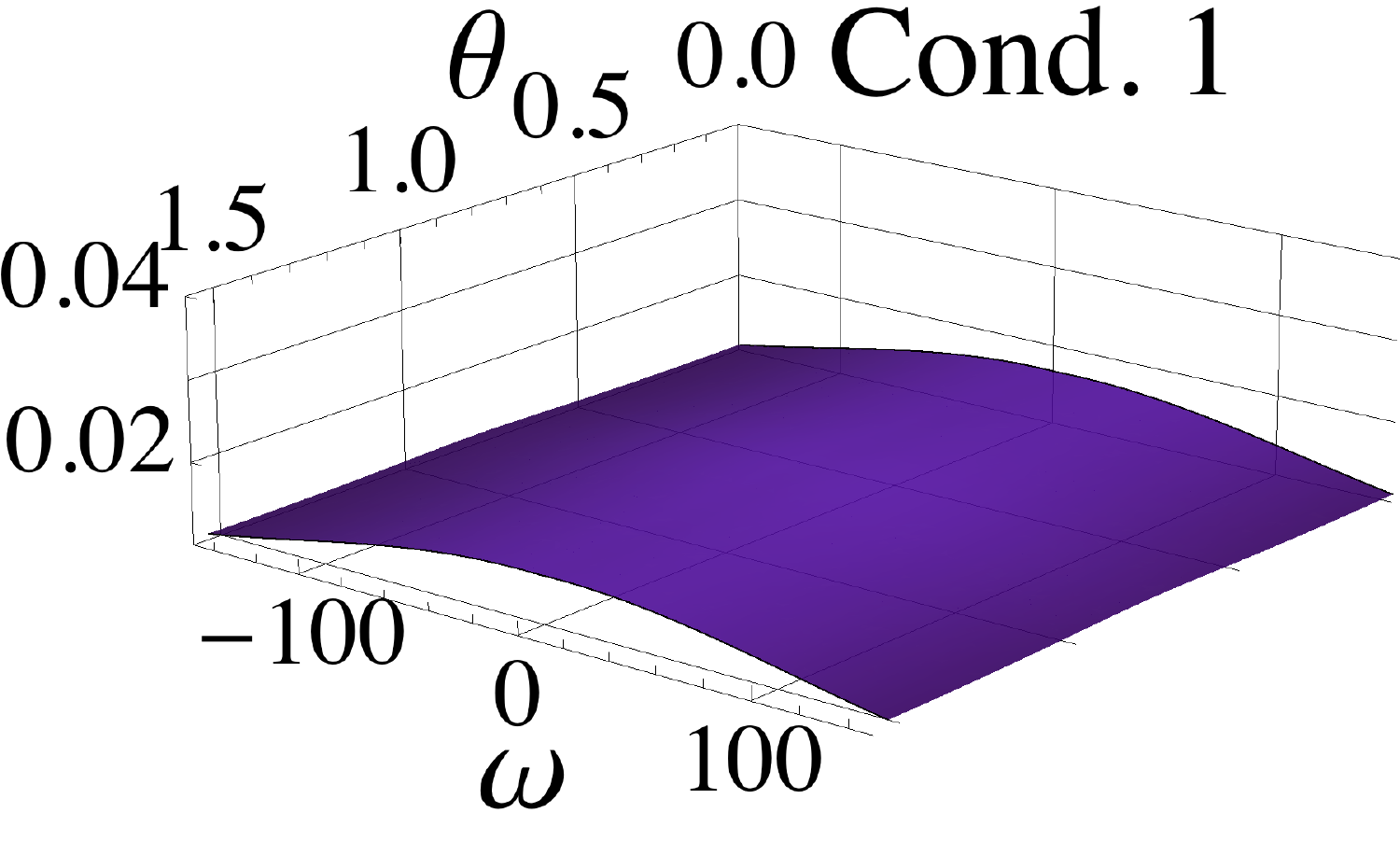} $a_{2}$) \includegraphics[scale=0.24]{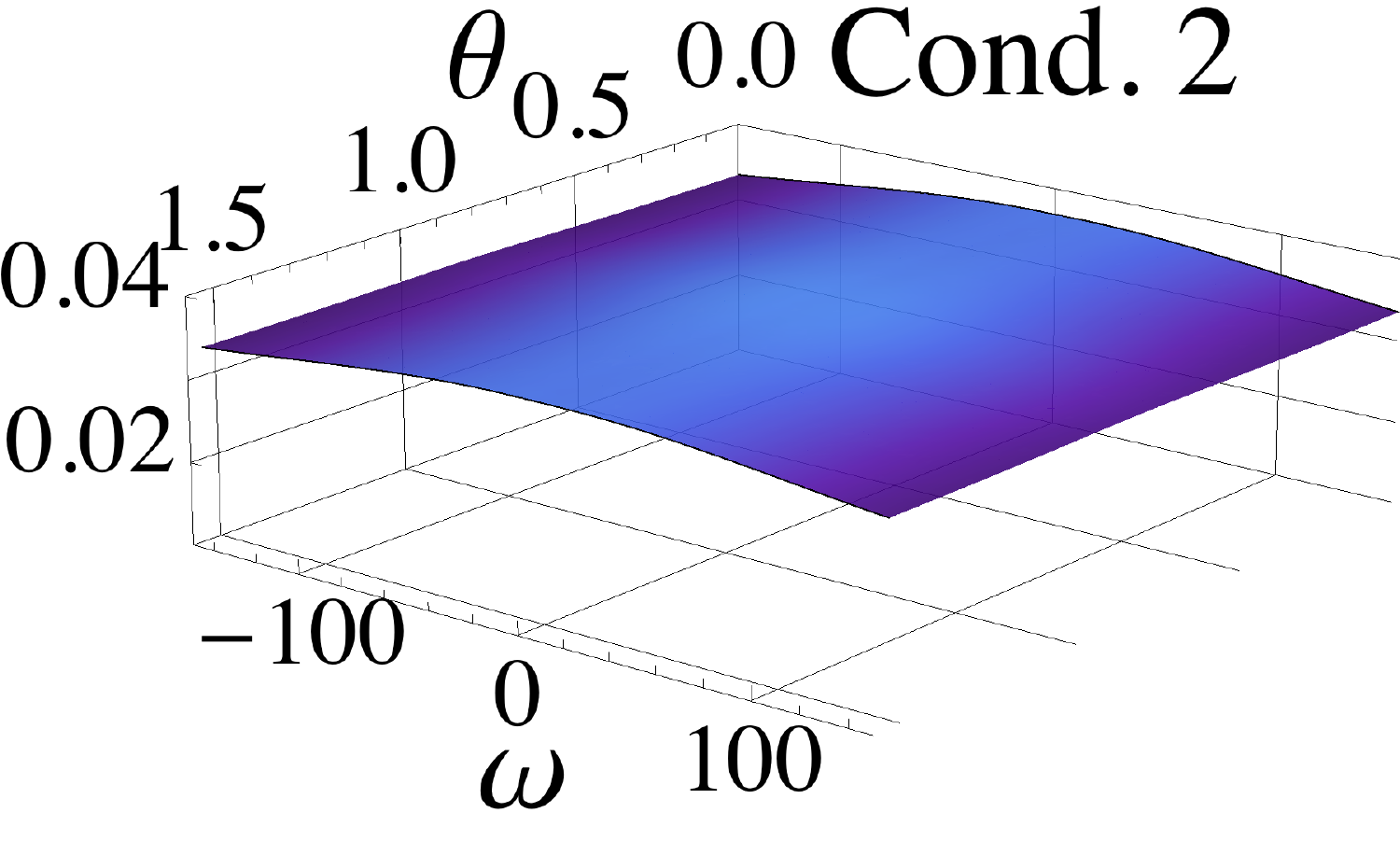}

$b_{1}$) \includegraphics[scale=0.24]{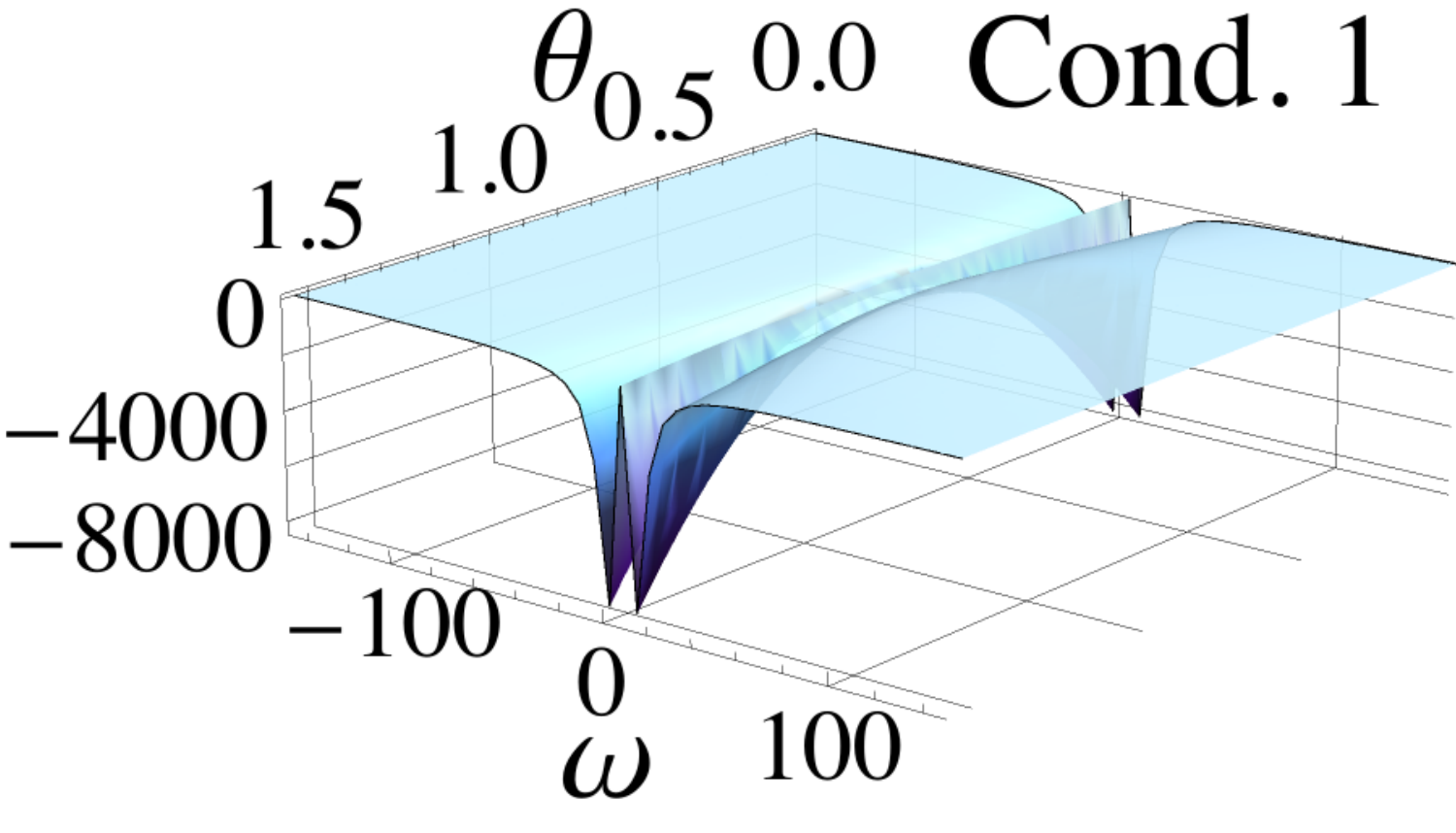} $b_{2}$) \includegraphics[scale=0.24]{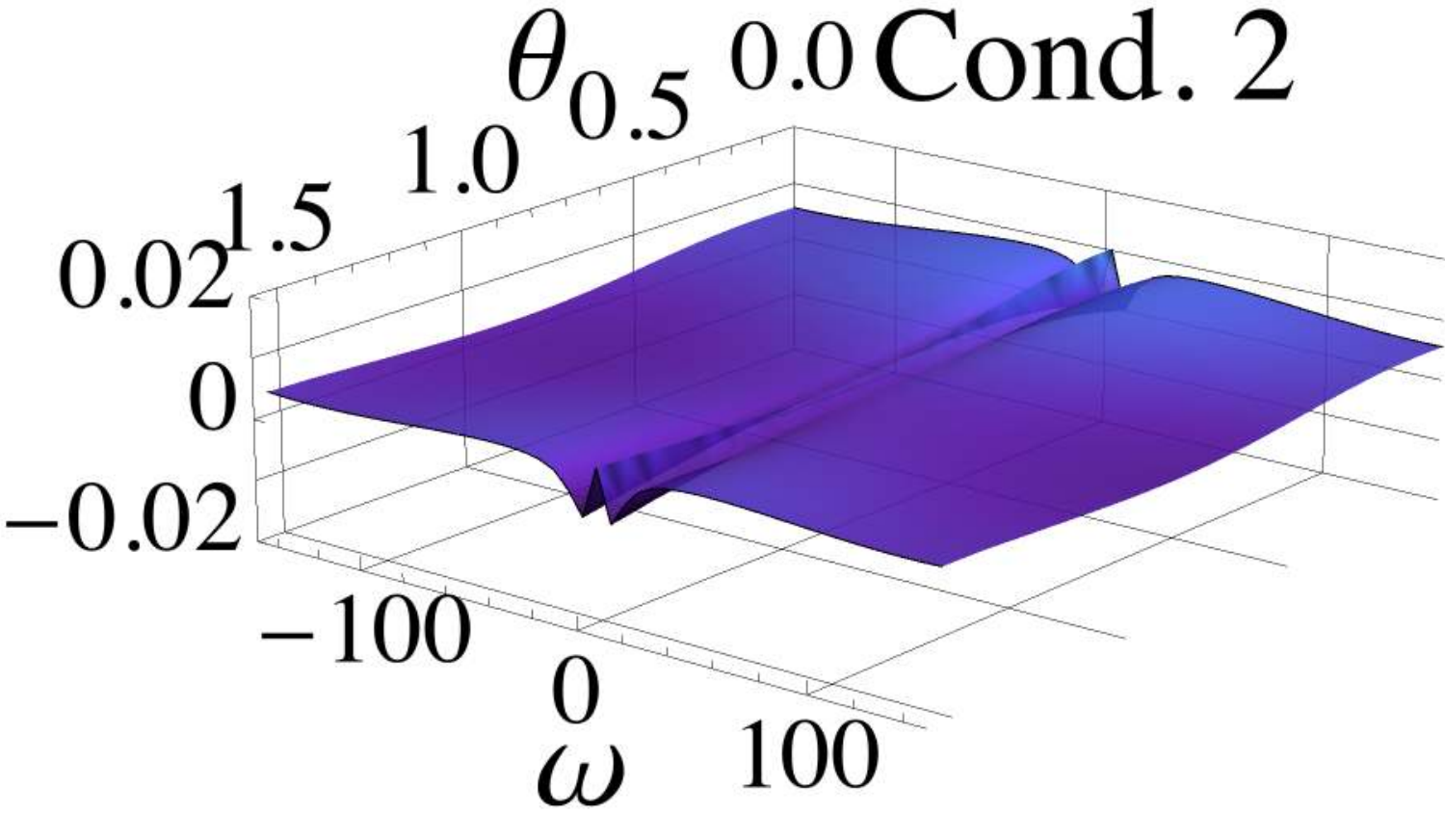}

$c_{1}$) \includegraphics[scale=0.24]{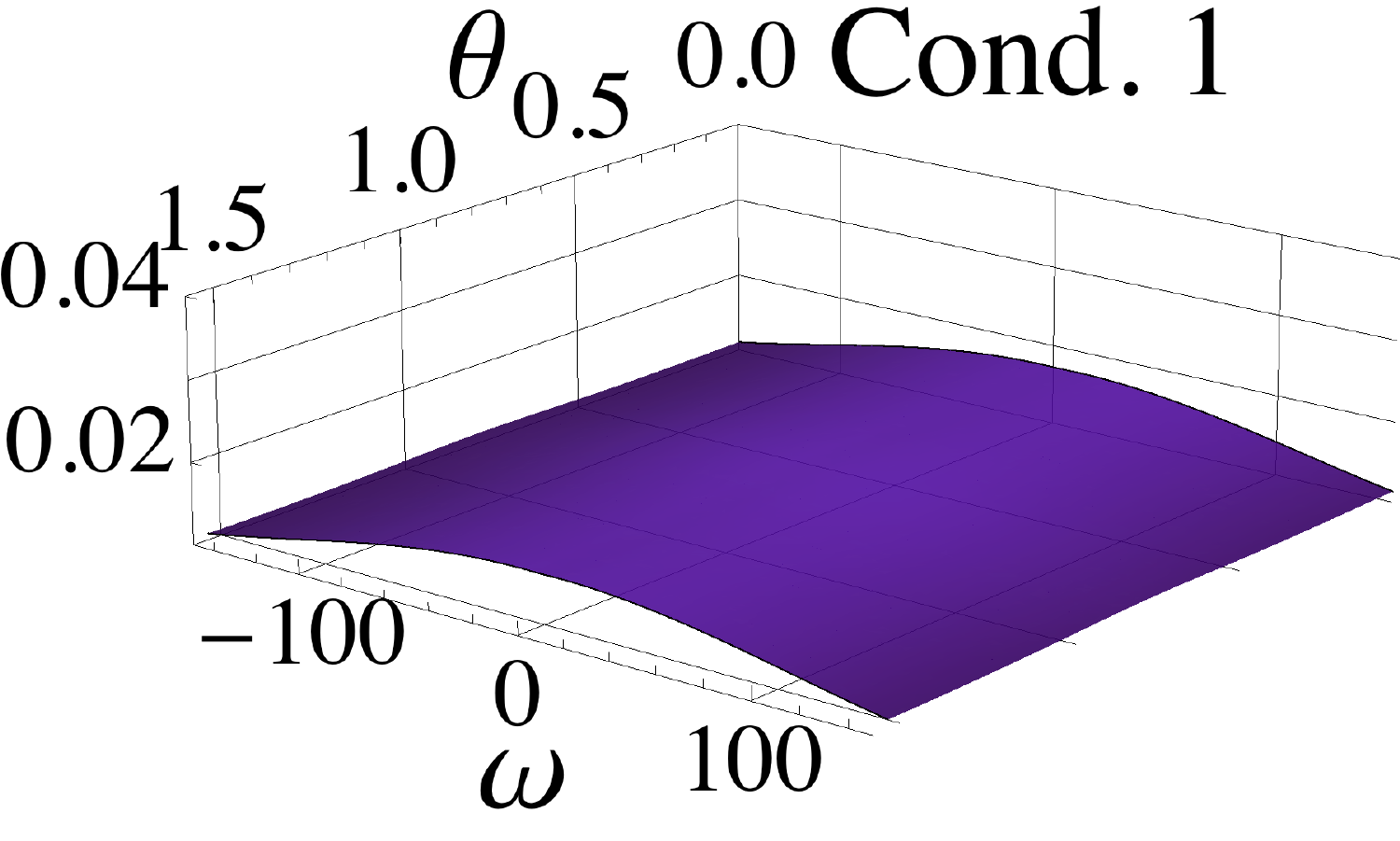} $c_{2}$) \includegraphics[scale=0.24]{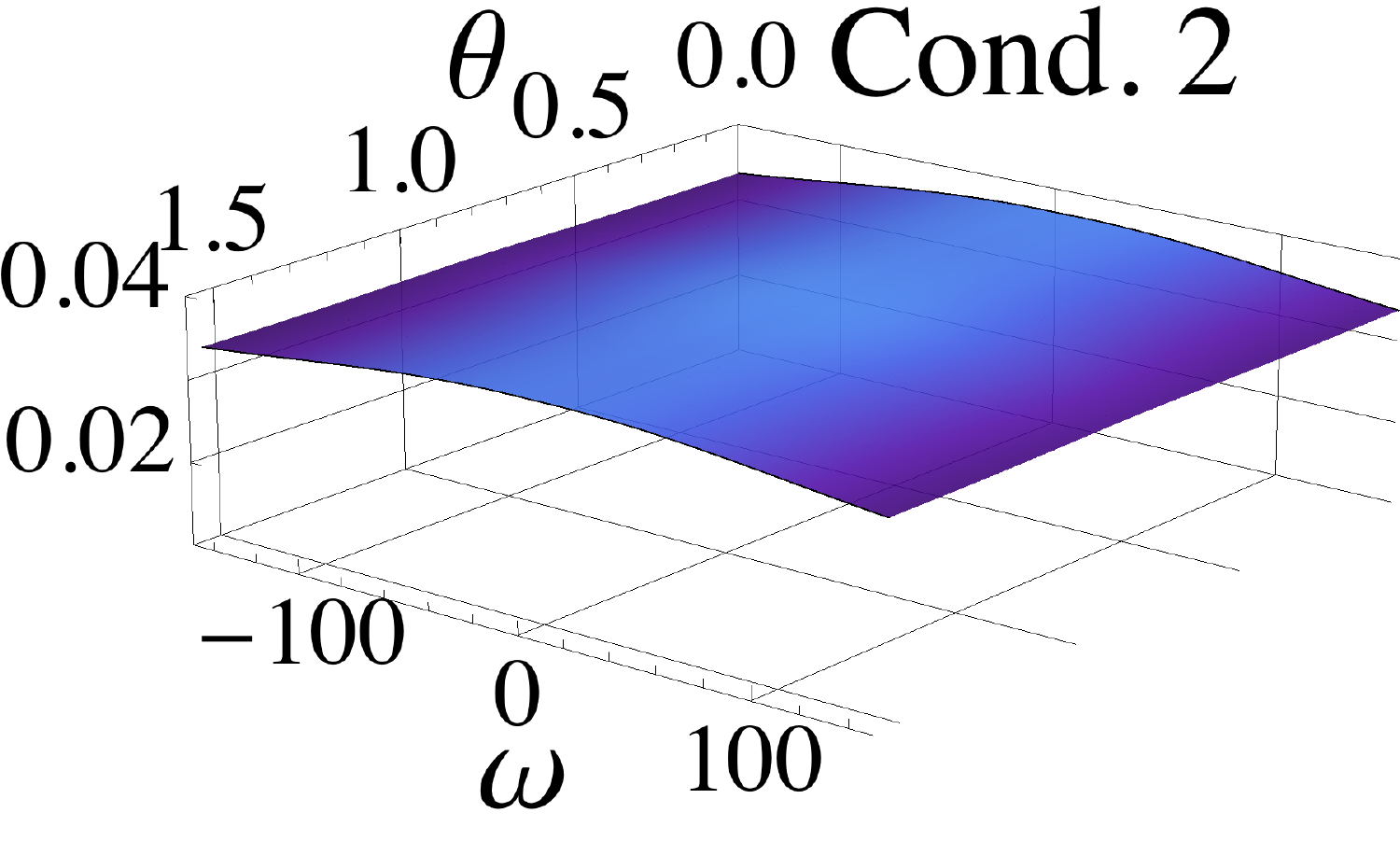}
\vspace{-2ex}
 \caption{\label{Fig4} (Color online) Stability conditions (l.h.s.\ of Eq.\ (\ref{cond0r}))
as a function of frequency $\omega$ and velocity angle $\theta$
for the three possible solutions in the directions of $B_{1}$ and
$B_{2}$: Pure QDW/SC for $a_{1}$) {[}dir. $B_{1}${]} and $a_{2}$)
{[}dir. $B_{2}${]}; pure CDW/PDW solution for $b_{1}$)
{[}dir. $B_{1}${]} and $b_{2}$) {[}dir. $B_{2}${]} and CE solution
for $c_{1}$) and $c_{2}$). Note that the pure CDW/PDW
solution is always unstable while the CE solution is stable. The parameters
are the same as in Fig.\ \ref{Fig3}.}
\end{figure}

To analyze the stability of the solutions, we include Gaussian fluctuations.
We expand Eq.\ (\ref{eq:8}) to the second order in the fluctuation
field $\delta B_{1x}$, $\delta B_{1y}$, $\delta B_{2x}$, $\delta B_{2y}$
and their conjugate. We find, see Appendix \ref{sec:gauss-fluct} for details, 
\begin{align}
\Delta F & =\frac{T}{V}\sum_{\epsilon}\sum_{\mathbf{k},\mathbf{k'}}\sum_{i=1}^{2}\Bigl[J_{\mathbf{k-}\mathbf{k'}}^{-1}\delta\overline{B}_{ix,\mathbf{k}}\delta B_{ix,\mathbf{k'}}\nonumber \\
 & \qquad\qquad\quad-\left(\frac{A_{ix,\mathbf{k}}-\delta A_{ix,\mathbf{k}}}{4}\right)\delta b_{ix,\mathbf{k}}^{2}\delta_{\mathbf{k,}\mathbf{k'}}\Bigr],\label{fluct}
\end{align}
with $\delta b_{ix}=\delta\overline{B}_{ix}+\delta B_{iy}$. In order
to study the stability of the various solution, we write the quadratic
form 
\begin{equation}
\Delta F=\frac{1}{2}\sum_{i=1}^{2}\Psi_{i}^{\dagger}M_{i}\Psi_{i},
\end{equation}
with $\Psi_{i}=\left(\delta B_{ix},\delta\overline{B}_{iy},\delta\overline{B}_{ix},\delta B_{iy}\right)^{T}$
and $\Psi_{i}^{\dagger}=\left(\delta B_{ix},\delta\overline{B}_{iy},\delta\overline{B}_{ix},\delta B_{iy}\right)$.
The stability matrix $M$ writes 
\begin{equation}
M_{i}=\left(\begin{array}{cccc}
J_{\mathbf{k-}\mathbf{k'}}^{-1} & 0 & \bar{A}_{ix} & \bar{A}_{ix}\\
0 & J_{\mathbf{k-}\mathbf{k'}}^{-1} & \bar{A}_{ix} & \bar{A}_{ix}\\
\bar{A}_{iy} & \bar{A}_{iy} & J_{\mathbf{k-}\mathbf{k'}}^{-1} & 0\\
\bar{A}_{iy} & \bar{A}_{iy} & 0 & J_{\mathbf{k-}\mathbf{k'}}^{-1}
\end{array}\right),
\end{equation}
with  $\bar{A}_{ix}=\left(A_{ix}-\delta A_{ix}\right)/2$
and $\bar{A}_{iy}=\left(A_{iy}-\delta A_{iy}\right)/2$. The stability
condition corresponds to the condition for which $M$ is positive
definite.This condition is equivalent to $\det M\geq0$,
which leads to 
\begin{equation}
J_{\mathbf{k-}\mathbf{k'}}^{-2}-4\bar{A}_{ix}\bar{A}_{iy}\geq0.\label{cond0r}
\end{equation}
Note that the free energy (\ref{fluct}) is always real. When $\delta A_{ix}=0$,
Eq.\ (\ref{cond0r}) is equivalent to the condition for existence
of MF solutions {[}Eqs.\ (\ref{mfeqs}). The stability condition
in the respective directions $B_{1x(y)}$ and $B_{2x(y)}$ are presented
in Fig.\ \ref{Fig4}. Typically, one observes that the pure QDW/SC
solution is stable {[}Fig.\ \ref{Fig4}, $a_{1}$) and $a_{2}$){]}
while the pure CDW/PDW solution is unstable {[}Fig.\ \ref{Fig4},
$b_{1}$) and $b_{2}$){]}. The CE solution is always stable at low
temperatures {[}Fig.\ \ref{Fig4}, $c_{1}$) and $c_{2}$){]}. 

\begin{figure}[H]
\begin{centering}
\includegraphics[scale=0.45]{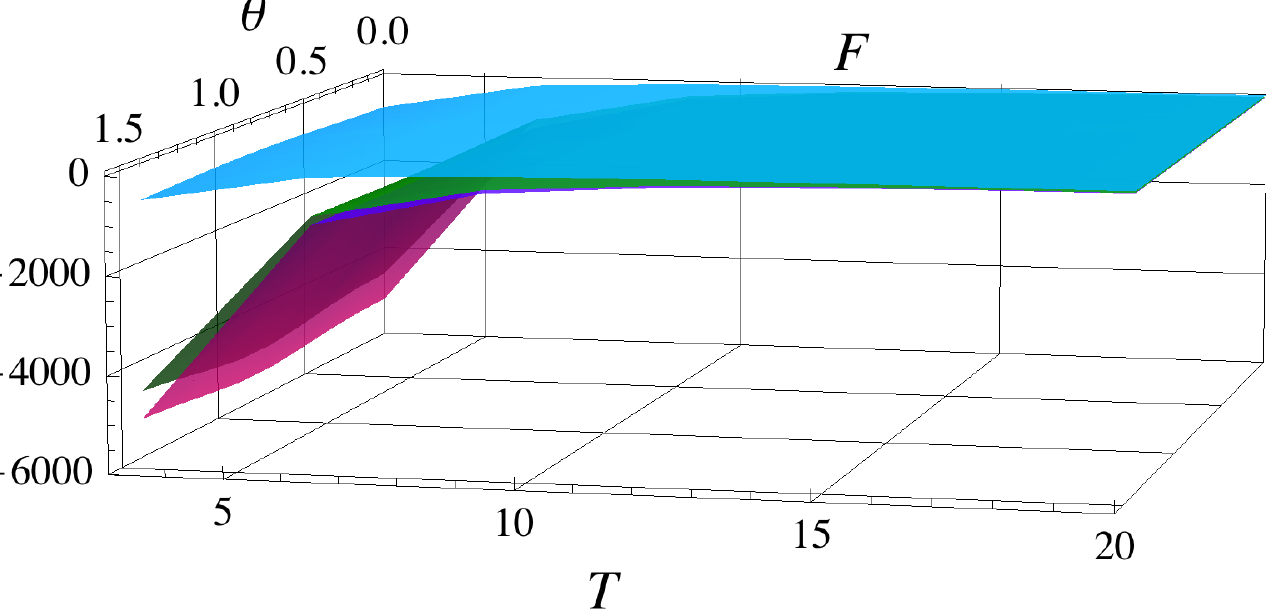} \vspace{-3ex}
 \par\end{centering}
\caption{\label{Fig5 } (Color online) Free energy $F$ of the three MF solutions
as a function of $\theta$ and $T$: pure CDW/PDW
(blue), pure QDW/SC (green) and CE (magenta). Note that the free energy
for the CE solution is slightly lower than for both the QDW/SC and
the CDW/PDW solutions. The parameters are the same
as in Fig.\ \ref{Fig2} }
\end{figure}

In order to differentiate between the two stable solutions {[}pure
QDW/SC and CE{]} we evaluate the free energy in Fig.\ \ref{Fig5 }.
We see that the CE solution is slightly lower than the pure QDW/SC
solution. This behavior is also observed in the limit $J_{2}\gg J_{1}$.
Our conclusion is that the generic tendency is a transition of slightly
first order towards the CE solution at lower temperatures.

\section{Discussion}

\begin{figure}
\includegraphics[scale=0.4]{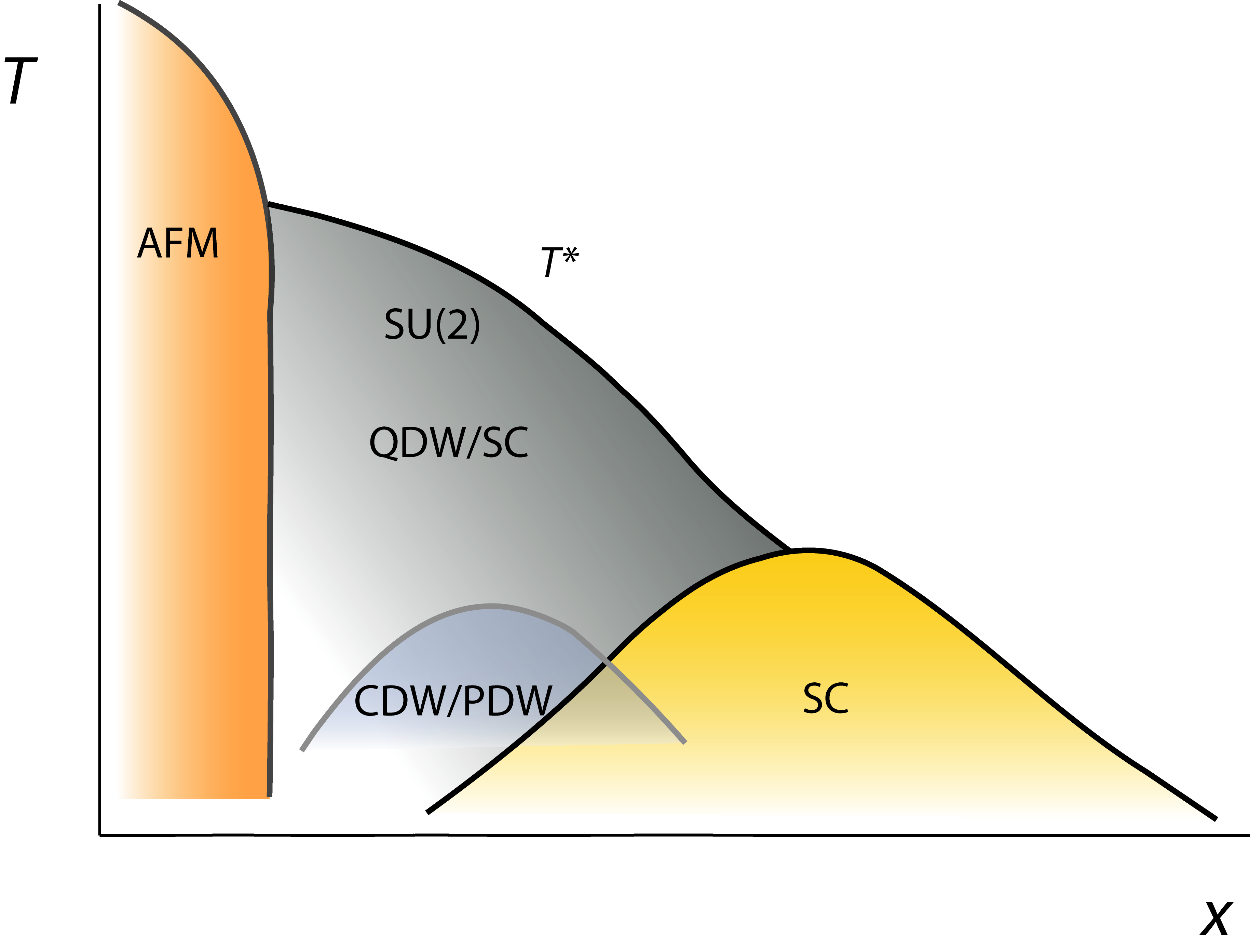}
\vspace{-3ex}
\caption{\label{Fig15} Schematic phase diagram for high temperature cuprate
superconductors as a function of temperature $T$ and hole doping $x$.
The different types of order are: Antiferromagnetic (AFM), quadrupole density wave (QDW), pairing density wave (PDW), charge density wave (CDW) and superconductivity (SC).}
\end{figure}

In the present work we have addressed the role and
presence of charge order in the pseudo-gap phase, in the context of
and emerging SU(2)- symmetry relating the charge and paring sectors.
The EHS model where the Fermi velocity has been linearized provides
a microscopic theory on which the SU(2) symmetry is verified at all
energies\cite{Efetov13,Metlitski10,Metlitski10b}. Namely in this
model the Peierls and the pairing channels are degenerate. Many effects
break the SU(2) symmetry in more realistic model for high temperature
superconductors. Curvature, on the one hand favors the emergence of
the superconducting state at low temperature, while on the other hand
application of an external magnetic field or also the growth of Umklapp
terms in the wake of the Mott-insulating transition act in favor of
the charge sector. The generic phase diagram envisioned for cuprates
superconductivity is one in which the SU(2) symmetry is ``lightly
broken'' in the sense that the splitting of the charge and pairing
sectors is much smaller than their mean value. While at low temperature
the lower energy state is for example the SC order, thermal effect
are then exciting the system to fill explore the higher energy state\cite{Efetov13}.
We identify temperature at which SU(2) symmetry exists between short
range SC pairing fluctuations and QDW order as the PG temperature
$T^{*}$, depicted on Fig.\ \ref{Fig15}. As a composite order parameter
made of two distinct short range components, this order has the potential
to gap out the anti-nodal part of the Fermi surface while leaving
``cold'' the nodal part. Such a treatment requires to start with
a more realistic model going beyond the EHS treatment of the Fermi
surface. The PG phase occurs as a preemptive instability around the
AFM QCP of itinerant electrons. The feed-back of such an instability
on the long range AFM modes is to self-consistently gap them out,
effectively producing short rang AFM correlation (hence pushing back
the QCP) to the left of the phase diagram. As a result the PG state
in this model is constituted of short range AFM correlation co-existing
with SU(2)-symmetric short range d-wave SC preformed pairs and charge
density wave. A recent two loops RG study precisely confirmed that
such a fixed point exists in the EHS model\cite{deCarvalho:9999ev}.

For a very long time three main players were identified
in the physics of cuprate superconductors : anti-ferromagnetism, d-wave
superconductivity and the Mott metal-insulator transition. Within
the SU(2) scenario a fourth player enters the game: charge order.
The assertion that d-wave charge order and pairing SC are quasi-degenerate
in the cuprate family has long shot consequences and enables us to
classify the compounds via their SU(2)-character. For example in the
La-based compounds due to the presence of the very strong Umklapp
interactions, the SU(2) symmetry is broken in favor of the charge
sector, whereas YBCO, Bi2212 and Hg-1201 see the balance towards SC
restored and are thus closer to the SU(2) symmetric case. Interestingly,
as the doping is decreased starting from the optimally doped case
in the SC state, the proliferation of Umklapp terms drive the systems
towards charge ordering, very likely producing a level crossing between
QDW and SC states. One can thus conjecture that an SU(2)-symmetric
point is naturally present in the under-doped compounds. 

We turn now to the discussion of the $\mathbf{Q}_{x}$/$\mathbf{Q}_{y}$
CDW recently observed experimentally. It is now a consensus in the
community that the $\mathbf{Q}_{x/y}$ CDW is distinct from the PG
phase and is ``not'' the gapping out factor of the Fermi surface
but rather a subsidiary instability occurring on top of an already
formed PG state\cite{Alloul14,LeTacon14}\cite{Alloul14,LeTacon14}.
This conclusion is in agreement with the findings of the present study
where the incipient CDW is obtained in co-existence, but at lower
temperature than the PG state (see Fig.\ \ref{Fig15}). It is found
that while the PG state is insensitive to the fermiology, the temperature
at which the CDW order occurs and its magnitude depends strongly on
its surroundings (disorder, orthorhombicity or peculiar form of the
Fermi surface). 

An important question raised within various models,
including the EHS model, is how to get a CDW ordering wave vector
parallel to the x and y-axes and not on the diagonal. Simple Hartree-Fock
evaluations generically produces charge ordering wave vectors on the
diagonal at $\left(\mathbf{Q}_{x}\pm\mathbf{Q}_{y}\right)/2$ \cite{Sachdev13,Allais14a}
and similarly with the solution of the gap equation for the QDW/SC
solution \cite{Efetov13,Meier13}, whereas nothing is observed in
this direction. To resolve this discrepancy, some works have introduced
Coulomb interactions \cite{Allais14b,Allais14c}, but the wave vector
is still a bit tilted. Another interesting proposal is to use the
three bands model and evaluate the charge response on top of an anti-ferromagnetic
PG \cite{Atkinson14}, which was recently reformulated for a spin
liquid type PG\cite{Chowdhury:2014ta}. We have also suggested to
use SC fluctuation to stabilize CDW with correct wave vector at the
anti-nodes\cite{Meier14}. Lately, a closely related work to ours
has suggested that CDW occurs directly at the hot spots\cite{Wang14}.
Our findings agree with Ref.\ [\onlinecite{Wang14}] in the understanding that
a competition exists at the hot spots between the CDW and the QDW/SC.
However, we find that even when the coupling constants are tuned so
that the CDW ordering is extremely favorable, at lower temperature
the QDW/SC re-emerges to form a CE solution. In our work, a small
breaking of tetragonal symmetry (orthorhombicity) is responsible in
the stabilization of a CDW/PDW as a subleading instability. It can
nicely be related to the observation that in compounds where the tetragonal
environment is present, like in Hg-1201, the CDW signal is very weak,
a;most undetectable, while in more orthorhombic compounds lime YBCO,
the CDW signal is stronger.

Lastly, a very interesting point of the emerging SU(2)
symmetry presented in this paper is that the CDW order occurs under
the SU(2) dome below $T^{*}$. As such it possesses an SU(2)-degenerate
counterpart in the form of a Pairing Density Wave (PDW) at finite
wave vector. Hence the denomination CDW/PDW order. This type of order
has been intensely studied recently as a potential candidate for the
PG phase, based on interpretation of ARPES data in Bi-2201\cite{Lee:2014ka,Agterberg:2014wf,Fradkin:2014wk}.
Within the SU(2) scenario, the PDW state is present but not as the
main instability producing the PG state, but as the smallest instability
in a ``logarithmic hierarchy''. From the itinerant electron point
of of view its logarithmic divergence is always cut off by the finite
pairing wave vector, hence even in the linearized EHS model the PDW
instability is subsidiary and not the leading one. It is interesting
to notice that in accordance with the observation of Refs.\ [\onlinecite{Agterberg:2014wf,Wang14,Agterberg:2014wf,Wang14}]
if the $\mathbf{Q}\longleftrightarrow-\mathbf{Q}$ symmetry is broken,
within the CDW/PDW state, then time reversal symmetry is broken, leading
to the possibility of observing a Kerr signal below $T_{CDW}$\cite{Xia08}
and possibly explain the presence of loop currents below $T^{*}$\cite{Fauque06,Varma06,Fauque06,Varma06}.

In conclusion, within a detailed investigation of
the eight hot spots model we show that charge ordering with the correct
wave vector can only occur on top of a pre-existing order which is
our candidate for the pseudo-gap. Our conclusion is that CDW/PDW order
can be stabilized at the hot spots of the SF model in co-existence
with the QDW/SC solution. The CDW/PDW can be considered as a bi-product
of the emergence of the QDW/SC order. Its magnitude is peaked at $T_{c}$
and non-linear $\sigma$-models uniting QDW and SC \cite{Efetov13,Meier13,Hayward14}
to explain sound experiments and X-rays findings are still valid. 
\begin{acknowledgments}
We are grateful for discussions with O.\ Babelon concerning the HS
transformation and the reduction of the free energy. Discussions with
M.\ Einenkel, H.\ Meier, K.\ Efetov, A.\ Chubukov, H.\ Alloul,
Y.\ Sidis, P.\ Bourges and S.S. Lee are also acknowledged. C.P.
has received financial support from the grant EXCELCIUS of the Labex
PALM of the Université Paris-Saclay, the project UNESCOS of the ANR,
as well as the grant Ph743-12 of the COFECUB which enabled frequent
visits to the IIP, Natal. V.S. de C. acknowledges the support of CAPES
and X.M. and T.K. founding from the IIP.


\end{acknowledgments}

\appendix

\global\long\def\theequation{A\arabic{equation}}
 \setcounter{equation}{0}

\section{\label{sec:APPENDIX-A:-Reduction} Reduction of the Free
Energy}

\label{sec:free-energy-red} We give here the essential steps in the
reduction of the free energy. For further details about the notations,
we refer the reader to the SI of our previous paper \cite{Efetov13}.
Using Eq.\ (\ref{eg6}) and setting $Q_{x}=i\hat{B}_{x}$, $Q_{y}=i\hat{B}_{y}$,
{[}resp. $\bar{Q}_{x}=-i\hat{B}_{x}$,$\bar{Q}_{y}=-i\hat{B}_{y}$
{]}, the free energy becomes \begin{subequations} 
\begin{align}
F & =F_{0}+F_{x}+F_{y},\\
F_{0} & =T\sum_{\epsilon}\int\frac{d\mathbf{p}}{(2\pi)^{2}}J_{\mathbf{x-x'}}^{-1}\left[\hat{\overline{B}}_{x}\hat{B}_{x}+\hat{\overline{B}}_{y}\hat{B}_{y}\right],\\
F_{x} & =-\frac{1}{2}T\sum_{\epsilon}\!\!\int\!\!\!\frac{d\mathbf{p}}{(2\pi)^{2}}\textrm{Tr}\ln\left(g_{0x}^{-1}\!-\!\hat{b}_{1x}\!-\!\hat{b}_{2x}\!\right),\\
F_{y} & =-\frac{1}{2}T\sum_{\epsilon}\!\!\int\!\!\!\frac{d\mathbf{p}}{(2\pi)^{2}}\textrm{Tr}\ln\left(g_{0xy}^{-1}\!-\!\hat{b}_{1y}\!-\!\hat{b}_{2y}\!\right),
\end{align}
\label{eq:7} \end{subequations} with $g_{0x}^{-1}=i\epsilon-\hat{x}_{\mathbf{p}}$,
$\hat{x}_{\mathbf{p}}=vp_{x}\cos\theta\Lambda_{3}L_{3}+vp_{y}\sin\theta\Lambda_{3}$,
$g_{0y}^{-1}=i\epsilon-\hat{y}_{\mathbf{p}}$, $\hat{y}_{\mathbf{p}}=-vp_{x}\sin\theta\Lambda_{3}L_{3}-vp_{y}\cos\theta\Lambda_{3}$,
$\hat{b}_{x}=\hat{\overline{B}}_{x}+\hat{B}_{y}$ and $\hat{b}_{y}=\hat{\overline{B}}_{y}+\hat{B}_{x}$.
Using $\mathrm{Tr}\ln G^{-1}=\ln\det G^{-1}$, where $G^{-1}=g_{0x}^{-1}\!-\!\hat{b}_{1x}\!-\!\hat{b}_{2x}$,
($G^{-1}=g_{0y}^{-1}\!-\!\hat{b}_{1y}\!-\!\hat{b}_{2y}$ ) and the
formula 
\begin{equation}
\det\left(M\right)=\det B\det\left(A-DB^{-1}C\right),\quad\mbox{ for all}\quad M=\left(\begin{array}{cc}
A & D\\
C & B
\end{array}\right),\label{magic}
\end{equation}
where $A,B,C,D$ are matrices, we are able to express the free energy
in scalar form. In the direction $\Sigma_{x}$ we find 
\begin{align}
M_{x} & =i\epsilon+\left(\mathbf{\hat{\mathbf{v}}p}\right)_{\Sigma_{x}}-b_{1x}\Lambda_{2}-b_{2x}L_{2}\Lambda_{3}\ \nonumber \\
 & =\left(\begin{array}{cc}
M_{1}^{a} & ib_{2x}\Lambda_{3}\\
-ib_{2x}\Lambda_{3} & M_{1}^{b}
\end{array}\right)_{L},
\end{align}
with \begin{subequations} 
\begin{align}
M_{1}^{a} & =i\epsilon+vp_{x}\cos\theta\Lambda_{3}+vp_{y}\sin\theta\Lambda_{3}-b_{1x}\Lambda_{2}\ ,\\
M_{1}^{b} & =i\epsilon-vp_{x}\cos\theta\Lambda_{3}+vp_{y}\sin\theta\Lambda_{3}-b_{1x}\Lambda_{2}\ ,
\end{align}
\end{subequations} which leads to \begin{subequations} 
\begin{align}
\det M & =\det M_{1}^{b}\det M_{1}\ ,\\
M_{1} & =M_{1}^{a}-b_{2x}^{2}\Lambda_{3}\left(M_{1}^{b}\right)^{-1}\Lambda_{3}\ .
\end{align}
\end{subequations}

We now decompose once again with \begin{subequations} 
\begin{align}
M_{1} & =\left(i\epsilon-b_{1x}\Lambda_{2}+vp_{x}\cos\theta\Lambda_{3}\right)\left(1+\bar{d}_{2}\right)\nonumber \\
 & \quad+vp_{y}\sin\theta\Lambda_{3}\left(1-\bar{d}_{2}\right)\ ,\\
 & \bar{d}_{2}=\frac{b_{2x}^{2}}{\epsilon^{2}+\left(vp_{x}\cos\theta-vp_{y}\sin\theta\right)^{2}+b_{1x}^{2}}\ ,\\
M_{1} & =\left(\begin{array}{cc}
M_{2}^{a} & ib_{1x}\left(1+\bar{d}_{2}\right)\\
-ib_{1x}\left(1+\bar{d}_{2}\right) & M_{2}^{b}
\end{array}\right)_{\Lambda},\mbox{ with }\\
M_{2}^{a} & =\left(i\epsilon+vp_{x}\cos\theta\right)\left(1+\bar{d}_{2}\right)+vp_{y}\sin\theta\left(1-\bar{d}_{2}\right),\\
M_{2}^{b} & =\left(i\epsilon-vp_{x}\cos\theta\right)\left(1+\bar{d}_{2}\right)-vp_{y}\sin\theta\left(1-\bar{d}_{2}\right).
\end{align}
\end{subequations} We get \begin{subequations} 
\begin{align}
\det M_{x} & =\det M_{1}^{b}\det M_{2}^{b}\det M_{2}\ ,\\
M_{2} & =M_{2}^{a}-\frac{b_{1x}^{2}\left(1+\bar{d}_{2}\right)^{2}}{\left(i\epsilon-vp_{x}\cos\theta\right)\left(1+\bar{d}_{2}\right)-vp_{y}\sin\theta\left(1+\bar{d}_{2}\right)}\ .
\end{align}
\end{subequations} We finally obtain \begin{subequations} 
\begin{align}
\det M_{x} & =\frac{\left(\epsilon^{2}+b_{1x}^{2}\right)d_{x}^{2}+\left(vp_{x}\cos\theta d_{x}+vp_{y}\sin\theta\left(d_{x}-2b_{2x}^{2}\right)\right)^{2}}{d_{x}-b_{2x}^{2}},\\
 & d_{x}=\epsilon^{2}+\left(vp_{x}\cos\theta-vp_{y}\sin\theta\right)^{2}+b_{1x}^{2}+b_{2x}^{2}.
\end{align}
\end{subequations} Reducing in the same manner the projection onto
the $\Sigma_{y}$ axis, and noticing that we get the right formulae
by shifting $\theta\rightarrow\pi/2-\theta$, we have \begin{subequations}
\begin{align}
\det M_{y} & =\frac{\left(\epsilon^{2}+b_{1y}^{2}\right)d_{y}^{2}+\left(vp_{x}\sin\theta d_{y}+vp_{y}\cos\theta\left(d_{y}-2b_{2y}^{2}\right)\right)^{2}}{d_{y}-b_{2y}^{2}},\\
 & d_{y}=\epsilon^{2}+\left(vp_{x}\sin\theta-vp_{y}\cos\theta\right)^{2}+b_{1y}^{2}+b_{2y}^{2},
\end{align}
\end{subequations} and we finally get Eq.\ (\ref{eq:8}) for the
free energy.

\section{\label{sec:APPENDIX-B:-Derivation}Derivation of the
mean-field equations}

\label{sec:mean-field-deriv} The Mean-Field Equations (MFEs) are
derived by differentiation of the free energy Eq.\ (\ref{eq:7})
with respect to $\bar{B}_{x,y}$ and $B_{x,y}$ successively. 
We get \begin{subequations} 
\begin{align}
J_{\mathbf{x-x'}}^{-1} \hat{B}_{x} & =-\frac{1}{2}\mathrm{Tr}\left[\hat{g}_{x}\right],\label{eq:34}\\
J_{\mathbf{x-x'}}^{-1} \hat{\overline{B}}_{x} & =-\frac{1}{2}\mathrm{Tr}\left[\hat{g}_{y}\right],\label{eq:35}\\
J_{\mathbf{x-x'}}^{-1} \hat{B}_{y} & =-\frac{1}{2}\mathrm{Tr}\left[\hat{g}_{y}\right],\label{eq:36}\\
J_{\mathbf{x-x'}}^{-1} \hat{\overline{B}}_{y} & =-\frac{1}{2}\mathrm{Tr}\left[\hat{g}_{x}\right],\label{eq:37}\\
\hat{g}_{x}=\left(g_{0x}^{-1}+\hat{b}_{x}\right)^{-1},\,\, & \,\,\hat{g}_{y}=\left(g_{0y}^{-1}+\hat{b}_{y}\right)^{-1}.
\end{align}
\end{subequations} We see that when the MFE do have a solution, Eq.\ (\ref{eq:34})
reduces identically to Eq.\ (\ref{eq:37}) and Eq.\ (\ref{eq:35})
reduces to Eq.\ (\ref{eq:36}). Two constraints are naturally obtained: 
\begin{align}
\hat{\overline{B}}_{x}=\hat{B}_{y},\qquad\hat{\overline{B}}_{y}=\hat{B}_{x}.\label{rep1}
\end{align}
The constraints (\ref{rep1}) correspond to a condition of reality
for the fields $\hat{B}_{x}$($\hat{B}_{y}$) within the conjugation
operation introduced in (\ref{eg6}). 
Multiplying both sides of Eqs.\ (\ref{eq:34}-\ref{eq:37})
by $J_{x-x'}$ we get 

\begin{align}
\hat{B}_{x} & =-\frac{1}{2}J_{x-x'}\mathrm{Tr}\left[\hat{g}_{x}\right],\label{eq:34-2}\\
\hat{\overline{B}}_{x} & =-\frac{1}{2}J_{x-x'}\mathrm{Tr}\left[\hat{g}_{y}\right],\label{eq:35-1}\\
\hat{B}_{y} & =-\frac{1}{2}J_{x-x'}\mathrm{Tr}\left[\hat{g}_{y}\right],\label{eq:36-1}\\
\hat{\overline{B}}_{y} & =-\frac{1}{2}J_{x-x'}\mathrm{Tr}\left[\hat{g}_{x}\right],\label{eq:37-1}
\end{align}
which after Fourier transforming, leads to

\begin{align}
T\sum_{\epsilon_{n},\mathbf{k}}\hat{B}_{x,\mathbf{k},\mathbf{k}+\mathbf{P}} & =-\frac{T^{2}}{2}\mathrm{Tr}\sum_{\epsilon_{n},\mathbf{k},\omega_{n},\mathbf{q}}J_{\omega_{n},\mathbf{q}}\left[\hat{g}_{x,\mathbf{k},\mathbf{k}+\mathbf{P}+\mathbf{q}}\right],\label{eq:34-2-1}\\
T\sum_{\epsilon_{n},\mathbf{k}}\hat{\bar{B}}_{x,\mathbf{k},\mathbf{k}+\mathbf{P}} & =-\frac{T^{2}}{2}\mathrm{Tr}\sum_{\epsilon_{n},\mathbf{k},\omega_{n},\mathbf{q}}J_{\omega_{n},\mathbf{q}}\left[\hat{g}_{y,\mathbf{k},\mathbf{k}+\mathbf{P}+\mathbf{q}}\right],\label{eq:35-1-1}\\
T\sum_{\epsilon_{n},\mathbf{k}}\hat{B}_{y,\mathbf{k},\mathbf{k}+\mathbf{P}} & =-\frac{T^{2}}{2}\mathrm{Tr}\sum_{\epsilon_{n},\mathbf{k},\omega_{n},\mathbf{q}}J_{\omega_{n},\mathbf{q}}\left[\hat{g}_{y,\mathbf{k},\mathbf{k}+\mathbf{P}+\mathbf{q}}\right],\label{eq:36-1-1}\\
T\sum_{\epsilon_{n},\mathbf{k}}\hat{\bar{B}}_{y,\mathbf{k},\mathbf{k}+\mathbf{P}} & =-\frac{T^{2}}{2}\mathrm{Tr}\sum_{\epsilon_{n},\mathbf{k},\omega_{n},\mathbf{q}}J_{\omega_{n},\mathbf{q}}\left[\hat{g}_{x,\mathbf{k},\mathbf{k}+\mathbf{P}+\mathbf{q}}\right].\label{eq:37-1-1}
\end{align}

In a similar way, differentiating Eq.\ (\ref{eq:8})
with respect to $b_{1x}$, $b_{1y}$, $b_{2x}$, $b_{2y}$ gives
four independent gap equations ; \begin{subequations} 
\begin{align}
\gamma_{1}^{-1}B_{1x} & =-T\sum_{\epsilon}\int\frac{d\mathbf{p}}{(2\pi)^{2}}\hat{D}\frac{\partial F_{x}}{\partial b_{1x}}\ ,\\
\gamma_{1}^{-1} B_{1y} & =-T\sum_{\epsilon}\int\frac{d\mathbf{p}}{(2\pi)^{2}}\hat{D}\frac{\partial F_{y}}{\partial b_{1y}}\ ,\\
\gamma_{2}^{-1} B_{2x} & =-T\sum_{\epsilon}\int\frac{d\mathbf{p}}{(2\pi)^{2}}\hat{D}\frac{\partial F_{x}}{\partial b_{2x}}\ ,\\
\gamma_{2}^{-1} B_{2y} & =-T\sum_{\epsilon}\int\frac{d\mathbf{p}}{(2\pi)^{2}}\hat{D}\frac{\partial F_{y}}{\partial b_{2y}}\ ,\\
 & \gamma_{1}=\frac{3g_{1}^{2}}{2},\quad\gamma_{2}=\frac{3g_{2}^{2}}{2}.
\end{align}
\label{mf1-1} \end{subequations} It is useful to introduce the notations
($x_{1}=b_{1x}^{2}$, $x_{2}=b_{2x}^{2}$, $y_{1}=b_{1y}^{2}$, $y_{2}=b_{2y}^{2}$)
\begin{subequations} 
\begin{align}
A_{1x} & =-\int\frac{d\mathbf{p}}{(2\pi)^{2}}\hat{D}\frac{\partial F_{x}}{\partial x_{1}}\ ,\\
A_{2x} & =-\int\frac{d\mathbf{p}}{(2\pi)^{2}}\hat{D}\frac{\partial F_{x}}{\partial x_{2}}\ ,\\
A_{1y} & =-\int\frac{d\mathbf{p}}{(2\pi)^{2}}\hat{D}\frac{\partial F_{y}}{\partial y_{1}}\ ,\\
A_{2y} & =-\int\frac{d\mathbf{p}}{(2\pi)^{2}}\hat{D}\frac{\partial F_{y}}{\partial y_{2}}\ .
\end{align}
\label{pref1} \end{subequations} The expressions for the partial
derivatives are given in Eq.\ (\ref{coefs}). With these notations,
the MFE write \begin{subequations} 
\begin{align}
\gamma_{1}^{-1} B_{1x} & =-2T\sum_{\epsilon}b_{1x}A_{1x}\ ,\\
\gamma_{1}^{-1} B_{1y} & =-2T\sum_{\epsilon}b_{1y}A_{1y}\ ,\\
\gamma_{2}^{-1} B_{2x} & =-2T\sum_{\epsilon}b_{2x}A_{2x}\ ,\\
\gamma_{2}^{-1} B_{2y} & =-2T\sum_{\epsilon}b_{2y}A_{2y}\ .
\end{align}
\label{mf_fin} \end{subequations} 
We can finally write the result in the form of Eq.\ (\ref{mfeqs}).

\section{Gaussian fluctuations}

\label{sec:gauss-fluct}

Let us explicitly derive the stability condition for one order parameter
$b_{x}$, $b_{y}$ in the presence of Gaussian fluctuations. Noting
$x=b_{x}^{2}$ and $y=b_{y}^{2}$, we have 
\begin{align}
F & =F_{0}+F_{x}(x)+F_{y}(y).
\end{align}
Using $B=B_{0}+\delta B$ and $\overline{B}=\overline{B}_{0}+\delta\overline{B}$
we get \begin{subequations} 
\begin{align}
F_{0} & =F_{0}^{(0)}+F_{0}^{(1)}+F_{0}^{(2)},\\
F_{0}^{(1)} & =J_{x-x'}^{-1} \left(\overline{B}_{x0}\delta B_{x}+B_{x0}\delta\overline{B}_{x}+\overline{B}_{y0}\delta B_{y0}+B_{y0}\delta\overline{B}_{y0}\right),\\
F_{0}^{(2)} & = J_{x-x'}^{-1} \left(\delta\overline{B}_{x}\delta B_{x}+\delta\overline{B}_{y}\delta B_{y}\right).
\end{align}
\end{subequations} From the MF relations $\overline{B}_{x0}=B_{y0}$
and $\overline{B}_{y0}=B_{x0}$ we obtain \begin{subequations} 
\begin{align}
F_{0}^{(1)} & =J_{x-x'}^{-1} \left(B_{xo}\delta b_{x}+B_{y0}\delta b_{y}\right),\\
\delta b_{x} & =\delta\overline{B}_{x}+\delta B_{y},\\
\delta b_{y} & =\delta\overline{B}_{y}+\delta B_{x}.
\end{align}
\end{subequations}

Now let us consider the second term in the Free energy: \begin{subequations}
\begin{align}
F_{x}^{(1)} & =2b_{x0}F_{x}^{\prime}\delta b_{x},\\
F_{x}^{(2)} & =\left(F_{x}^{\prime}+2b_{x0}^{2}F_{x}^{\prime\prime}\right)\left(\delta b_{x}\right)^{2},\\
F_{y}^{(1)} & =2b_{y0}F_{y}^{\prime}\delta b_{y},\\
F_{y}^{(2)} & =\left(F_{y}^{\prime}+2b_{y0}^{2}F_{y}^{\prime\prime}\right)\left(\delta b_{y}\right)^{2}.
\end{align}
\end{subequations} Hence we get for the factors $\delta A_{i}$ (
notation: $x_{1}=b_{1x}^{2}$, $x_{2}=b_{2x}^{2}$, $y_{1}=b_{1y}^{2}$,
$y_{2}=b_{2y}^{2}$ ) \begin{subequations} 
\begin{align}
\delta A_{1x} & =2\int\frac{d\mathbf{q}}{\left(2\pi\right)^{2}}D\left(\omega,\mathbf{q}\right)b_{1x}^{2}\frac{\partial^{2}F}{\partial x_{1}^{2}}\ ,\\
\delta A_{2x} & =2\int\frac{d\mathbf{q}}{\left(2\pi\right)^{2}}D\left(\omega,\mathbf{q}\right)b_{2x}^{2}\frac{\partial^{2}F}{\partial x_{2}^{2}}\ ,\\
\delta A_{1y} & =2\int\frac{d\mathbf{q}}{\left(2\pi\right)^{2}}D\left(\omega,\mathbf{q}\right)b_{1y}^{2}\frac{\partial^{2}F}{\partial y_{1}^{2}}\ ,\\
\delta A_{2y} & =2\int\frac{d\mathbf{q}}{\left(2\pi\right)^{2}}D\left(\omega,\mathbf{q}\right)b_{2y}^{2}\frac{\partial^{2}F}{\partial y_{2}^{2}}\ ,
\end{align}
\end{subequations} with \begin{subequations} 
\begin{align}
\frac{\partial^{2}F}{\partial x_{1}^{2}} & =\frac{d_{numx1}}{d_{denx}},\\
d_{denx} & =\left(d{}_{x}^{2}+4vq_{x}vq_{y}\cos\theta\sin\theta d_{x}-4b_{2x}^{2}v^{2}q_{y}^{2}\sin^{2}\theta\right)^{2},\\
d_{numx1} & =d_{x}^{2}+4vq_{x}vq_{y}\cos\theta\sin\theta\left(d_{x}+2vq_{x}vq_{y}\cos\theta\sin\theta\right)\nonumber \\
 & \quad+2b_{2x}^{2}v^{2}q_{y}^{2}\sin^{2}\theta,
\end{align}
\end{subequations} and \begin{subequations} 
\begin{align}
\frac{\partial^{2}F}{\partial y_{1}^{2}} & =\frac{d_{numy1}}{d_{deny}},\\
d_{deny} & =\left(d{}_{y}^{2}+4vq_{x}vq_{y}\cos\theta\sin\theta d_{y}-4b_{2x}^{2}v^{2}q_{y}^{2}\cos^{2}\theta\right)^{2},\\
d_{numy1} & =d_{y}^{2}+4vq_{x}vq_{y}\cos\theta\sin\theta\left(d_{y}+2vq_{x}vq_{y}\cos\theta\sin\theta\right)\nonumber \\
 & \quad+2b_{2y}^{2}v^{2}q_{y}^{2}\cos^{2}\theta,
\end{align}
\end{subequations} and \begin{subequations} 
\begin{align}
\frac{\partial^{2}F}{\partial x_{2}^{2}} & =\frac{d_{numx2}}{d_{denx}},\\
d_{denx} & =\left(d{}_{x}^{2}+4vq_{x}vq_{y}\cos\theta\sin\theta d_{x}-4b_{2x}^{2}v^{2}q_{y}^{2}\sin^{2}\theta\right)^{2},\\
d_{numx2} & =d_{x}^{2}+4vq_{x}vq_{y}\cos\theta\sin\theta d_{x}\nonumber \\
 & \quad-4b_{2x}^{2}v^{2}q_{y}^{2}\sin^{2}\theta\left(2b_{1x}^{2}+2\epsilon^{2}+b_{2x}^{2}\right),
\end{align}
\end{subequations} and \begin{subequations} 
\begin{align}
\frac{\partial^{2}F}{\partial y_{2}^{2}} & =\frac{d_{numy2}}{d_{deny}},\\
d_{deny} & =\left(d{}_{y}^{2}+4vq_{x}vq_{y}\cos\theta\sin\theta d_{y}-4b_{2x}^{2}v^{2}q_{y}^{2}\cos^{2}\theta\right)^{2},\\
d_{numy2} & =d_{y}^{2}+4vq_{x}vq_{y}\cos\theta\sin\theta d_{y}\nonumber \\
 & \quad-4b_{2y}^{2}v^{2}q_{y}^{2}\cos^{2}\theta\left(2b_{1y}^{2}+2\epsilon^{2}+b_{2y}^{2}\right).
\end{align}
\end{subequations} 

\section{Structure of the Mean-field solutions}

\label{sec:Structure-of-the}

Let us give some more numerical solutions of the Mean field equations.
In Fig.\ \ref{Fig2} are depicted the typical form of the QDW/SC and
CDW/PDW components of the CE solution. Despite the
CE solution, the MFEs (\ref{mf1}-\ref{mf4}) admits two other solutions
that we describe here.

The pure QDW/SC solution is depicted on Fig.\ \ref{Fig6} below. The
most noticeable fact about this solution is the very feeble dependence
on the Fermi velocity angle $\theta$. This solution is very robust
to changes of the shape of the Fermi surface at the hot-spots and
at the anti-nodes, characterized by $\theta=0$ for flat portions
of the Fermi surface at the anti-nodes and $\theta=\pi/4$ for the
generic case. The pure QDW/SC solution is very similar to the observed
PG of cuprate superconductors in that respect.

\begin{figure}[H]
$a_{1}$) \includegraphics[scale=0.24]{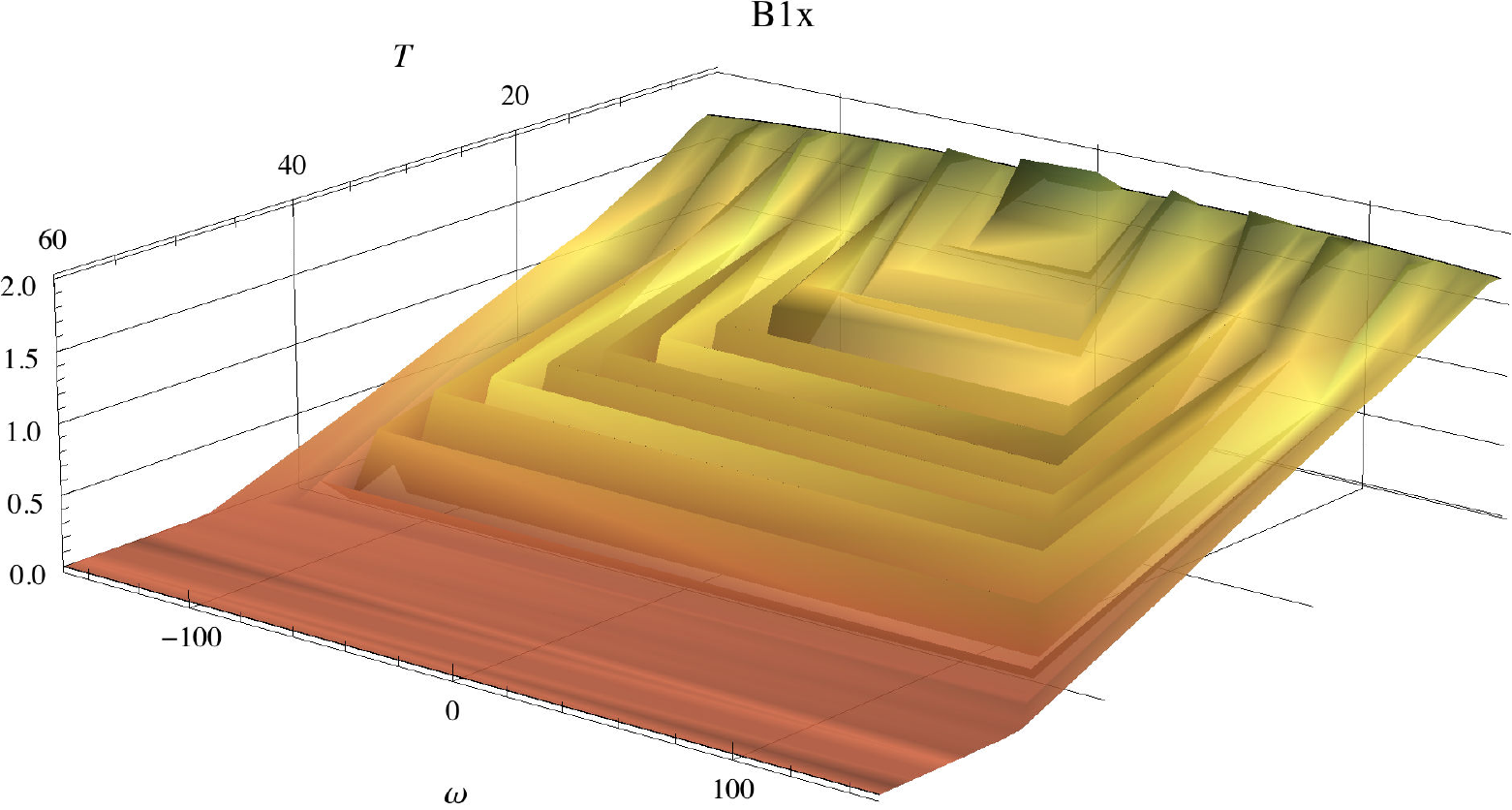} $a_{2}$) \includegraphics[scale=0.24]{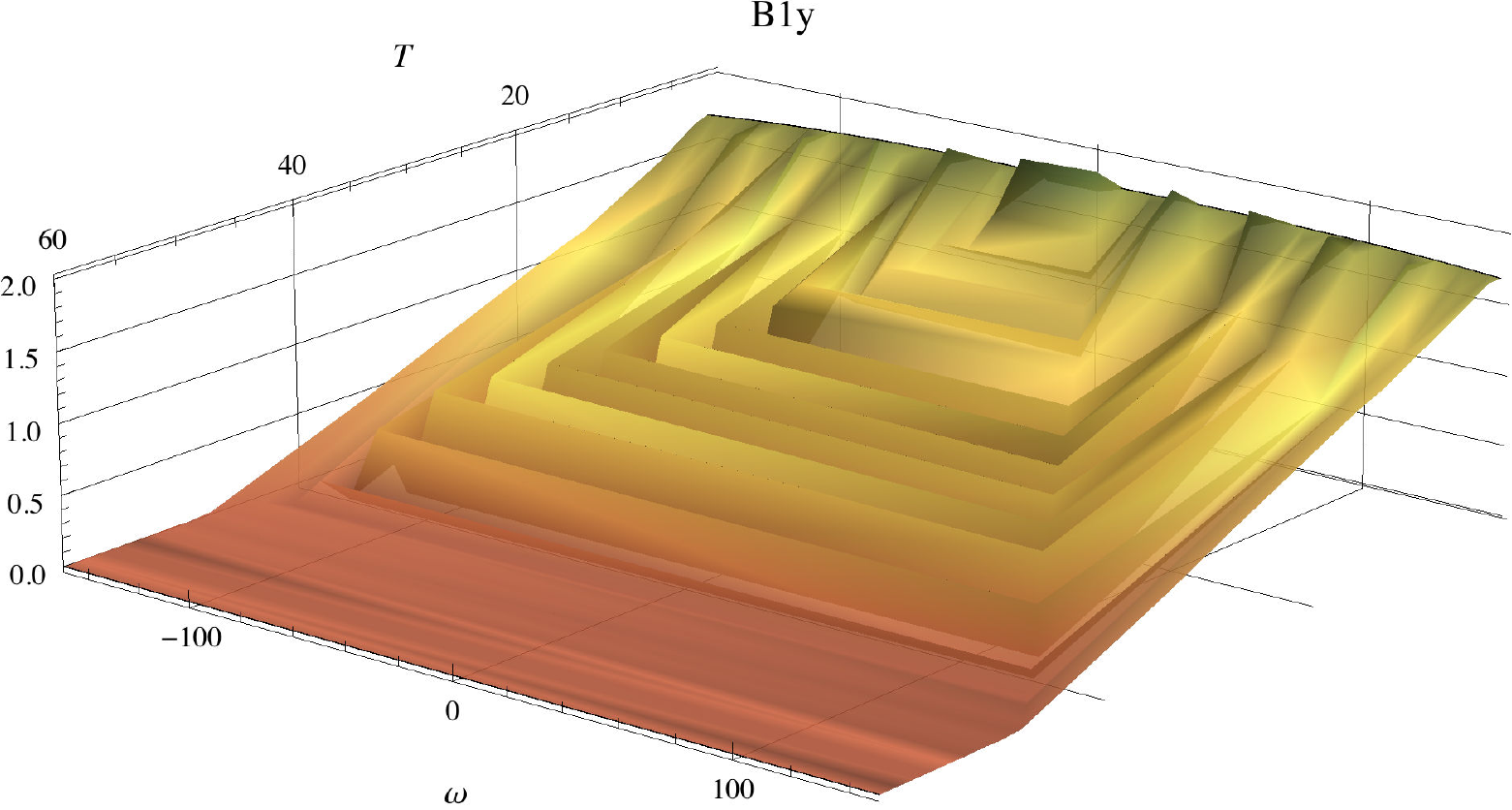}

$b_{1}$) \includegraphics[scale=0.24]{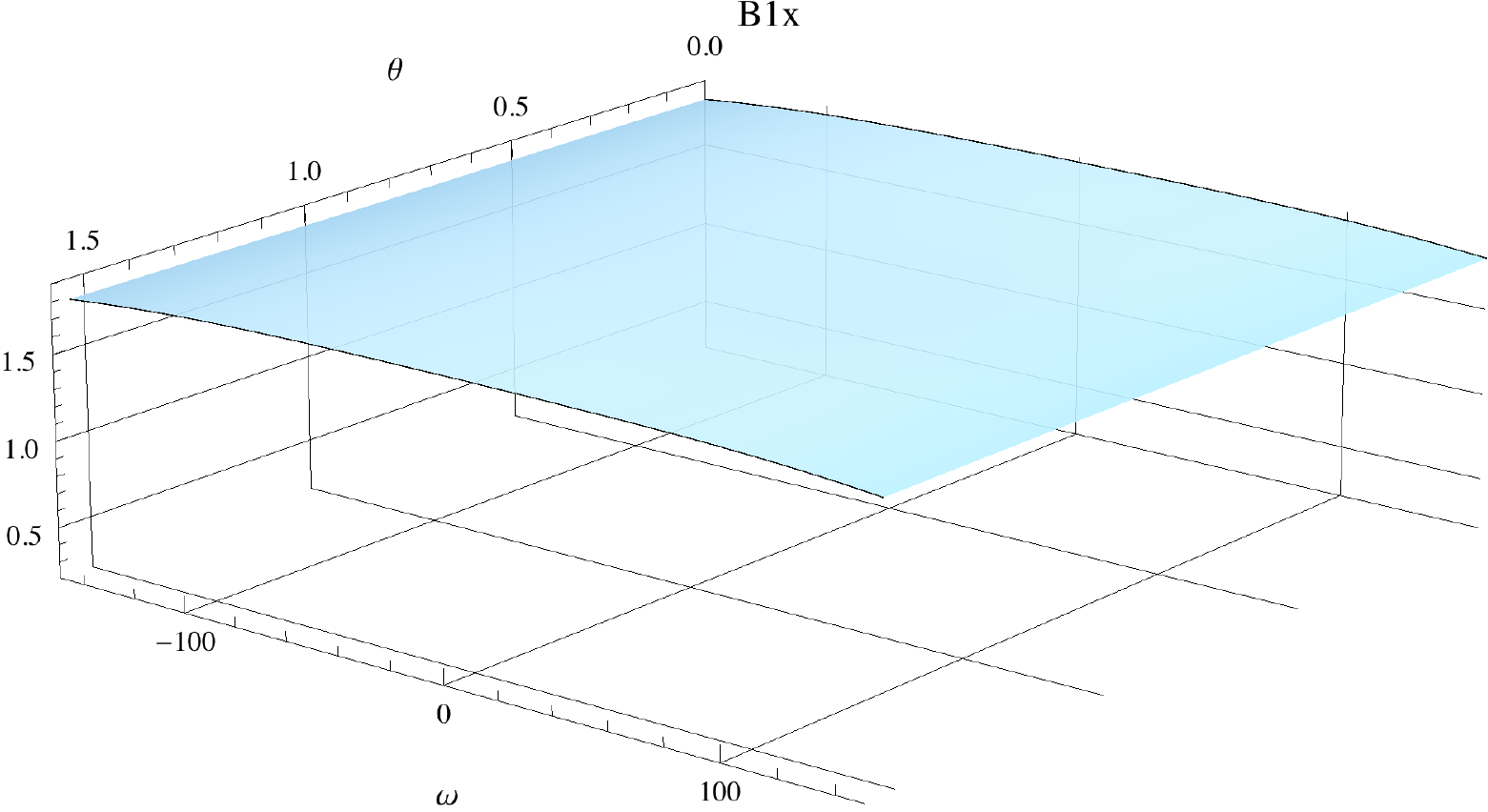} $b_{2}$) \includegraphics[scale=0.24]{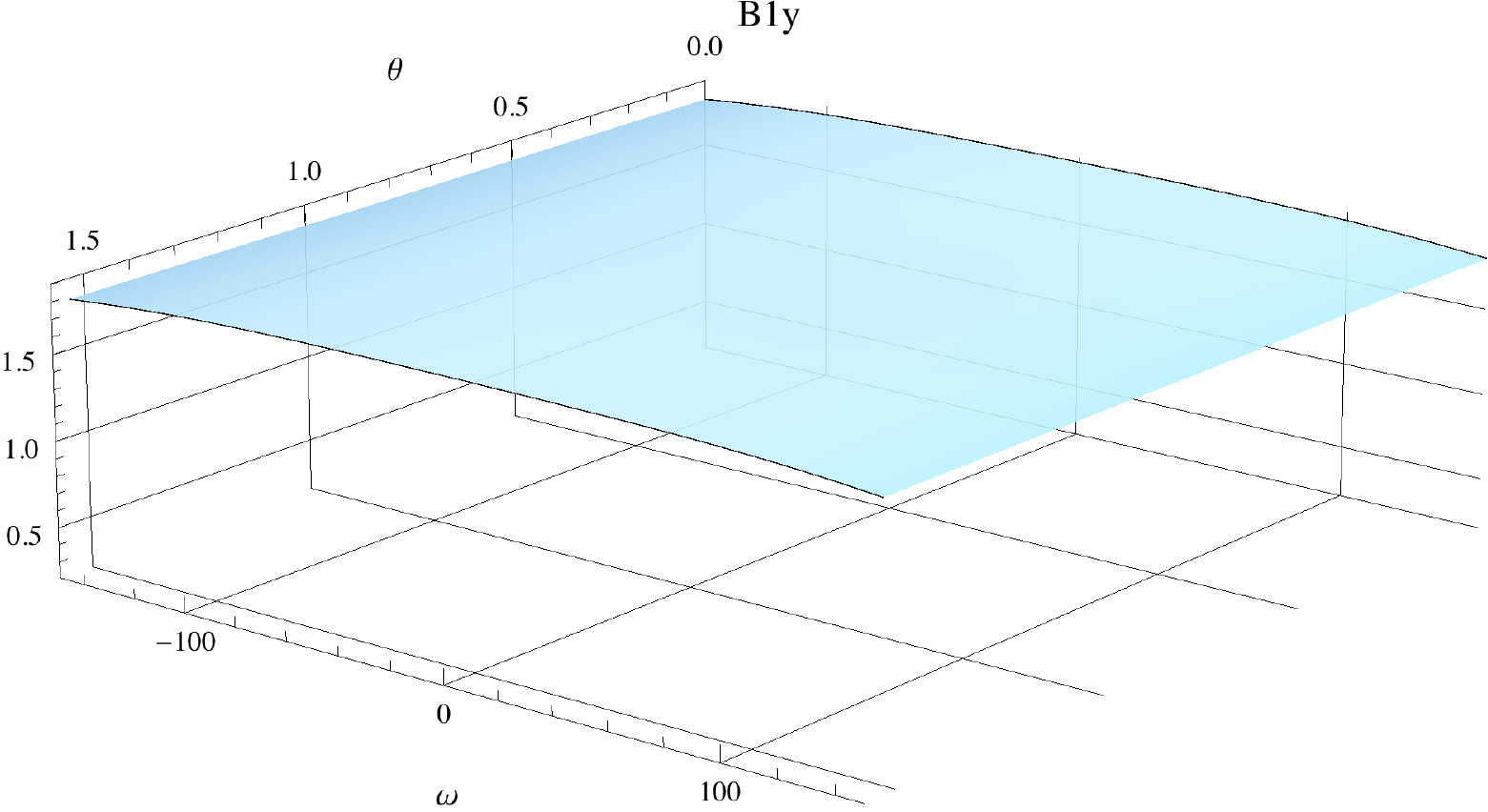}

\caption{\label{Fig6} (Color online) Typical solution for the pure QDW/SC
solution, as a function of $\left(\epsilon_{n},T\right)$ for $a_{1}$)
and $a_{2}$), and as a function of $\left(\epsilon_{n},\theta\right)$
for $b_{1}$) and $b_{2}$). The values of the parameters are $g_{1}=20$,
$g_{2}=30$, $v=6$, $m_{a}=0.1$, $\gamma=3$, $W=2\pi$.
The velocity angle is $\theta=0.1$ for $a_{1}$) and $a_{2}$) whereas
the temperature is $T=1$ for $b_{1}$) and $b_{2}$). Note the very
feeble dependence in the Fermi angle for this solution. }
\end{figure}

\begin{figure}[H]
$a_{1}$) \includegraphics[scale=0.24]{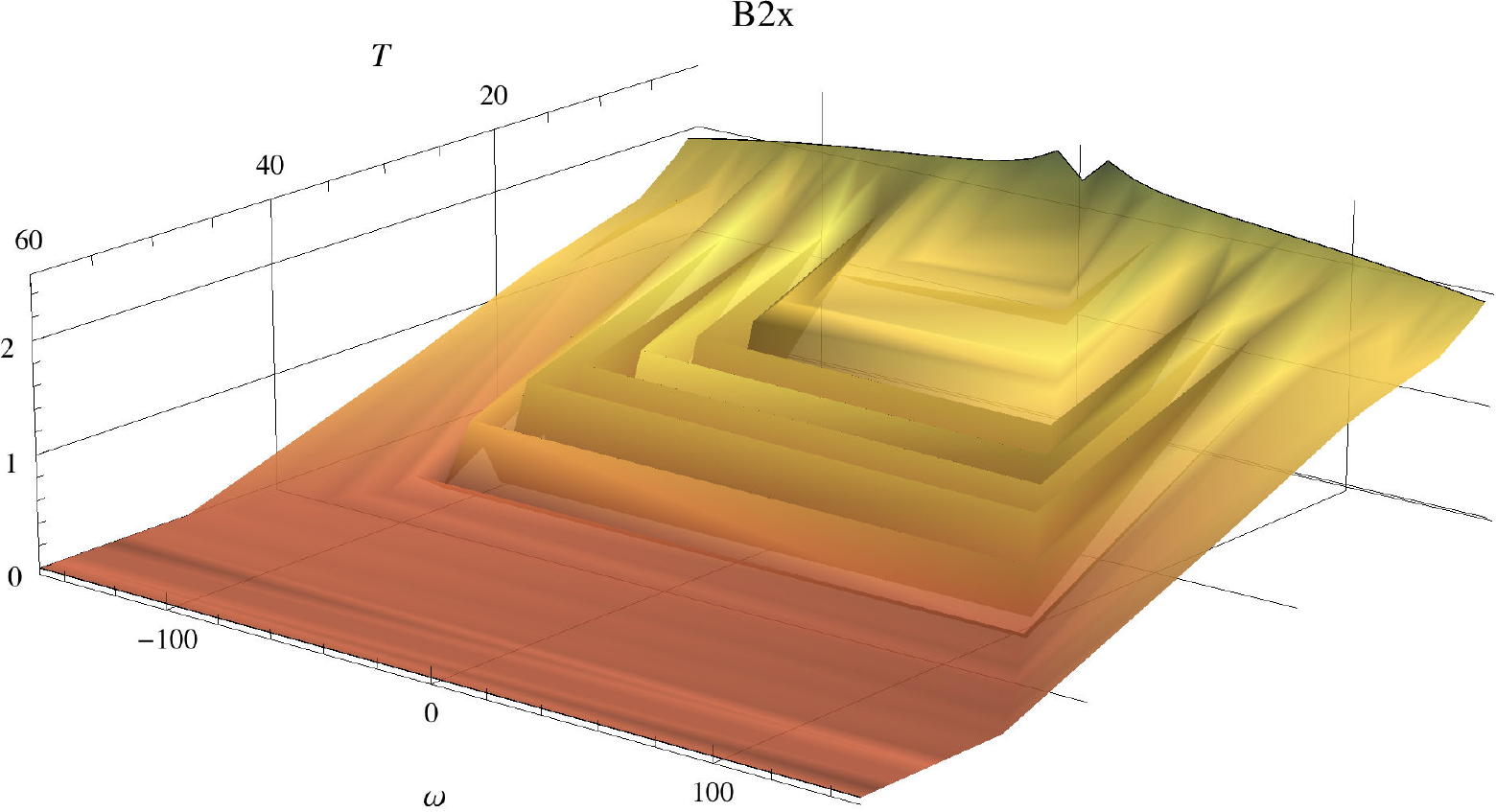} $a_{2}$) \includegraphics[scale=0.24]{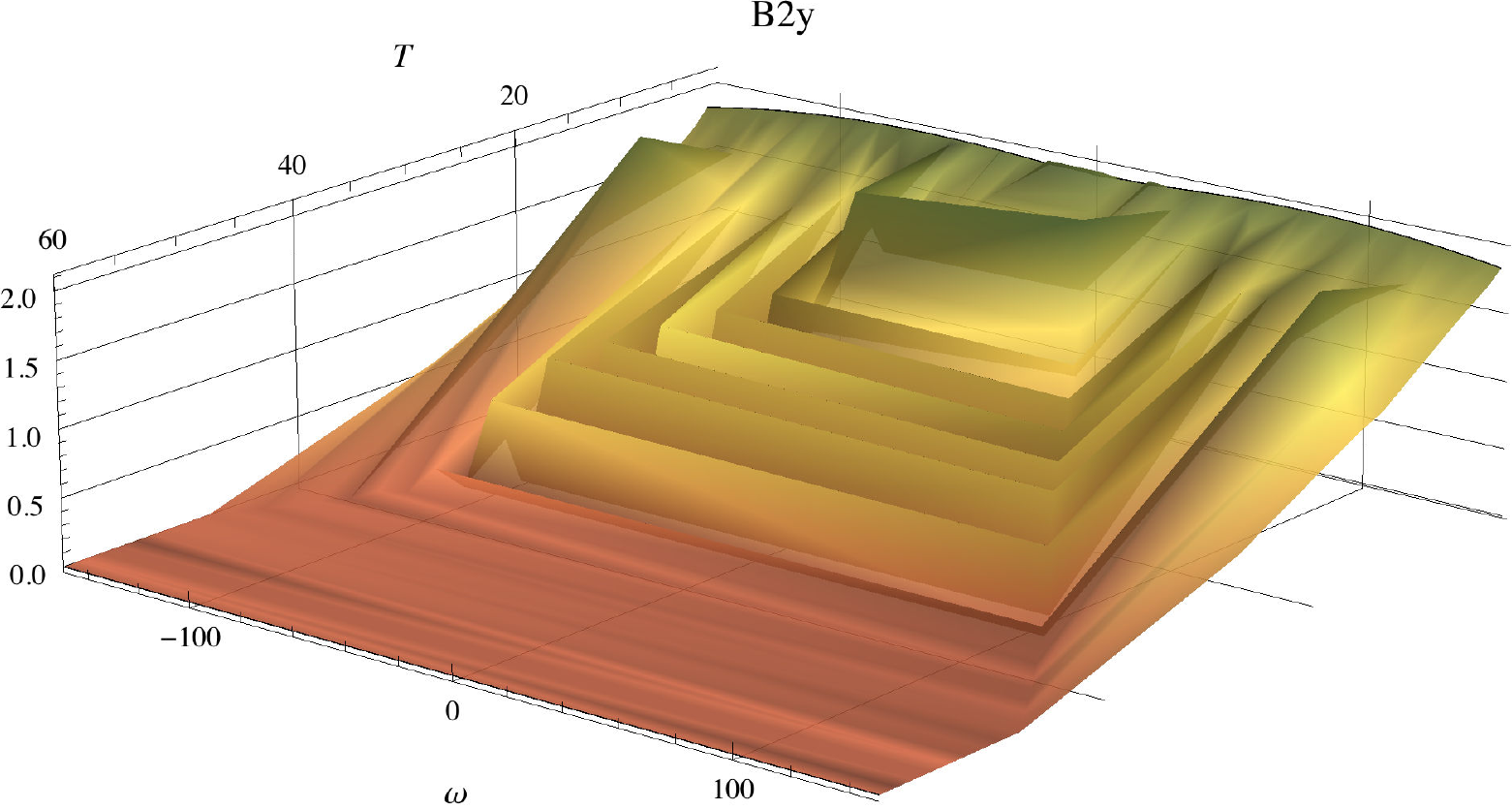}

$b_{2}$) \includegraphics[scale=0.24]{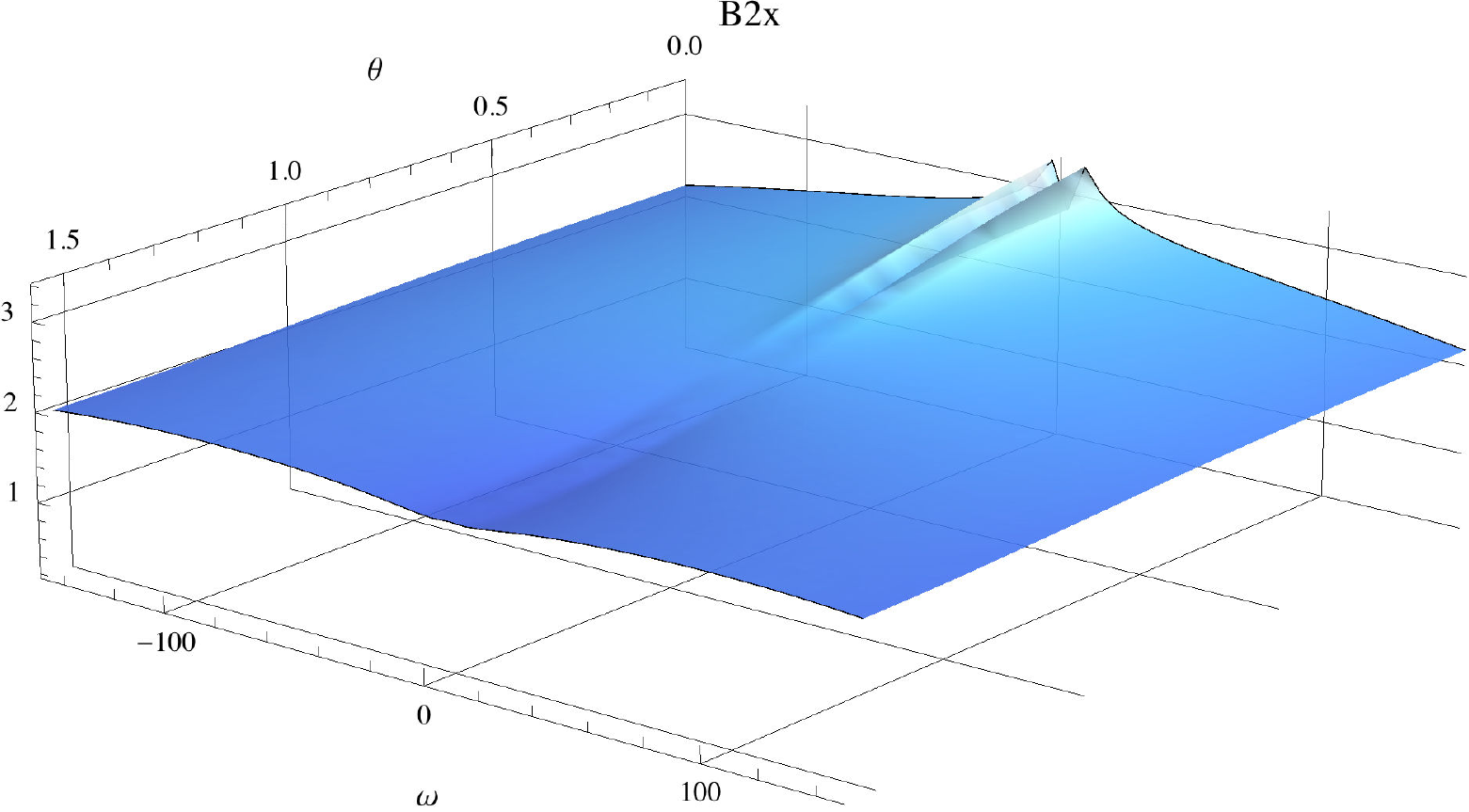} $b_{2}$) \includegraphics[scale=0.24]{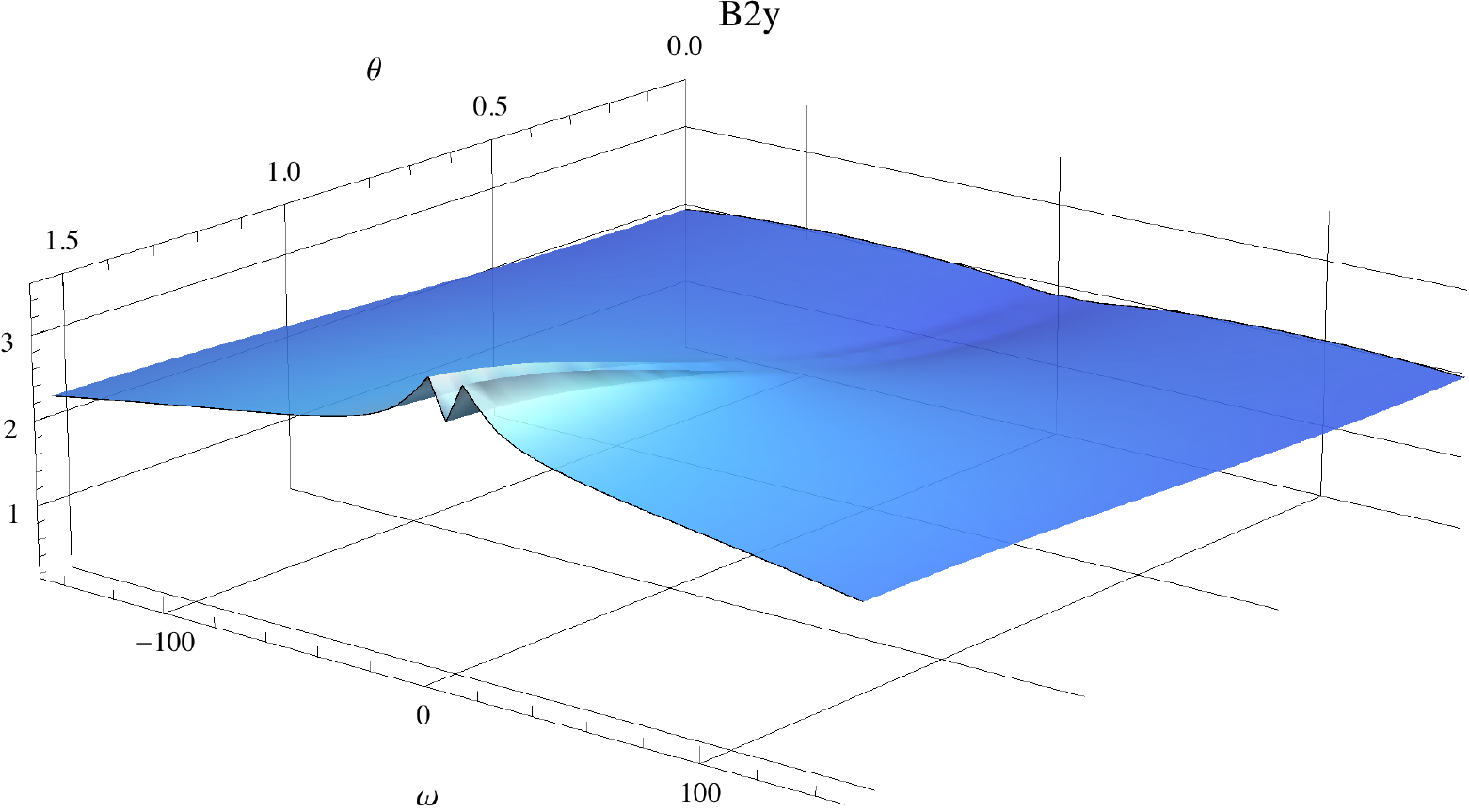}

\caption{\label{Fig7} (Color online) Typical solution for the pure CDW solution,
as a function of $\left(\epsilon_{n},T\right)$ for $a_{1}$) and
$a_{2}$), and as a function of $\left(\epsilon_{n},\theta\right)$
for $b_{1}$) and $b_{2}$). The values of the parameters are $g_{1}=20$,
$g_{2}=30$, $v=6$, $m_{a}=0.1$, $\gamma=3$, $W=2\pi$.
The velocity angle is $\theta=0.1$ for $a_{1}$) and $a_{2}$) whereas
the temperature is $T=1$ for $b_{1}$) and $b_{2}$). Note that although
the $T$-dependence of the solution is different from the one on Fig.\ \ref{Fig7},
especially as the magnitude of the solution is concerned, its angular
dependence is very similar to the one depicted in Fig.\ \ref{Fig7}.}
\end{figure}

In contrast, the pure CDW/PDW solution depicted
in Fig.\ \ref{Fig7} is much more dependent on the angle $\theta$ of
the Fermi velocity at the anti-nodes.

\section{Extreme limit where $\gamma_{1}\gg\gamma_{0}$}

\label{sec:Extreme-limit-where} For completeness, let us discuss
the extreme limit where $J_{2}\gg J_{1}$ in Eqs.\ (\ref{mf1}-\ref{mf4}).
Recently, an interesting work \cite{Wang14} has proposed the pure
CDW solution as a candidate for the PG phase. This solution is pre-empted
by the formation of a $q=0$-bond state at the PG temperature $T^{*}$,
which has the property to give a non zero Kerr signal \cite{Xia08}.
It is interesting to see what happens within the study of co-existence
when the quadratic coupling constant $J_{2}$ favoring
the CDW order is pushed to a very high limit compared to $J_{1}$.
This study is presented below for $J_{2}=10\ J_{1}$.

First, it is worth noticing that the three solutions {[}pure CDW/PDW,
pure QDW/SC and the CE solution{]} are still present in the extreme
limit where $J_{2}\gg J_{1}$, which is extremely
favorable to the pure CDW/PDW order. It
happens that in this limit, the pure CDW/PDW solution
becomes unstable towards the CE solution. Comparison of the free energies
shows that the CE solution has a slightly lower energy than the pure
CDW/PDW solution in this case, while the splitting
with the QDW/SC solution is higher (Fig.\ \ref{Fig14}).

\subsection{MF solutions}

\begin{figure}[H]
$a_{1}$) \includegraphics[scale=0.24]{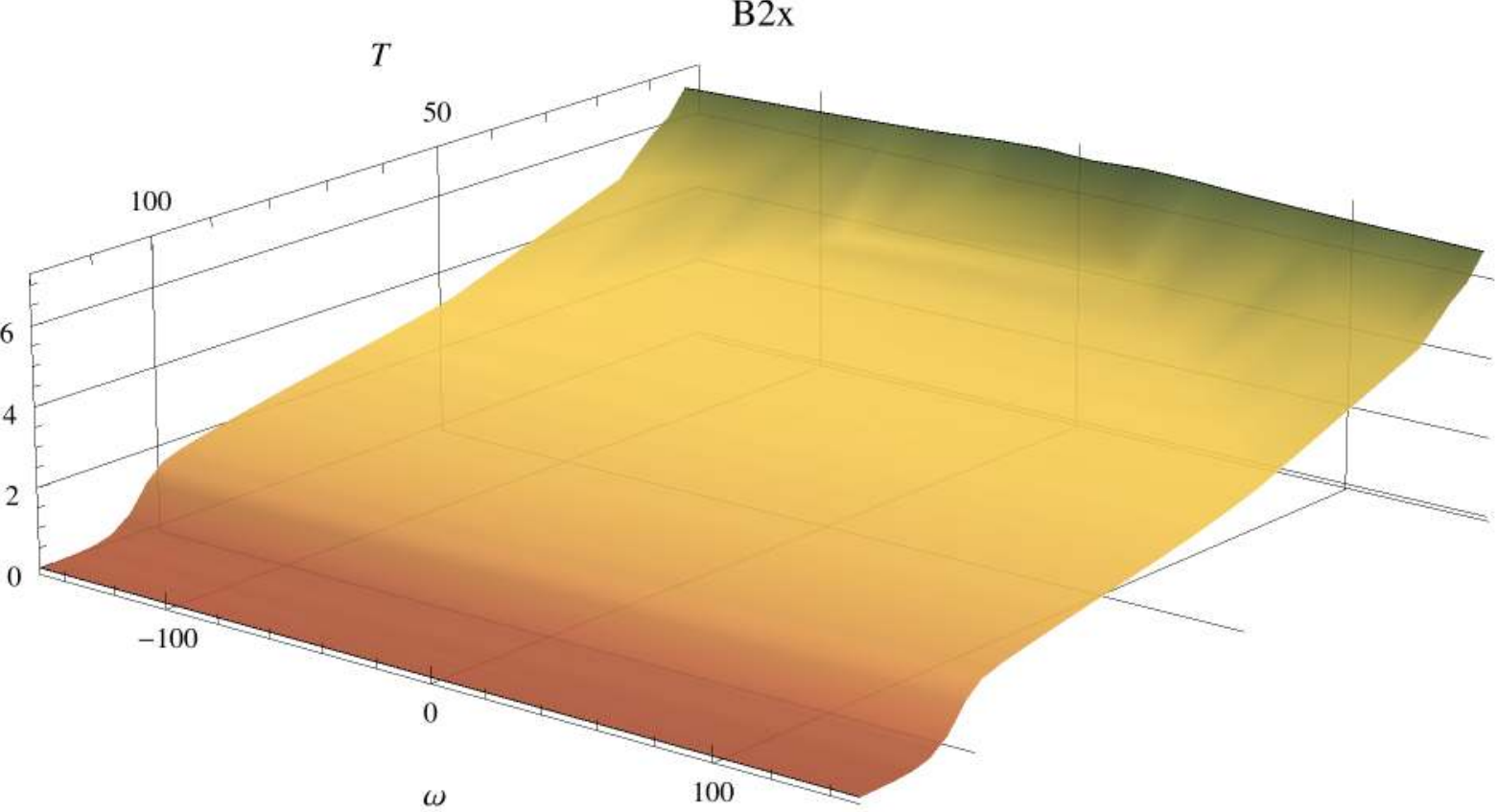} $a_{2}$) \includegraphics[scale=0.24]{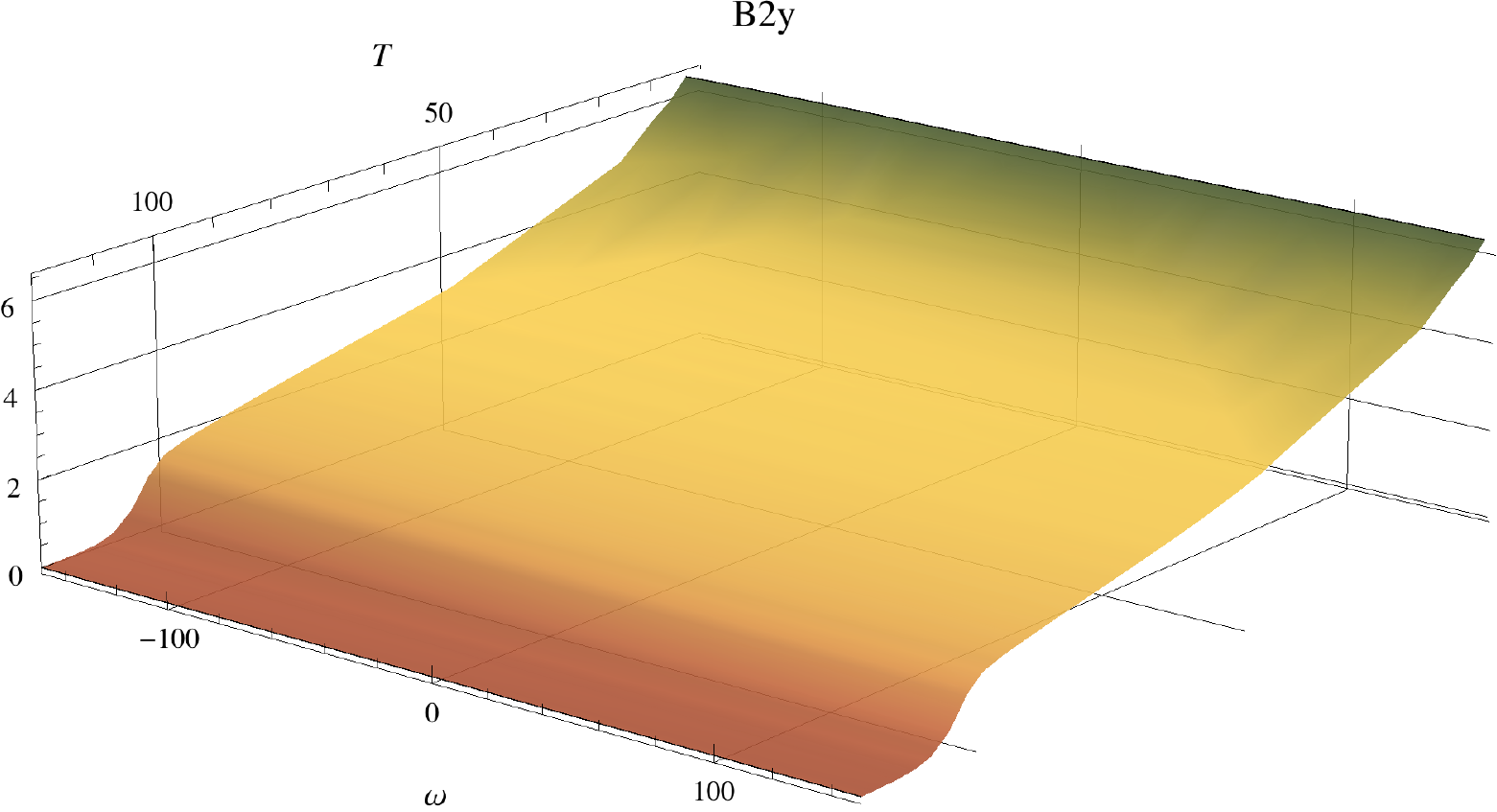}

$b_{1}$) \includegraphics[scale=0.24]{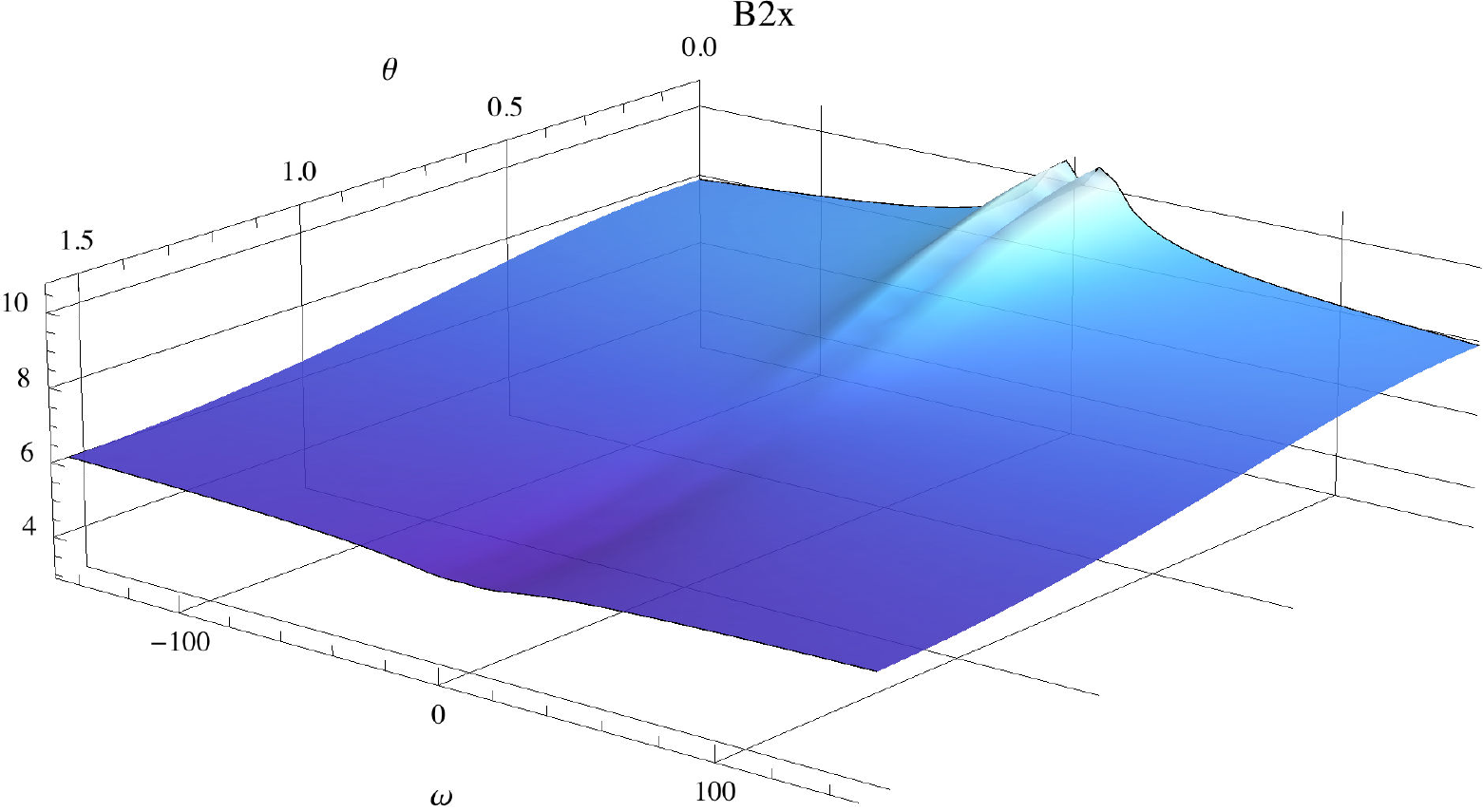} $b_{1}$) \includegraphics[scale=0.24]{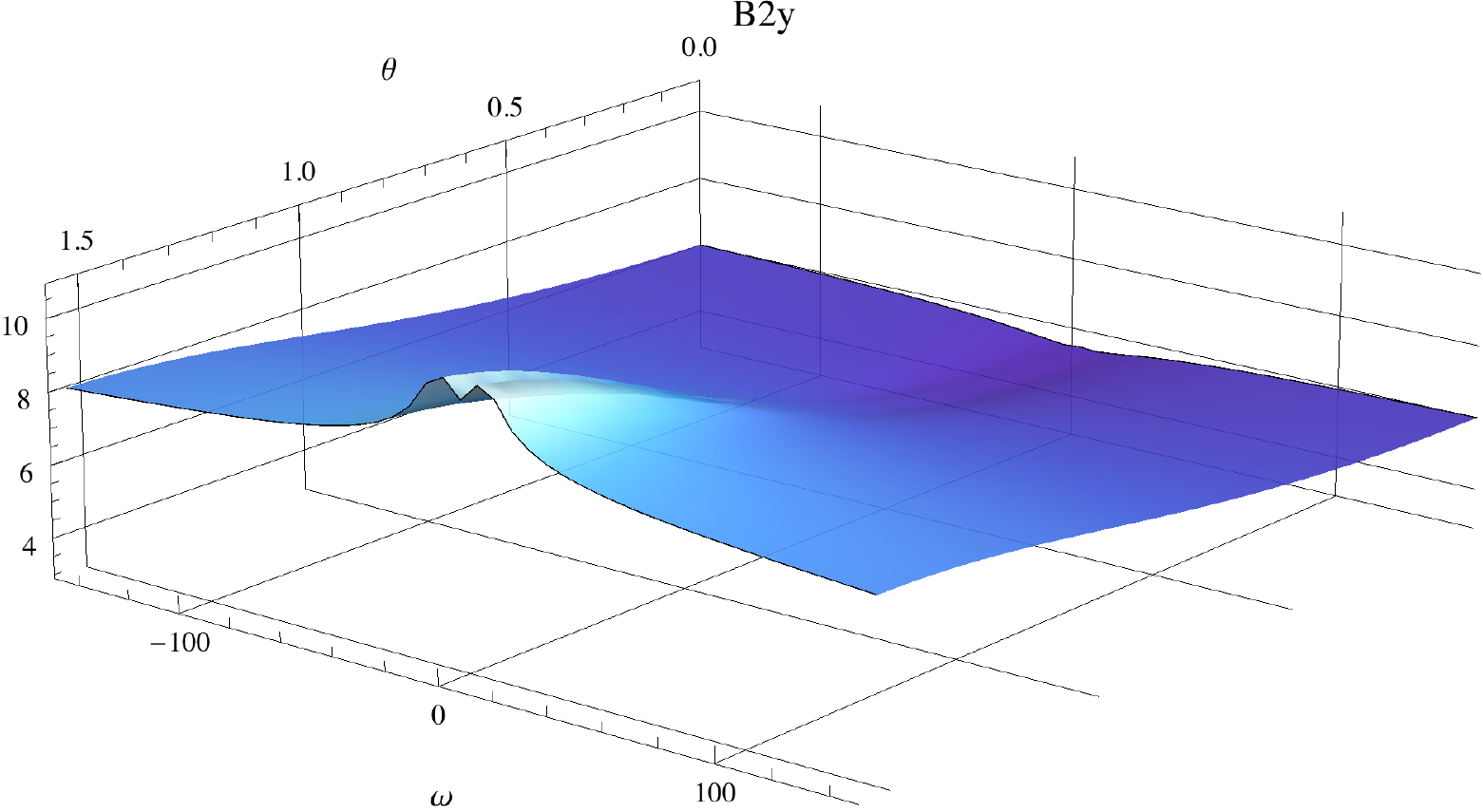}

\caption{\label{Fig8-1} (Color online) Typical solution for the pure CDW/PDW  solution, as a function of $\left(\epsilon_{n},T\right)$
for $a_{1}$) and $a_{2}$), and as a function of $\left(\epsilon_{n},\theta\right)$
for $b_{1}$) and $b_{2}$). The values of the parameters are $g_{1}=20$,
$g_{2}=200$, $v=6$, $m_{a}=0.1$, $\gamma=3$, $W=2\pi$. The velocity
angle is $\theta=0.1$ for $a_{1}$) and $a_{2}$) whereas the temperature
is $T=1$ for $b_{1}$) and $b_{2}$). Note the strong dependence
in the Fermi angle for this solution. }
\end{figure}

We start with the pure CDW/PDW solution depicted
in Fig.\ \ref{Fig8-1}. On can observe the large magnitude of the pure
CDW/PDW solution {[}graphs $a_{1}$) and $a_{2}$){]}
whereas the $\theta$-dependence of the solution {[}graphs $b_{1}$)
and $b_{2}$){]} has not changed compared to the one in Fig.\ \ref{Fig7}.

\begin{figure}[H]
$a_{1}$) \includegraphics[scale=0.24]{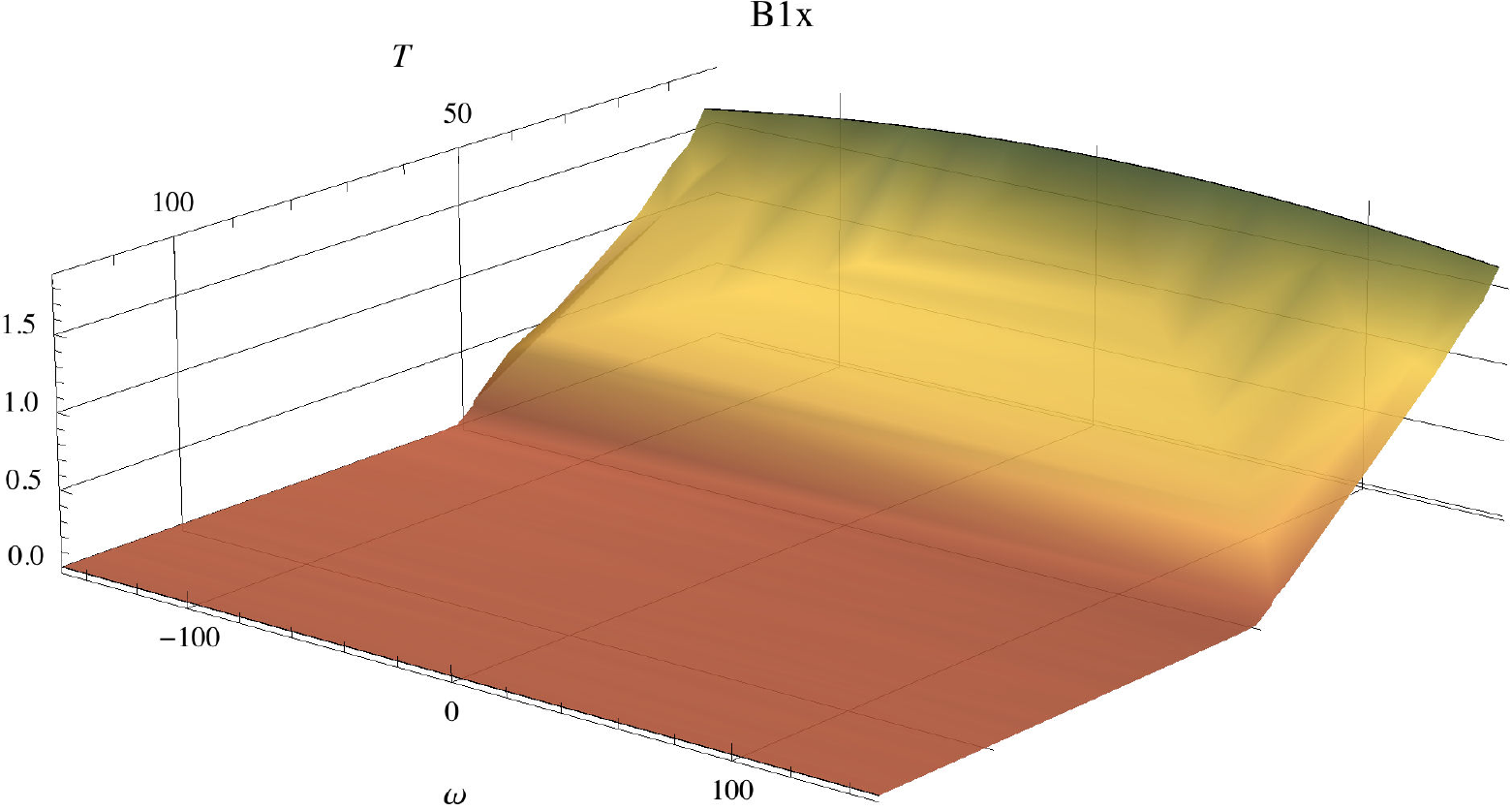} $a_{2}$) \includegraphics[scale=0.24]{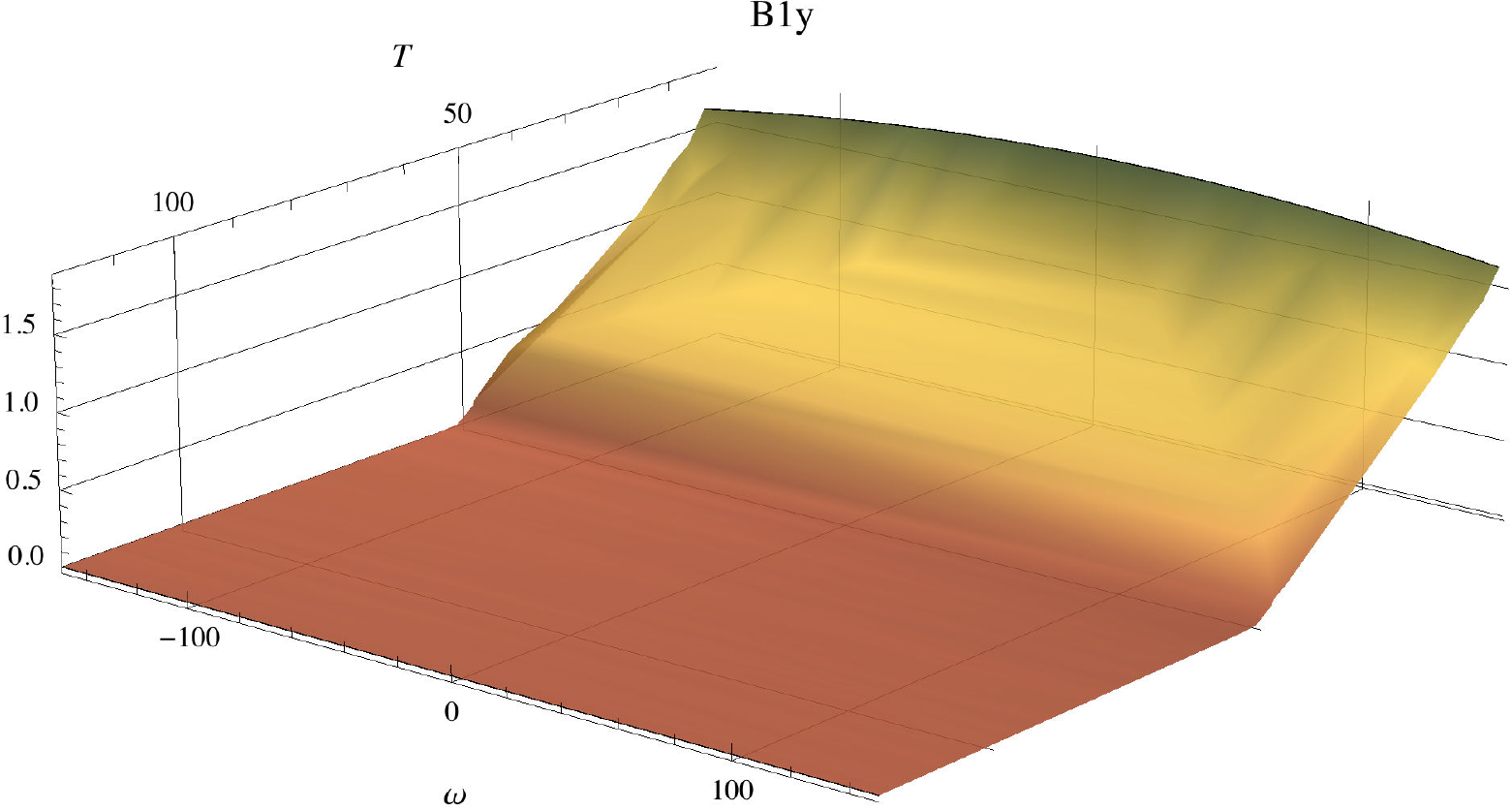}

$b_{1}$) \includegraphics[scale=0.24]{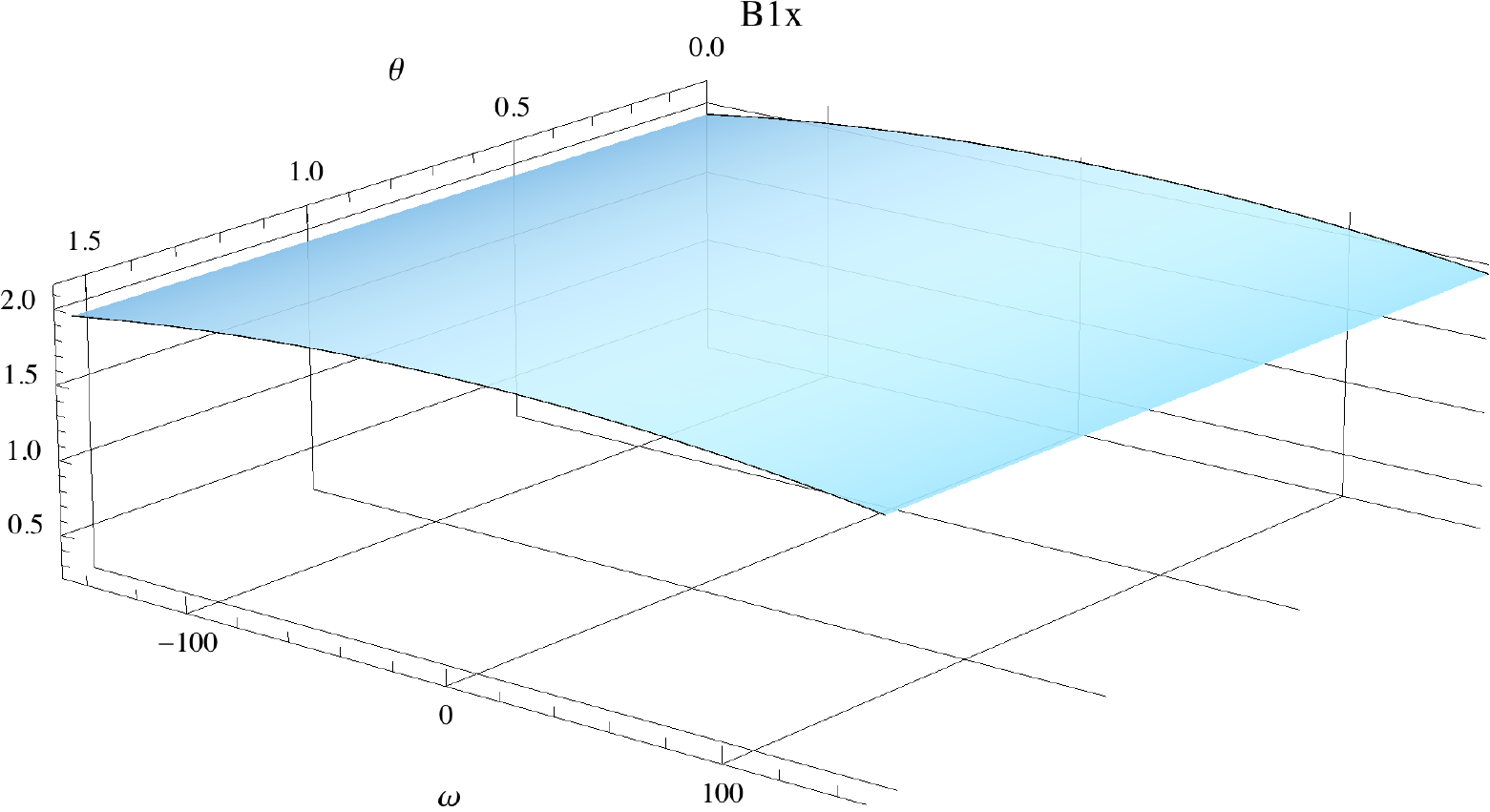} $b_{2}$) \includegraphics[scale=0.24]{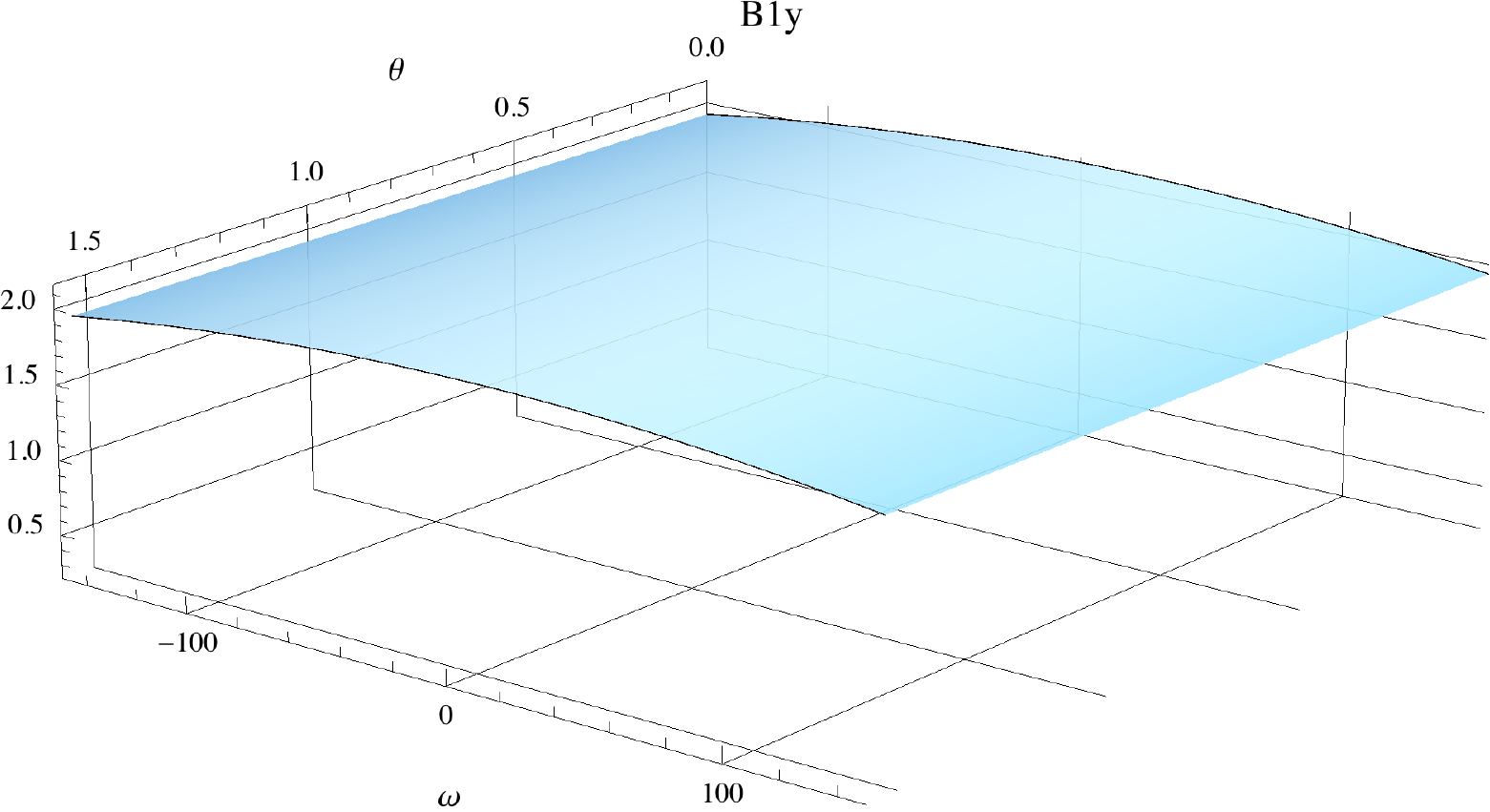}

\caption{\label{Fig9} (Color online) Typical solution for the pure QDW/SC
solution, as a function of $\left(\epsilon_{n},T\right)$ for $a_{1}$)
and $a_{2}$), and as a function of $\left(\epsilon_{n},\theta\right)$
for $b_{1}$) and $b_{2}$). The values of the parameters are $g_{1}=20$,
$g_{2}=200$, $v=6$, $m_{a}=0.1$, $\gamma=3$,
$W=2\pi$. The velocity angle is $\theta=0.1$ for $a_{1}$) and $a_{2}$)
whereas the temperature is $T=1$ for $b_{1}$) and $b_{2}$). Note
the striking similarities between this solution and the one in Fig.\ \ref{Fig6}.}
\end{figure}

We turn now to the pure QDW/SC solution, which was introduced in Ref.\ 
[\onlinecite{Efetov13}] as a good candidate for the PG at $T^{*}$. Comparing
Fig.\ \ref{Fig9} to Fig.\ \ref{Fig6}, we see the similarity between
the two solutions. It is to be expected since only the parameter $J_{2}$
has been increased between the two figures and the pure QDW/SC solution
is insensitive to the value of $J_{2}$.

\begin{figure}[H]
$a_{1}$) \includegraphics[scale=0.24]{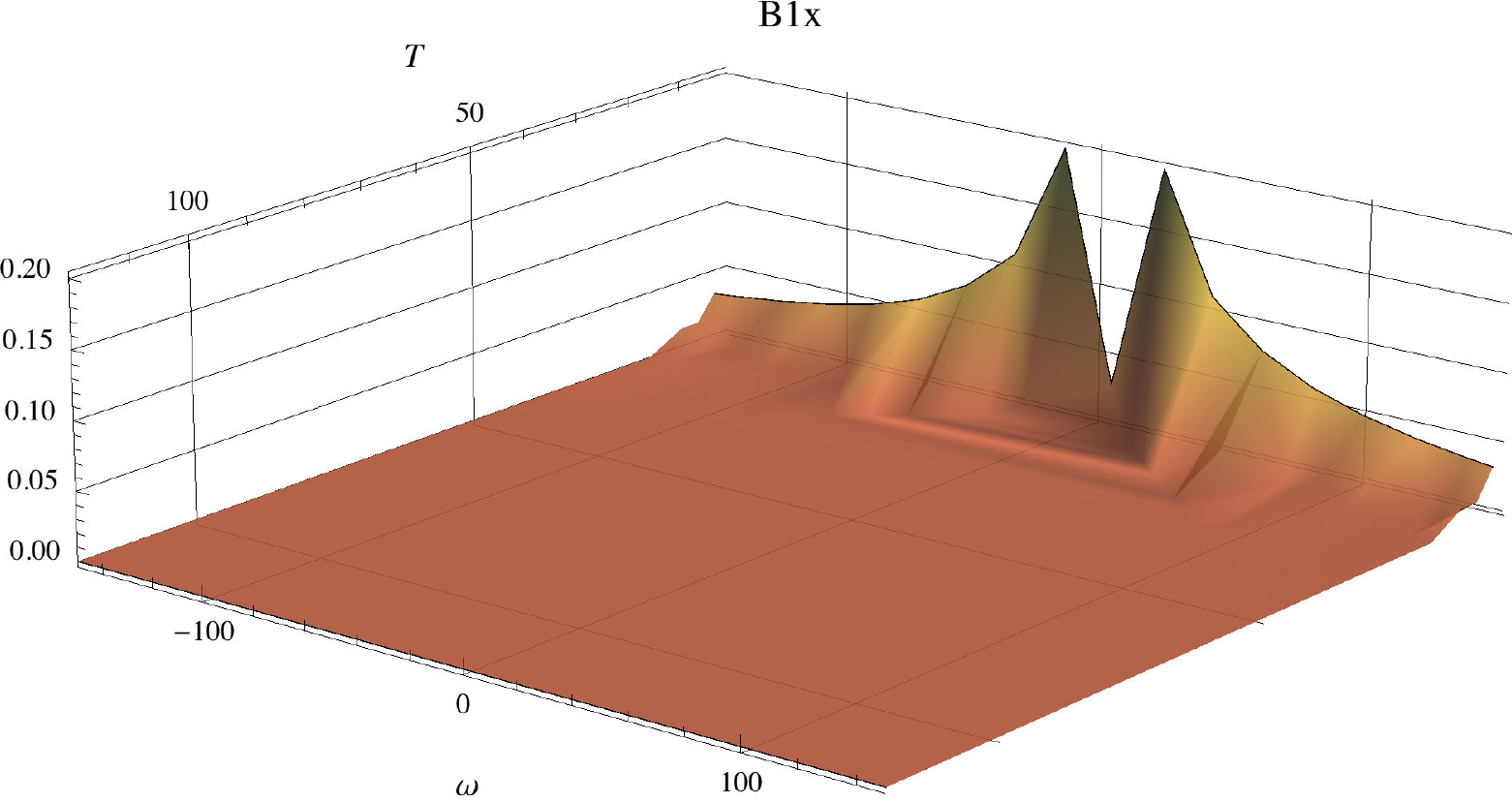} $a_{2}$) \includegraphics[scale=0.24]{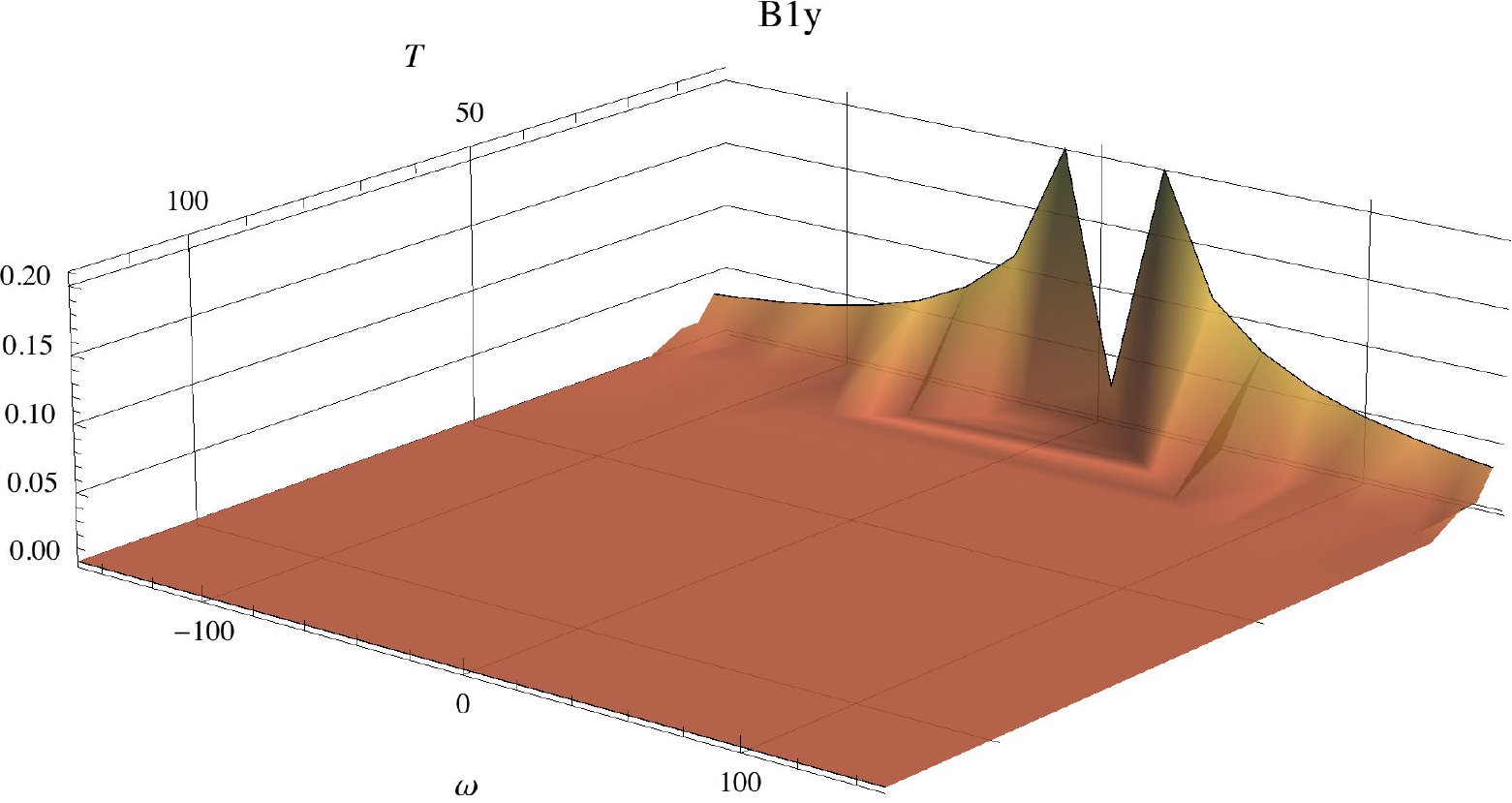}

$b_{1}$) \includegraphics[scale=0.24]{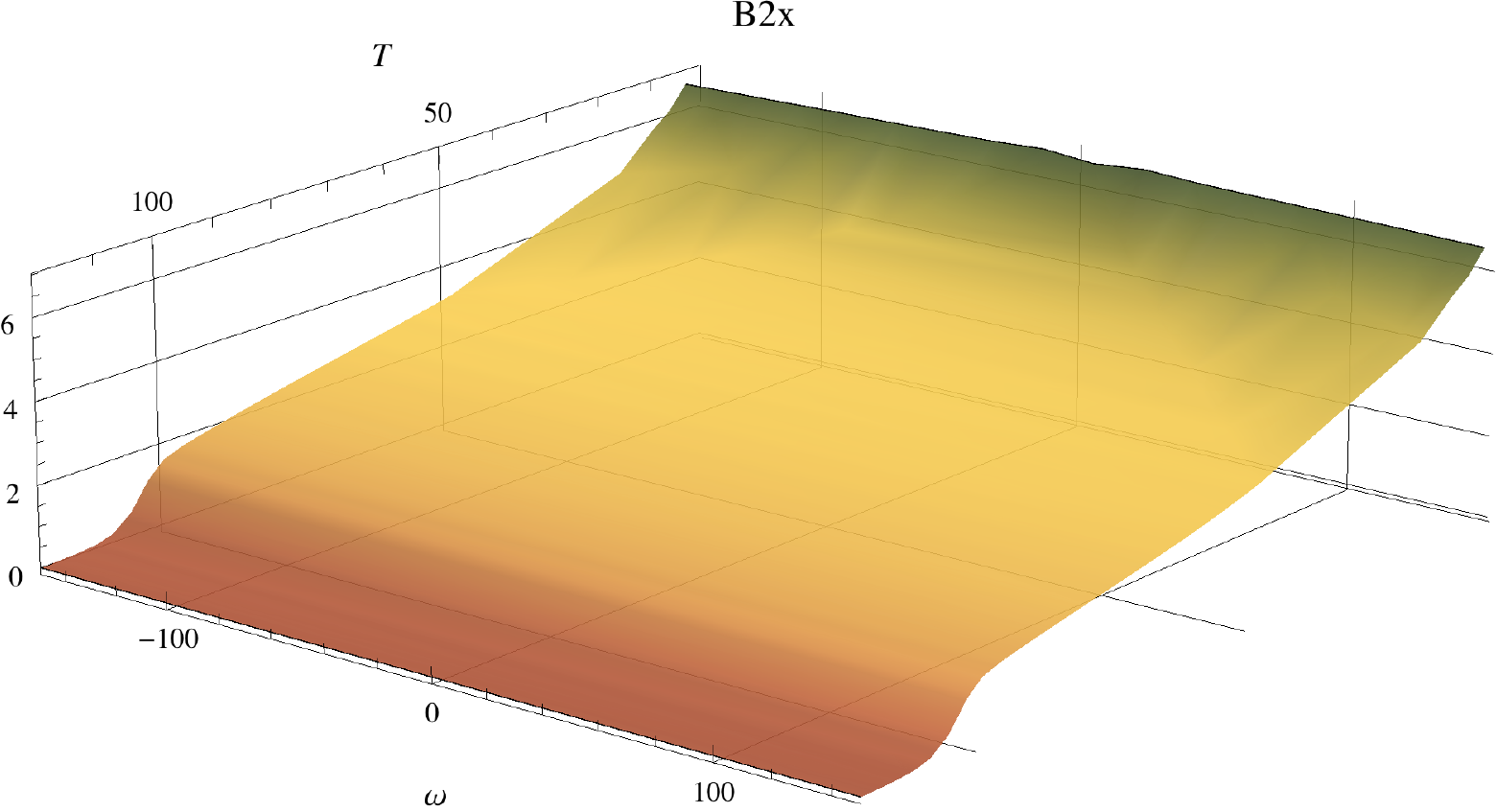} $b_{2}$) \includegraphics[scale=0.24]{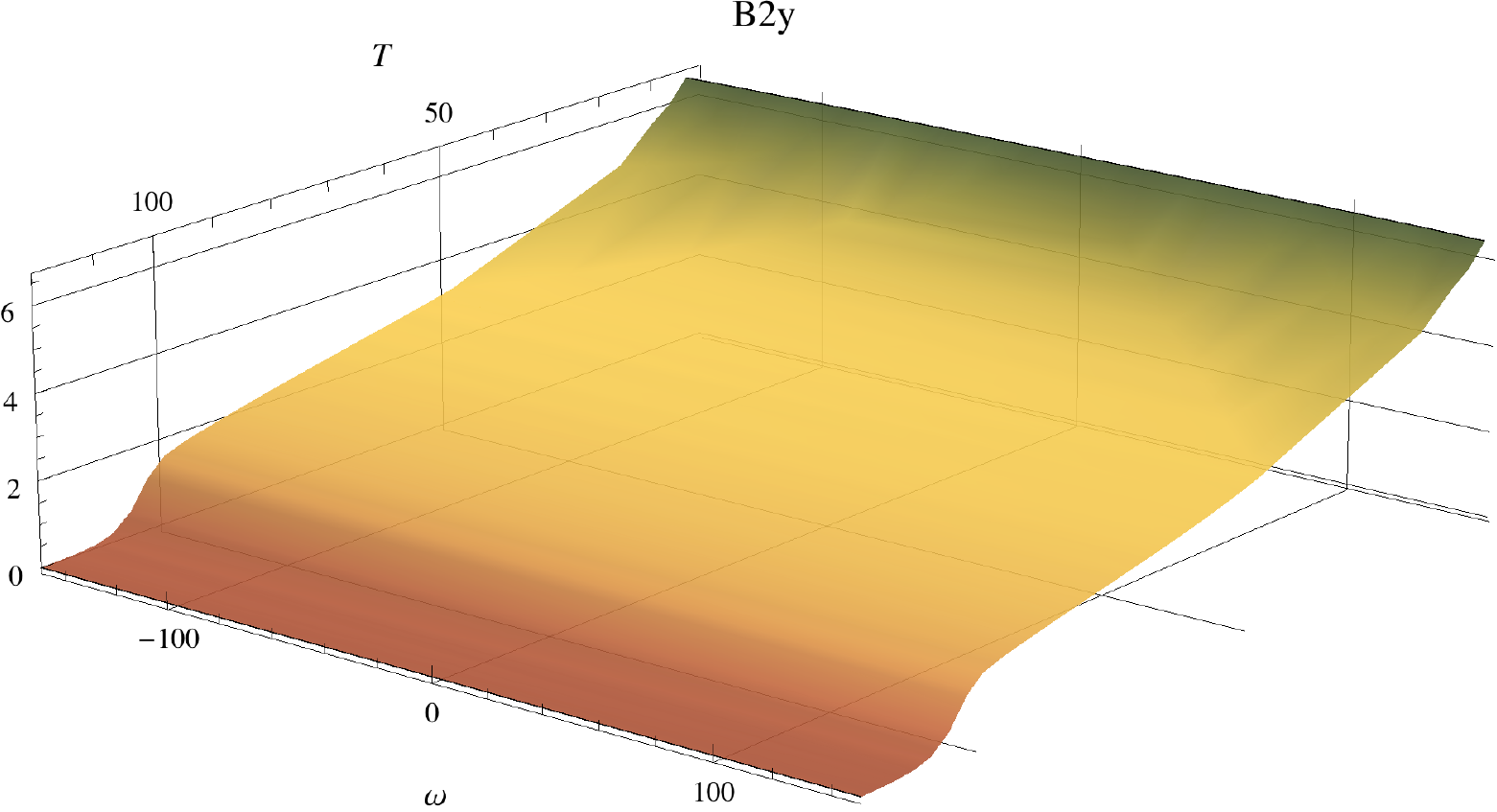}

$c_{1}$) \includegraphics[scale=0.24]{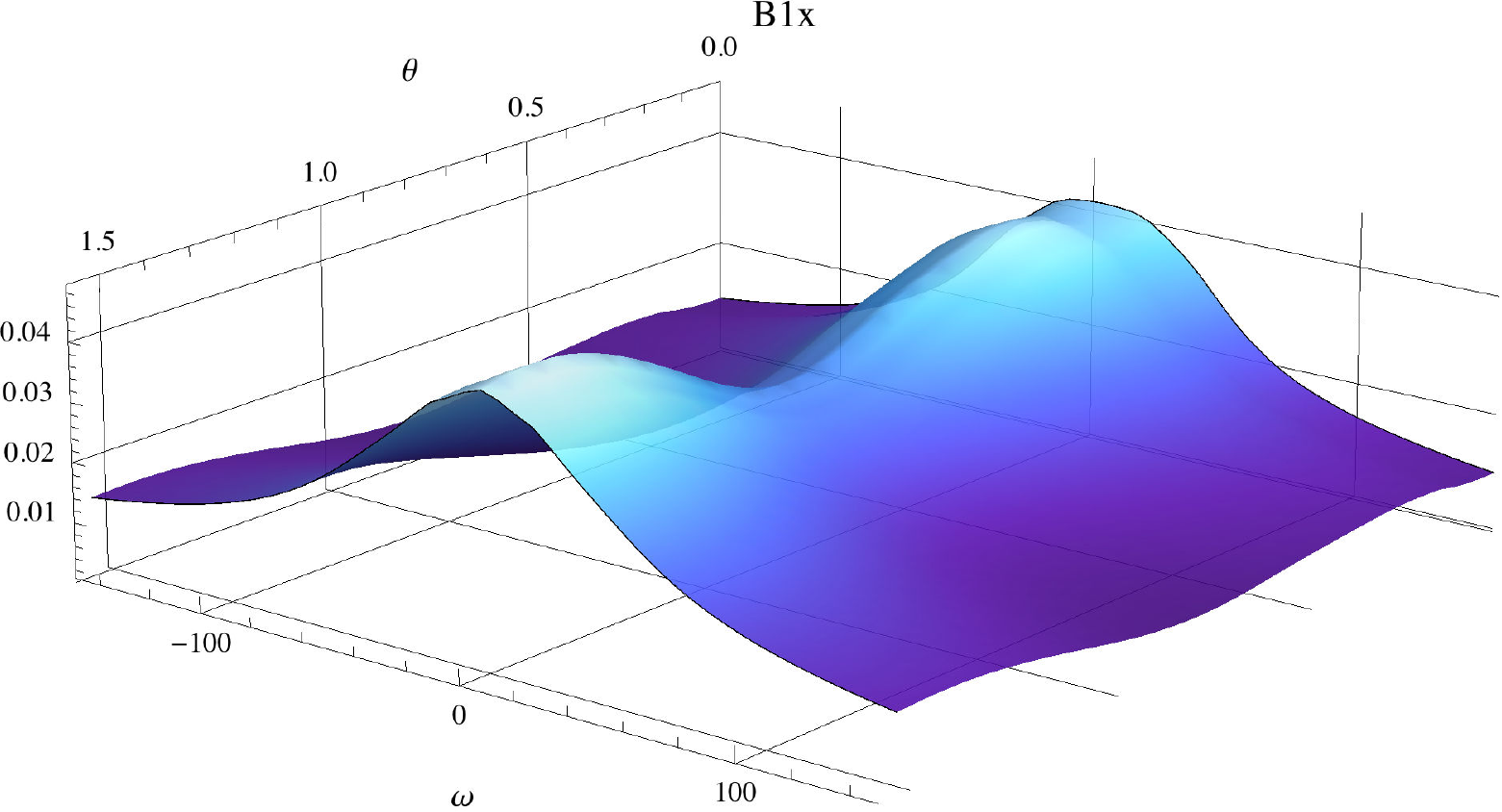} $c_{2}$) \includegraphics[scale=0.24]{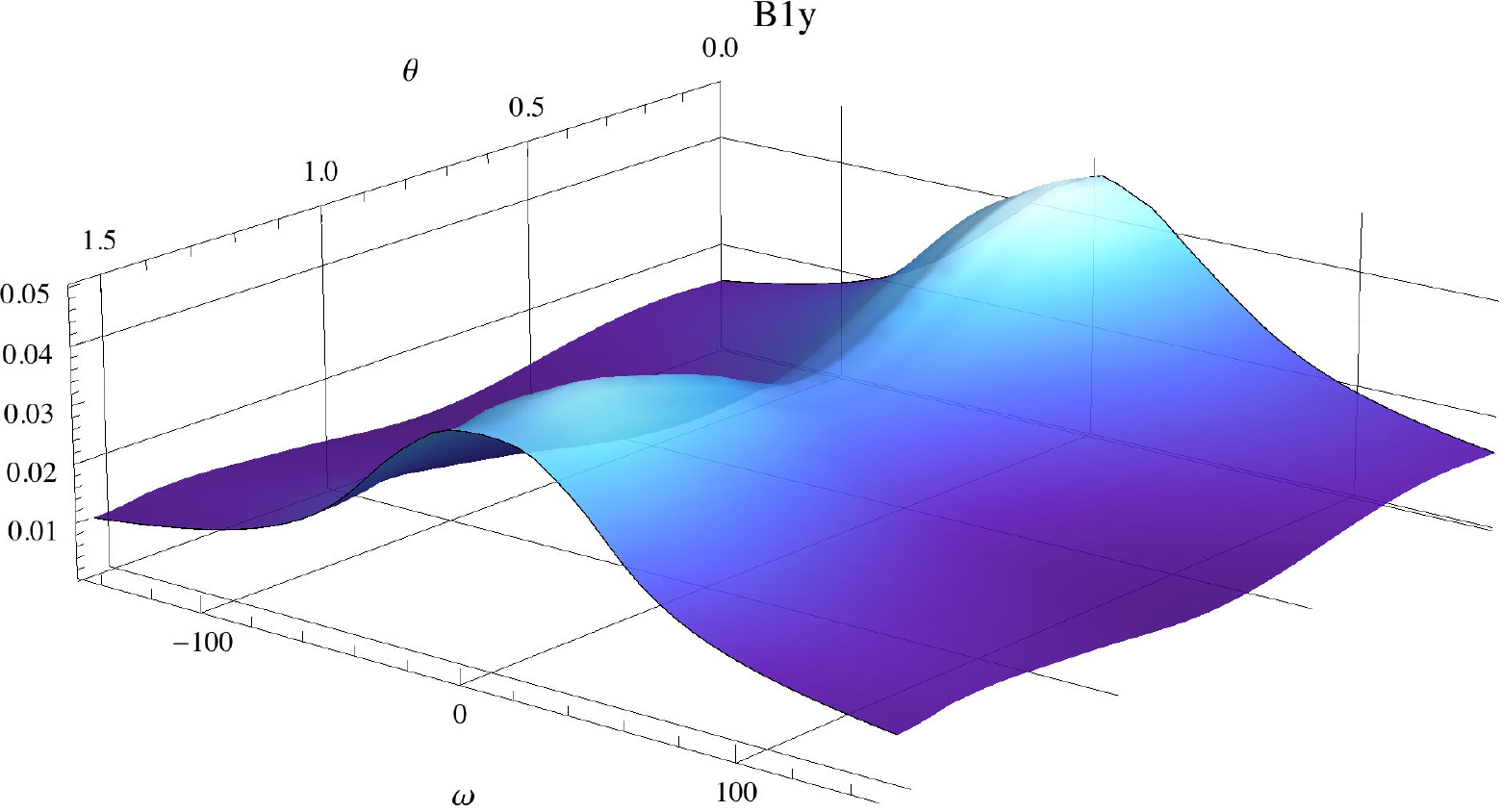}

$d_{1}$) \includegraphics[scale=0.24]{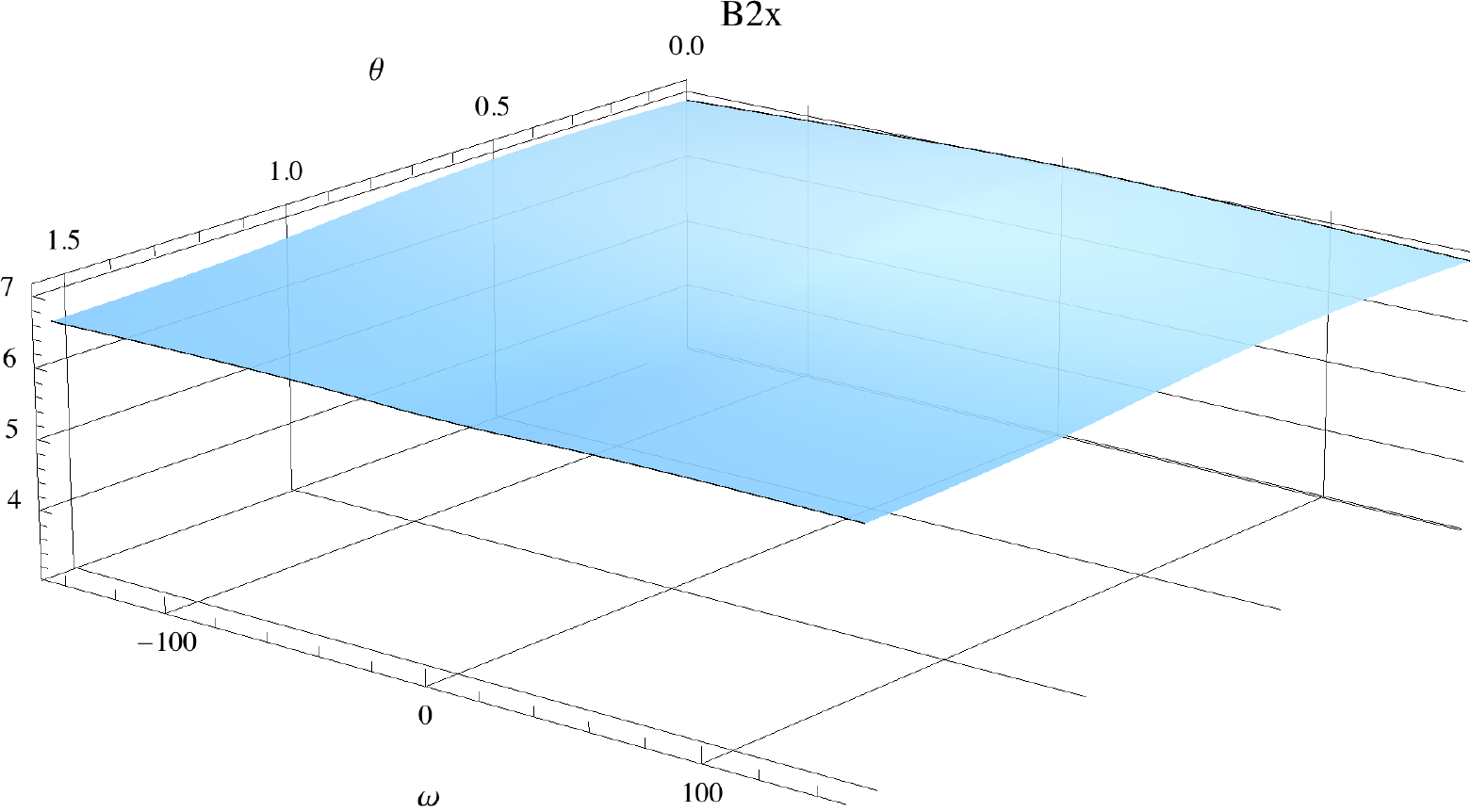} $d_{2}$) \includegraphics[scale=0.24]{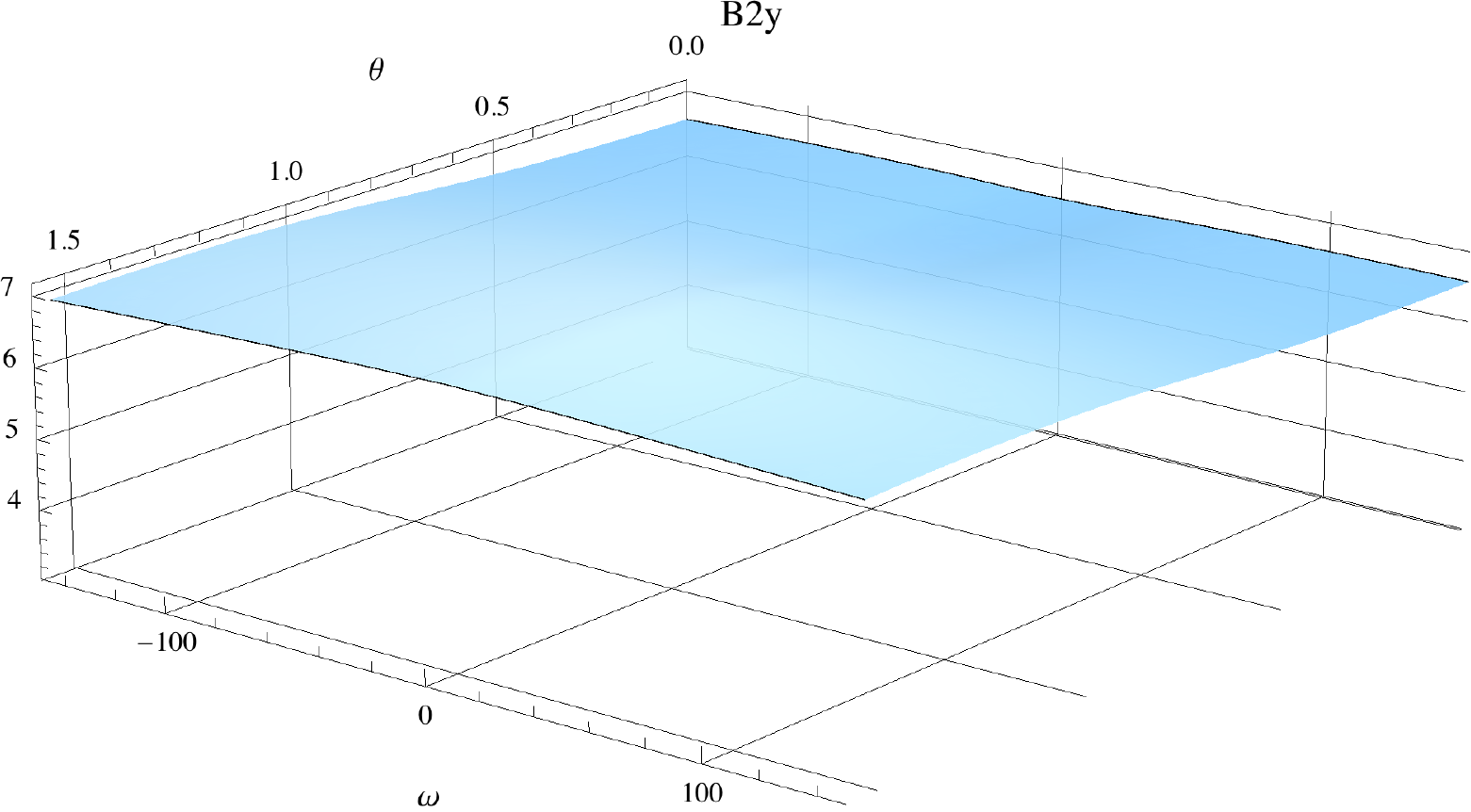}

\caption{\label{Fig8} (Color online) Typical solutions for the QDW/SC ($B_{1x}$
and $B_{1y}$) and CDW ($B_{2x}$ and $B_{2y}$ ) in the CE phase.
The values of the parameters are $g_{1}=20$, $g_{2}=200$,
$v=6$, $m_{a}=0.1$, $\gamma=3$, $W=2\pi$. The velocity angle is
$\theta=0.1$ for $a_{1}$), $a_{2}$), $b_{1}$) and $b_{2}$) whereas
the temperature is $T=1$ for $c_{1}$), $c_{2}$), $d_{1}$) and
$d_{2}$). Note the contrast with the CE solution of Fig.\ \ref{Fig2}.
Here the CDW/PDW component is one order of magnitude
larger than the QDW/SC component and the $\theta$ dependence of the
CDW/PDW component {[}$d_{1}$) and $d_{2}$){]}
is minimal compared to the one of the QDW/SC component {[}$c_{1})$
and $c_{2}$){]}.}
\end{figure}

The CE solution is depicted in Fig.\ \ref{Fig8}. The trend has been
inverted compared to Fig.\ \ref{Fig2}. Here the CDW/PDW 
component of the solution {[}$b_{1})$and $b_{2})$ in Fig.\ \ref{Fig8}{]}
is one order of magnitude larger than the pure QDW/SC component {[}$a_{1})$
and $a_{2}$){]} , in proportion to $J_{2}/J_{1}$.
One can also note the pronounced $\theta$- dependence of the QDW/SC
solution {[}$c_{1}$) and $c_{2}$){]} compared to the CDW/PDW
component {[}$d_{1}$) and $d_{2}$){]}. This seems to confirm the
intuition of Ref.\ [\onlinecite{Wang14}] that it is possible to stabilize the
CDW/PDW solution compared to the pure QDW/SC solution.
The price to be paid however is to enforce $J_{2}\gg J_{1}$
to such an extend that seems rather artificial for high $T_{c}$-cuprates.
Moreover, even when a giant CDW/PDW solution is stabilized,
we observed a re-entrance of the QDW/SC component at lower temperatures.
The conclusion is that it is very difficult to get rid totally of
the QDW/SC component.

\subsection{Stability conditions}

\begin{figure}[H]
$a_{1}$) \includegraphics[scale=0.24]{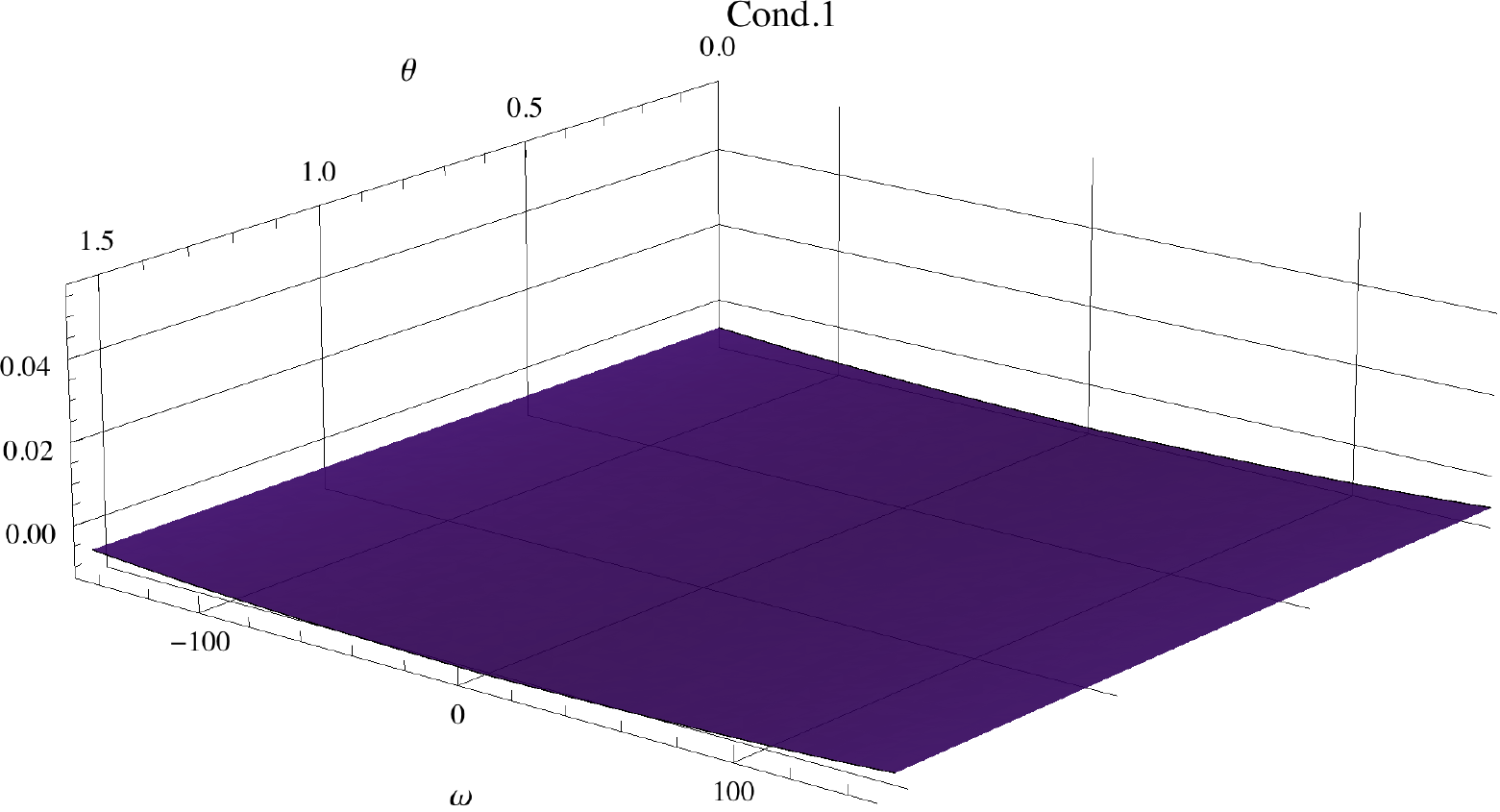} 
$a_{2}$) \includegraphics[scale=0.24]{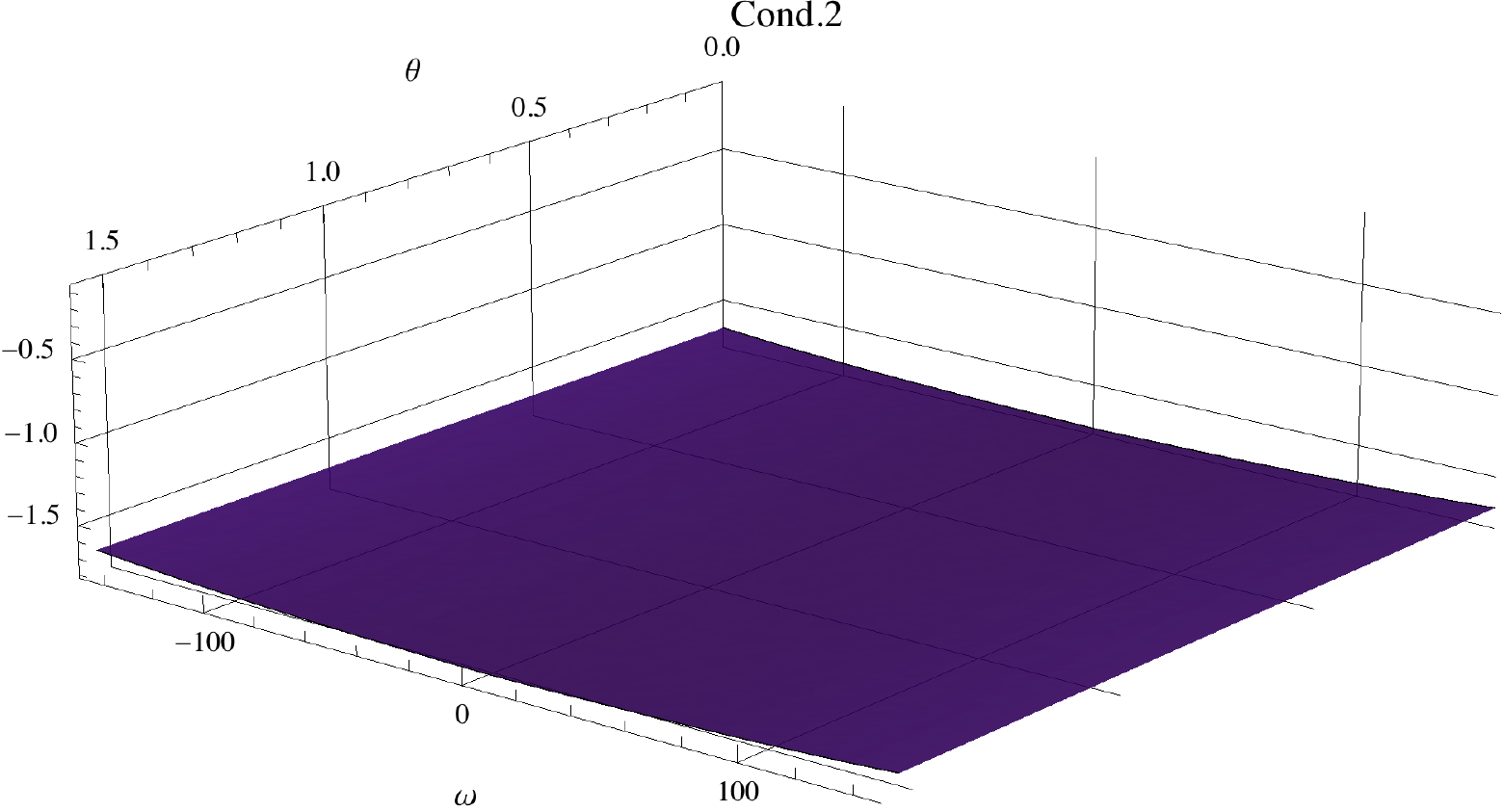}

\caption{\label{Fig11} (Color online) Stability conditions (\ref{cond0r})
for the pure QDW/SC solution as a function of $\left(\epsilon_{n},\theta\right)$
for $T=1K$. $a_{1}$) {[}dir. $B_{1}${]} and $a_{2}$) {[}dir. $B_{2}${]}.
Note that although the limit $J_{2}\gg J_{1}$ is
very defavorable to this solution, it is still stable.}
\end{figure}

We give now the stability conditions for the various solution in the
limit $J_{2}\gg J_{1}$, in analogy with Fig.\ \ref{Fig4}.
The results of this investigation are quite unexpected. Although the
limit $J_{2}\gg J_{1}$ is extremely favorable to
the pure CDW/PDW solution, we can see that at low
temperatures (here the study is made at $T=1K$) this solution becomes
unstable in the direction of the QDW/SC {[}dir. $B_{1}$ in these
notations{]}, indicating an instability towards co-existence at low
temperature. This observation corroborates the results of section
\ref{sec:Structure-of-the} where it was concluded that it is very
difficult to get rid completely of the QDW/SC solution. Notice that
the pure QDW/SC solution (Fig.\ \ref{Fig11}) is now stable in one
direction but becomes unstable in the direction of the CDW/PDW
solution, due to the favorable ratio $J_{2}/J_{1}\gg1$.
Lastly, the CE solution becomes stable in the two directions (Fig.
\ref{Fig13}).

\begin{figure}[H]
$a_{1}$) \includegraphics[scale=0.24]{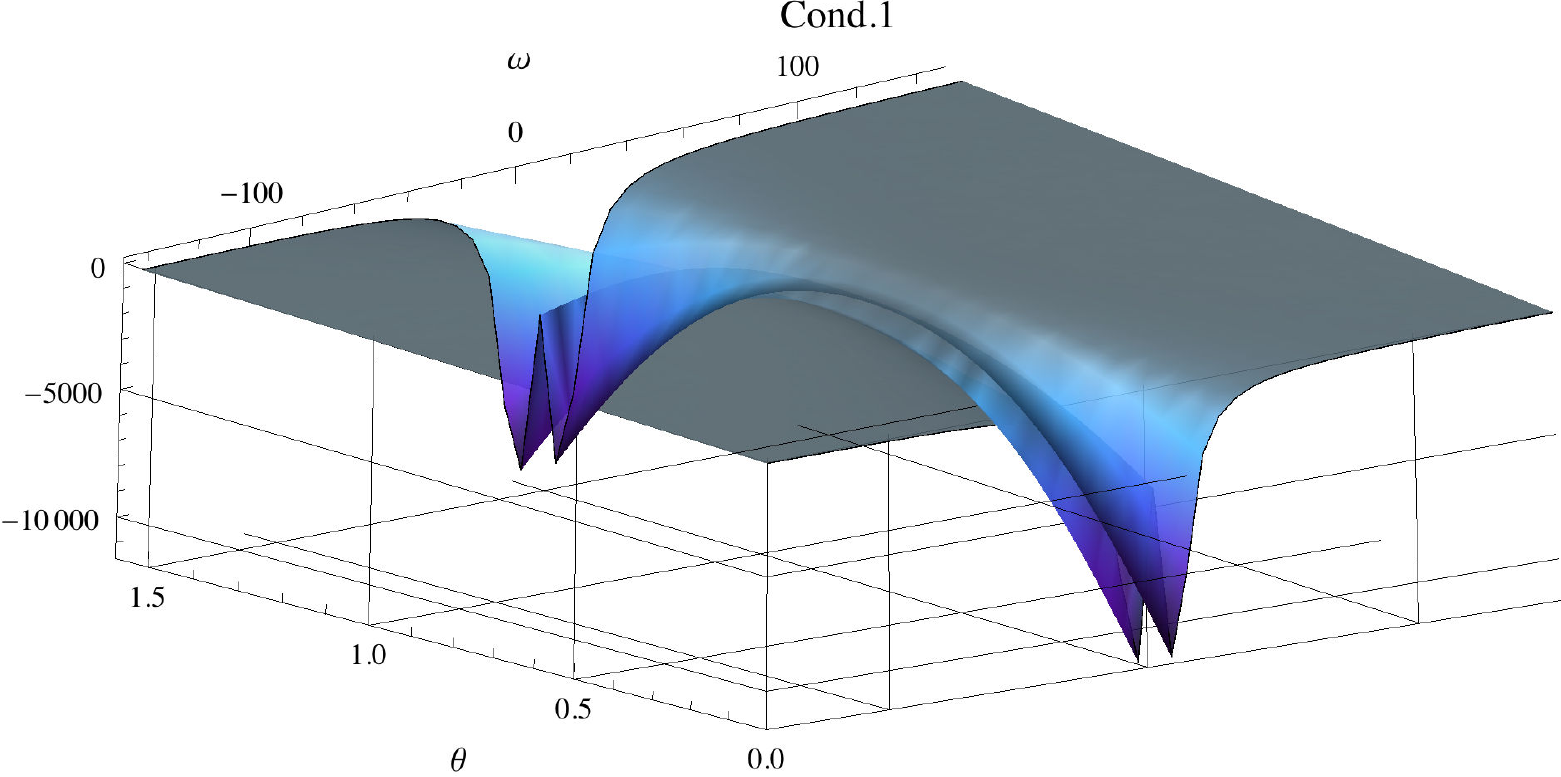} 
$a_{2}$) \includegraphics[scale=0.24]{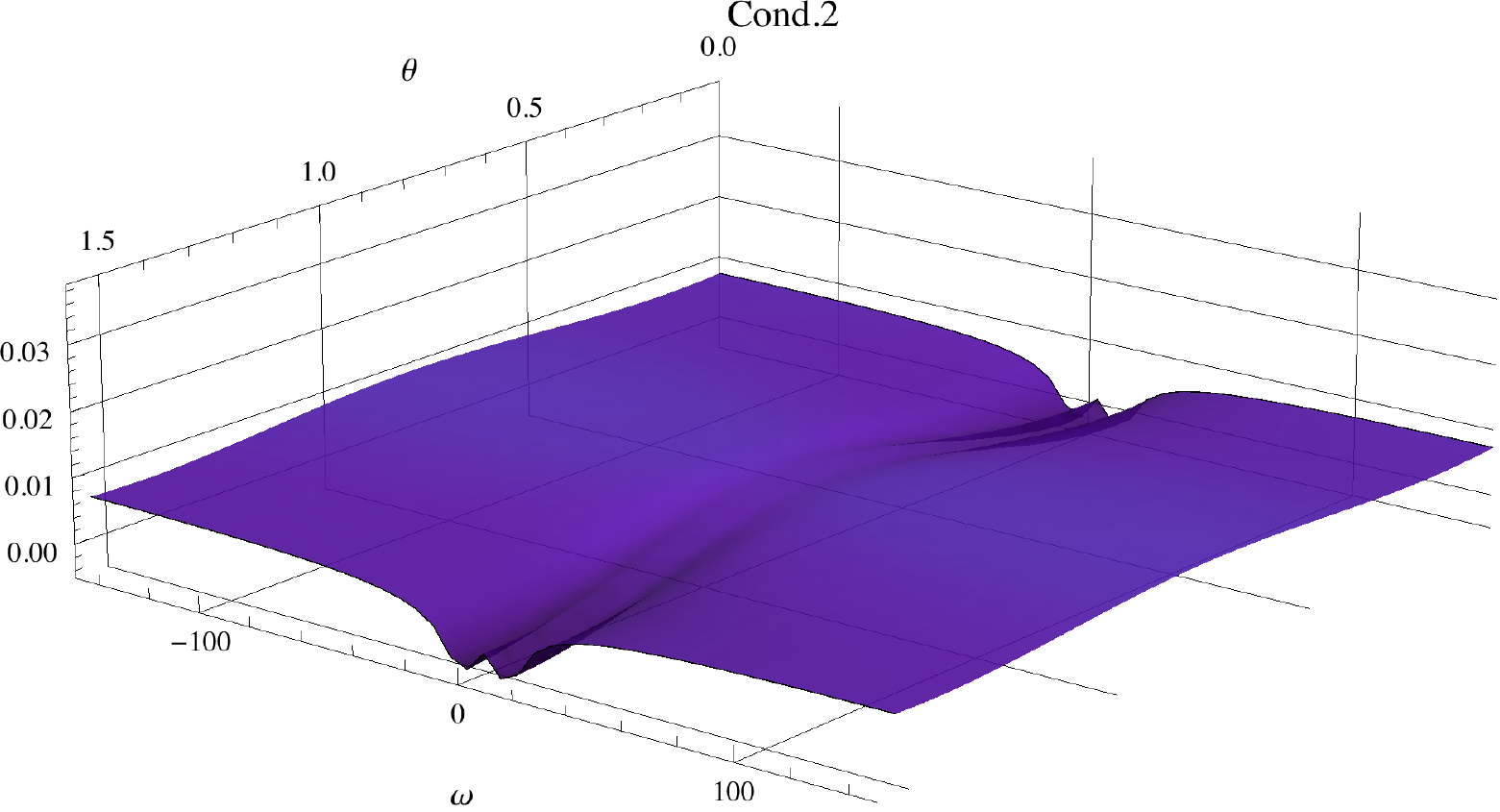}

\caption{\label{Fig12} (Color online) Stability conditions (\ref{cond0r})
for the pure CDW/PDW solution as a function of $\left(\epsilon_{n},\theta\right)$
for $T=1K$. $a_{1}$) {[}dir. $B_{1}${]} and $a_{2}$) {[}dir. $B_{2}${]}.
Note that although the limit $J_{2}\gg J_{1}$ is
energetically very favorable to this solution, there is a direction
of instability {[}dir. $B_{1}${]} at low temperature, indicating
an instability at lower temperatures towards the CE solution.}
\end{figure}

\begin{figure}[H]
$a_{1}$) \includegraphics[scale=0.24]{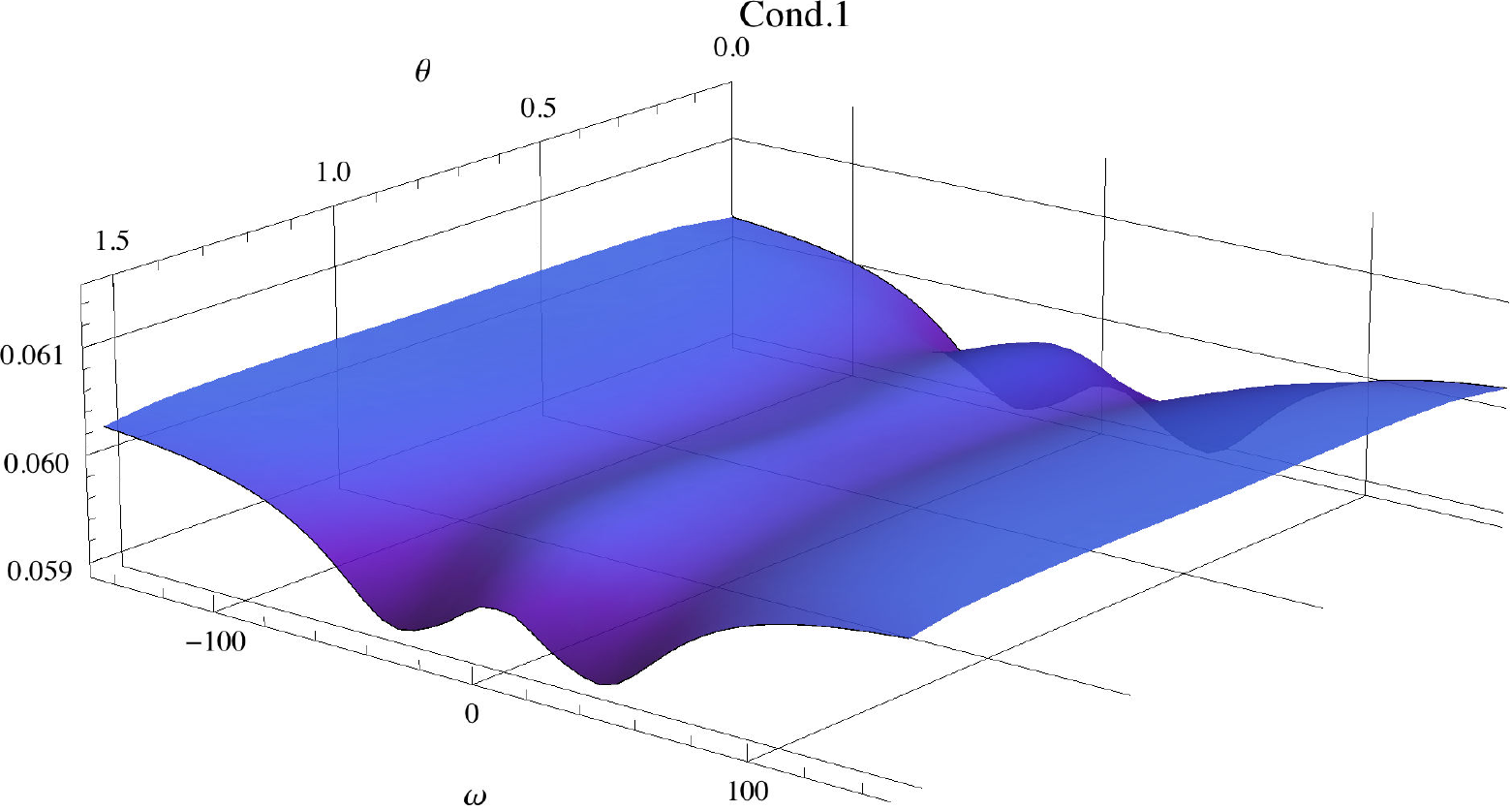} 
$a_{2}$) \includegraphics[scale=0.24]{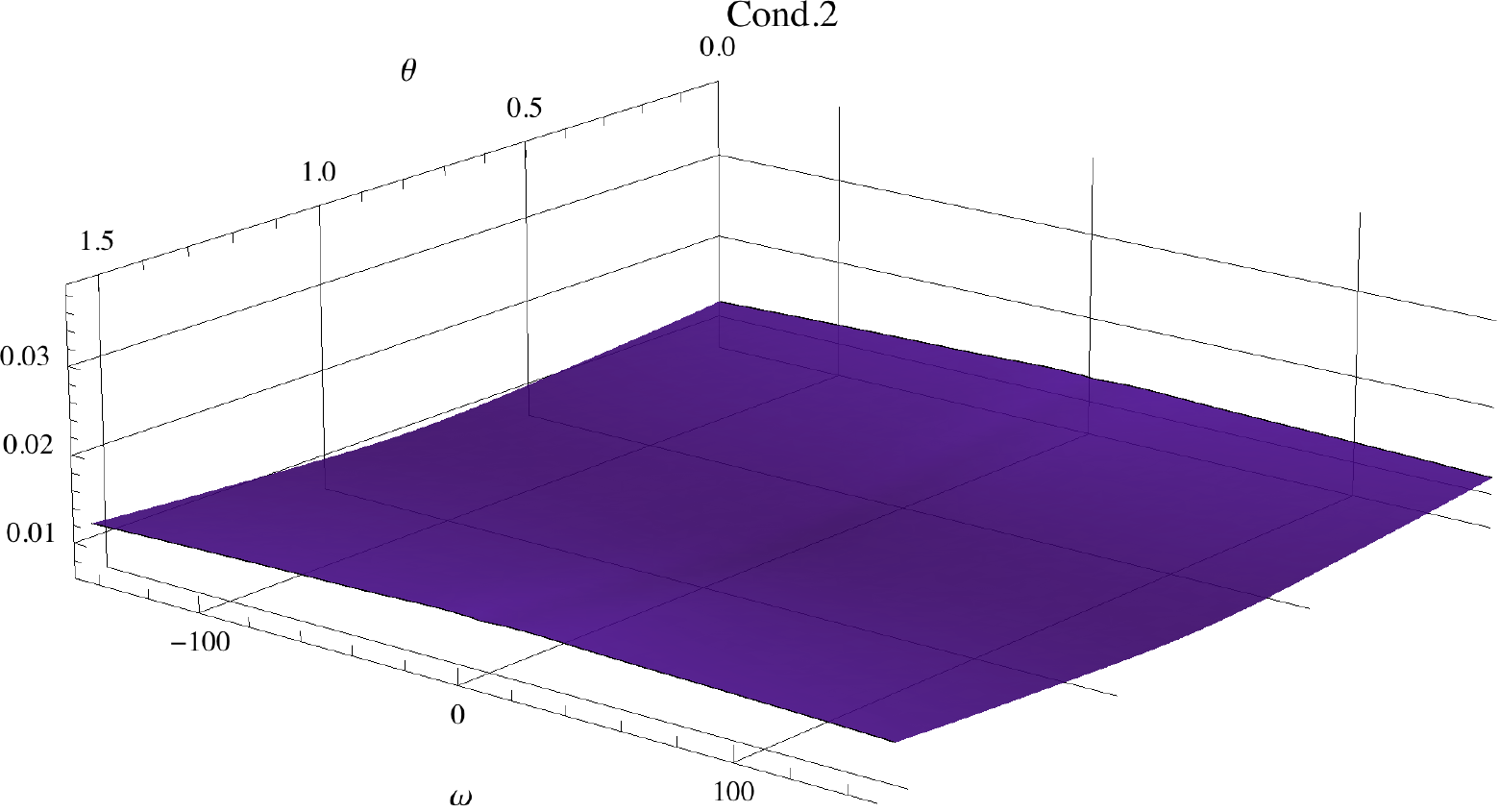}
\caption{\label{Fig13} (Color online) Stability conditions (\ref{cond0r})
for the pure CE solution as a function of $\left(\epsilon_{n},\theta\right)$
for $T=1K$. $a_{1}$) {[}dir. $B_{1}${]} and $a_{2}$) {[}dir. $B_{2}${]}.
Note that the CE solution is stable in both directions at low temperatures.}
\end{figure}

\subsection{Free energy}

\begin{figure}[h]
\includegraphics[scale=0.45]{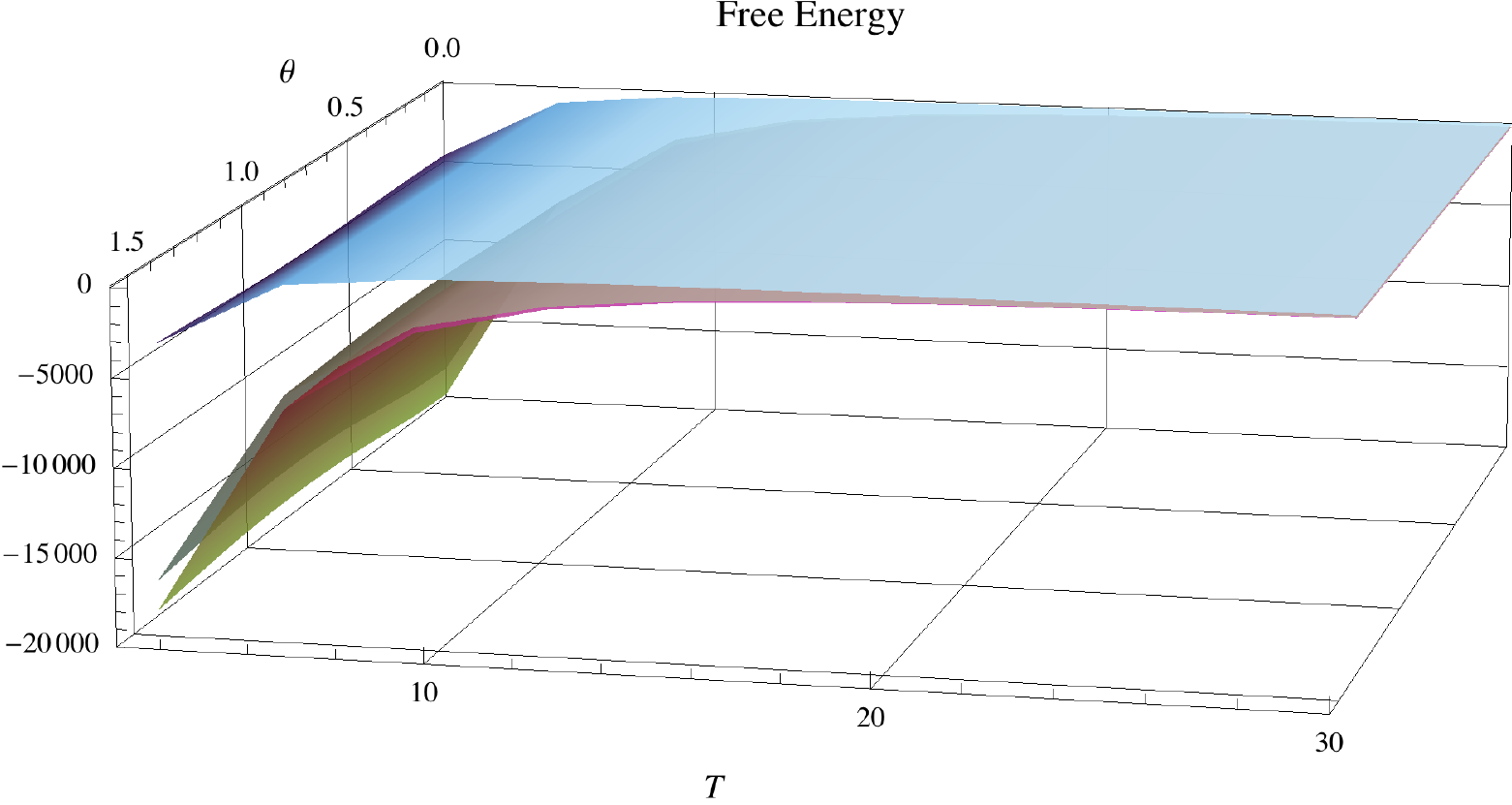}

\caption{\label{Fig14} (Color online) Free energy at of the three MF solutions.
Pure CDW/PDW (brown), pure QDW/SC (dark blue) and
CE (neon colors). The values of the parameters are $g_{1}=20$,
$g_{2}=200$, $v=6$, $m_{a}=0.1$, $\gamma=3$,
$W=2\pi$ and $\theta=0.1$.}
\end{figure}

Lastly we turn to the comparison of the free energy for the three
solution in the limit $J_{2}\gg J_{1}$. The result
is shown in Fig.\ \ref{Fig14}. By comparison with Fig.\ \ref{Fig5 }
we see that the energy of the pure QDW/SC solution is now higher than
the one of the pure CDW/PDW solution. The co-existence
solution, however, is always the lowest one, although quite close
in energy to the pure CDW/PDW solution. This gives
support to our conclusion that the system is in fine unstable towards
the CE solution.

\clearpage{}

\bibliography{Cuprates}
 \bibliographystyle{apsrev4-1} 

\end{document}